\documentclass[letterpaper,11pt]{article}
\usepackage{amsthm,amsmath,amssymb,setspace}
\usepackage[dvips]{graphicx}
\usepackage[letterpaper,nohead,left=1.25in,right=1.25in,top=1.0in,bottom=1.25in ]{geometry}
\spacing{1.12} \oddsidemargin=0.0cm \evensidemargin=0.0cm \textwidth=16.2cm \textheight=22cm

\begin{document}

\def\color#1#2{\mbox{Color}_{#1}\left[ #2 \right]} \def\vector#1#2{\mbox{Vec}_{#1}\left[ #2 \right]}
\def\size#1{\mbox{\rm Size}\left[#1\right]} \def\bdry#1{\mathop{\mathbb{\partial}}\left(#1  \right)}
\def\vs#1#2#3{#1_{#2},\dots,#1_{#3}} \def\sizeof#1{\left|#1  \right|}
\def\setof#1{\left\{{\let\st\colon #1 }\right\}}
\def\pnorm#1#2{\left\| #2 \right\|_{#1}}
\def\orig#1{\bar{#1}} \def\calF{\mathcal{F}} \def\calG{\mathcal{G}} \def\calD{\mathcal{D}}
\def\calC{\mathcal{C}} \def\calA{\mathcal{C}_A} \def\calI{\mathcal{C}_I} \def\calS{\mathcal{S}}
\def\calP{\mathcal{P}} \def\calT{\mathcal{T}}

\def\smoothed#1#2{\mbox{Smoothed}_{#1}\left[#2 \right]} \def\Reals#1{\mathbb{R}^{#1}}
\def\expec#1#2{\mbox{\rm E}_{#1}\left[ #2 \right]} \def\orig#1{\bar{#1}}
\def\form#1#2{\left\langle#1 | #2 \right\rangle} \def\xxi{\mathbf{\xi}}

\def\smoothed#1#2{\mbox{Smoothed}_{#1}\left[#2 \right]}
\def\expec#1#2{\mbox{\rm E}_{#1}\left[ #2 \right]} \def\BOO={=^{\hspace{0.06cm}\epsilon}_B}

\def\oxx{\overline{\xx}} \def\oyy{\overline{\yy}}
\def\oxxc{\overline{\xx}_C} \def\oyyc{\overline{\yy}_C}
\def\00{\mathbf{0}}
\def\aa{\mathbf{a}} \def\bb{\mathbf{b}} \def\cc{\mathbf{c}} \def\ee{\mathbf{e}}
\def\lll{\mathbf{l}}
\def\uu{\mathbf{u}} \def\vv{\mathbf{v}} \def\xx{\mathbf{x}} \def\yy{\mathbf{y}}
\def\AA{\mathbf{A}} \def\BB{\mathbf{B}} \def\LL{\mathbf{L}} \def\UU{\mathbf{U}}
\def\pp{\mathbf{p}} \def\qq{\mathbf{q}} \def\rr{\mathbf{r}} \def\ss{\mathbf{s}}
\def\ttt{\mathbf{t}} \def\xx{\mathbf{x}} \def\yy{\mathbf{y}} \def\ZZ{\mathbb{Z}}
\def\zz{\mathbf{z}} \def\LL{\mathbf{L}} \def\RR{\mathbf{R}} \def\MM{\mathbf{M}}
\def\NN{\mathbf{R}} \def\calG{\mathcal{G}} \def\calA{\mathcal{C}_A} \def\calI{\mathcal{C}_I}
\def\calS{\mathcal{S}} \def\calP{\mathcal{P}} \def\calC{\mathcal{C}} \def\symP{\mathbb{P}}
\def\calL{\mathcal{L}}

\newtheorem{theo}{Theorem}[section] \newtheorem{coro}[theo]{Corollary} \newtheorem{defi}[theo]{Definition}
\newtheorem{exam}[theo]{Example} \newtheorem{lemm}[theo]{Lemma} \newtheorem{prop}[theo]{Proposition}
\newtheorem{prope}{Property} \newtheorem{remar}[theo]{Remark} \newtheorem{notat}[theo]{Notation}
\newtheorem{conj}{Conjecture} \newtheorem{stat}{Statement}

\title{\rm Settling the Complexity of Computing Two-Player Nash~Equilibria }
\author{
    Xi Chen\thanks{Department of Computer Science, Tsinghua University, Beijing, P.R.China.
    \texttt{email:\hspace{0.05cm}csxichen@gmail.com}}
    \and
    Xiaotie Deng\thanks{Department of Computer Science, City University
    of Hong Kong, Hong Kong SAR, P.R.\hspace{0.08cm}China. \texttt{email: deng@cs.cityu.edu.hk}}
\and
    Shang-Hua Teng\thanks{Department of Computer Science, Boston
University, Boston and Akamai Technologies Inc., Cambridge, MA, USA.
    \texttt{email:\hspace{0.05cm}steng@cs.bu.edu}}}
\date{}\maketitle\vspace{0.2cm}

\begin{abstract}
We settle a long-standing open question in algorithmic game theory.
We prove that {\sc Bimatrix}, the problem of finding a Nash equilibrium
  in a two-player game, is complete for
  the complexity class \textbf{PPAD} (\hspace{0.03cm}Polynomial Parity
  Argument, Directed version\hspace{0.03cm}) introduced by Papadimitriou in 1991.

This is the first of
   a series of results concerning the complexity
   of
   Nash equilibria. In particular, we prove the following theorems:
\vspace{0.15cm}
\begin{itemize}
\item {\sc Bimatrix}
   does not have a fully polynomial-time approximation
   scheme unless every problem in \textbf{PPAD}
   is solvable in polynomial time.

\item The smoothed complexity of the classic Lemke-Howson
  algorithm and, in fact, of
   any algorithm for {\sc Bimatrix} is not polynomial
   unless every problem in $\textbf{PPAD}$ is solvable
   in randomized polynomial time.\vspace{0.15cm}
 \end{itemize}

 Our results demonstrate that, even
   in the simplest form of non-cooperative games,
  equilibrium computation and approximation are polynomial-time equivalent
  to fixed point computation.
Our results also have two broad complexity
  implications in mathematical
  economics and operations research:
\begin{itemize}\vspace{0.15cm}
  \item Arrow-Debreu market equilibria are \textbf{PPAD}-hard to compute.

\item The P-Matrix Linear Complementary Problem
    is computationally harder
    than convex programming unless
    every problem in \textbf{PPAD} is solvable in polynomial time.
\end{itemize}
\end{abstract}

\newpage

\section{Introduction}


In 1944, Morgenstern and von Neumann
  \cite{MOR47} initiated the study of game theory
  and its applications
  to economic behavior.
At the center of their study
  was von Neumann's minimax equilibrium solution for two-player
   zero-sum games~\cite{vN1928}.
In a two-player zero-sum game,  one player's gain is equal
  to the loss of the other.
They observed that any general $n$-player (non-zero-sum) game
  can be reduced to an $(n+1)$-player zero-sum game.
Their work went on to introduce the notion of cooperative games
  and the solution concept of stable sets.



In 1950, following the original spirit of Morgenstern and
   von Neumann's work on two-player zero-sum games,
    Nash \cite{NashNonCooperative,NAS50} 
  formulated a solution concept for
  non-cooperative games among multiple players.
In a non-cooperative game, the zero-sum condition is relaxed
  and no communication and coalition among players are allowed.
Building on the notion of mixed strategies of~\cite{vN1928},
  the solution concept, now commonly
  referred to as the {\em Nash equilibrium},
  captures the notion of the individual rationality of players
  at an equilibrium point.
In a Nash equilibrium, each player's strategy
  is a best response to other players' strategies.
Nash proved that every $n$-player, finite, non-cooperative game
  has an equilibrium point.
His original proof
  \cite{NashNonCooperative,Leonard}
  was based on Brouwer's Fixed Point Theorem~\cite{BRO10}.
David Gale suggested the use of Kakutani's Fixed Point
  Theorem~\cite{KAKU} to simplify the proof.
Mathematically, von Neumann's Minimax Theorem for
  two-player zero-sum games can be
  proved by linear programming duality.
In contrast, the fixed point approach to Nash's
  Equilibrium Theorem seems to be necessary:
  even for two-player non-cooperative games, linear programming duality
  is no longer applicable.

Nash's equilibrium concept has had a tremendous influence on economics,
 as well as in other social and
  natural science disciplines \cite{NashSocial}.
Nash's approach to non-cooperative games
   has played an essential role in shaping
   mathematical economics, which consider agents with
   competing individual interests.
A few years after Nash's work, Arrow and
  Debreu~\cite{AD}, also applying fixed point theorems,
  proved a general existence theorem for market equilibria.
Since then, various forms of equilibrium theorems
  have been established via fixed point theorems.

However, the existence proofs based on fixed point theorems do
  not usually lead to efficient algorithms for
  finding equilibria.
In fact, in spite of many remarkable breakthroughs
  in algorithmic game theory and  mathematical programming,
  answers to several fundamental questions
  about the computation of Nash and Arrow-Debreu
  equilibria remain elusive.
The most notable open problem is that of deciding
  whether the problem of finding an equilibrium point in a two-player
  game is solvable in polynomial time.

In this paper, we settle the complexity of
  computing a two-player Nash equilibrium
 and answer two central questions regarding the approximation
  and smoothed complexity of this game theoretic problem.
In the next few subsections, we will review previous
  work on the computation of Nash equilibria, state our main results,
  and discuss their extensions to the computation of market equilibria.

\subsection{Finite-Step Equilibrium Algorithms}

Since Nash and Arrow-Debreu's pioneering work,
   great progress has been made
  in the effort to find constructive and algorithmic
  proofs of equilibrium theorems.
The advances for equilibrium computation
  can be chronologically classified according to
  the following two periods:
\begin{itemize}
\item {\bf Finite-step period}: In this period,
  the main objective was to design
  equilibrium algorithms that terminate in
  a finite number of steps and to understand for which
  equilibrium problems finite-step algorithms do not exist.
\item {\bf Polynomial-time period}: In this
  period, the main objective has been to develop
  polynomial-time algorithms for computing equilibria and
  to characterize the complexity of equilibrium computation.
\end{itemize}

The duality-based proof of the minimax theorem
  leads to a linear programming formulation 
  of the problem of finding an equilibrium in a two-player zero-sum
  game.
One can apply the simplex algorithm, in a finite number of steps
   in the Turing model\footnote{The simplex algorithm also terminates
   in a finite number of steps in various computational
   models invol\-ving real numbers, such as the Blum-Shub-Smale model
  \cite{BSS}.},
   to compute an equilibrium in a two-player zero-sum game with
   rational payoffs.
A decade or so after Nash's seminal work,
    Lemke and Howson~\cite{LemkeHowson}
    developed a path-following, simplex-like
    algorithm for
    finding a Nash equilibrium in a general two-player game.
Like the simplex algorithm,
  their algorithm terminates in a finite number of
  steps for a two-player game with rational payoffs.

The Lemke-Howson algorithm has been extended to
  non-cooperative games with more than two players \cite{Wilson}.
However, due to Nash's observation that
   there are rational three-player games all of whose
   equilibria are irrational,
   finite-step algorithms become harder to obtain for games with
   three or more players.
For those multi-player games, no finite-step algorithm exists
  in the classical Turing model.

Similarly, some exchange economies do not have
  any rational Arrow-Debreu equilibria.
The absence of a rational equilibrium underscores the continuous
  nature of equilibrium computation.
Brouwer's Fixed Point Theorem --- that any continuous map $f$ from
  a convex compact body, such as a simplex or a hypercube, to itself has a
  fixed point --- is inherently continuous.
Mathematically, the continuous nature does not hinder the
  definition of search problems for finding equilibria and fixed points.
But to measure the computational complexity of these continuous
  problems in the classical Turing model,
  some imprecision or inaccuracy must be introduced to ensure the
  existence of a solution with a finite description
  \cite{Scarf,Scarfprice,PAP91,HPV,DPS}.
For example, one possible definition of an approximate fixed point
  of a continuous map $f$ is a point
  $\xx$ in the convex body such that $||f(\xx)-\xx|| \leq \epsilon$
  for a given $\epsilon> 0$ \cite{Scarf}.

In 1928, Sperner \cite{SPE28} discovered a
  discrete fixed point theorem that led to one of the most elegant
  proofs of the Brouwer's Fixed Point Theorem.
Suppose that $\Omega$ is a $d$-dimensional simplex with vertices
  $v_1,v_2, ..., v_{d+1}$, and that
  ${\cal S}$ is a simplicial decomposition of $\Omega$.
Suppose $\Pi$ assigns to each vertex of ${\cal S}$ 
  a color from $\{1,2,...,d+1\}$
  such that, for every vertex 
  $v$ of ${\cal S}$, $\Pi(v) \neq i$ if the $i^{th}$
  component of the barycentric coordinate of $v$, in terms of
  $v_1,v_2, ..., v_{d+1}$, is 0.
Then, Sperner's Lemma asserts that there exists a
  simplex cell in ${\cal S}$ that contains all colors.
This fully-colored simplex cell is often referred to as a {\em panchromatric simplex}
  or a {\em Sperner simplex} of $({\cal S}, \Pi)$.
Consider a Brouwer map $f$ with Lipschitz constant $L$ over the simplex $\Omega$.
Suppose further that the diameter of each simplex cell in $\cal S$ is at most $\epsilon/L$.
Then, one can define a color assignment $\Pi_f$  such that
  each panchromatric simplex in $({\cal S}, \Pi_f)$ must have a vertex $\vv$ satisfying
  $||f(\vv)-\vv|| \leq \Theta(\epsilon)$.
Thus, a panchromatic simplex of $({\cal S}, \Pi_f)$
  can be viewed as an approximate, discrete fixed point of $f$.

Inspired by the Lemke-Howson algorithm,
  Scarf  developed a path-following algorithm, using
  simplicial subdivision, for computing approximate fixed points
  \cite{Scarf} and competitive
  equilibrium prices~\cite{Scarfprice}.
The path-following method has also had extensive
  applications to mathematical programming and has since grown
 into an algorithm-design paradigm in optimization and equilibrium analysis.
One can take a similar approximation approach to study the complexity of Nash equilibria,
  especially for games involving three or more players.

\subsection{Computational Complexity of Nash Equilibria}

Since 1960s, the theory of computation
  has shifted its focus from whether
  problems can be solved on a computer to
  how efficiently problems can be solved on a computer.
The field has gained maturity
  with rapid advances in algorithm design, algorithm analysis,
   and complexity theory.
Problems are categorized into complexity classes,
  capturing the potential difficulty
  of decision, search, and optimization problems.
The complexity classes \textbf{P}, \textbf{RP}, and \textbf{BPP},
and their search counterparts such as \textbf{FP},
  have become the standard classes
  for characterizing computational problems that are
  tractable\footnote{\textbf{FP} stands for Function Polynomial-Time.
In this paper, as we only consider search problems, without further notice,
  we will (ab)use \textbf{P}
  and \textbf{RP} to denote the classes
  of search problems that can be solved in
  polynomial time or in randomized polynomial-time, respectively. We believe
  doing so will help more general readers.
}.

The desire to find fast and polynomial-time algorithms for computing
equilibria has been greatly enhanced with the rise of the Internet
  \cite{PapadimitriouInternet}.
The rise has created
  a surge of human activities that make computation,
  communication and optimization of participating agents
  accessible at microeconomic levels.
Efficient computation is instrumental to
  support the basic operations, such as pricing, in
  this large scale on-line market~\cite{Sandholm2000}.
Many new game and economic problems have been introduced,
  and in the meantime, classical game and economic problems
  have become the subjects for active complexity studies
  \cite{PapadimitriouInternet}.
Algorithmic game theory has grown into a highly interdisciplinary field
  intersecting economics, mathematics, operations research,
  numerical analysis, and computer science.

In 1979,  Khachiyan made a ground-breaking
    discovery that the
    ellipsoid algorithm can solve a linear program
    in polynomial time~\cite{KHA79}.
Shortly after, Karmarkar improved the complexity for solving linear programming
  with his path-following,  interior-point algorithm \cite{Karmarkar}.
His work initiated the implementation of
  theoretically-sound linear programming algorithms.
Motivated by a grand challenge in Theory of Computing \cite{NSFWorkshop1999},
  Spielman and Teng \cite{SpielmanTengSimplex} introduced
  a new algorithm analysis framework, {\em smoothed analysis},
  based on perturbation theory,
  to provide rigorous complexity-theoretic justification for the good
  practical performance of the simplex algorithm.
They proved that although almost all known simplex algorithms have
    exponential worst-case complexity \cite{KleeMinty},
    the smoothed complexity of the simplex
    algorithm with the shadow-vertex pivoting rule is polynomial.
As a result of these developments in linear programming,
  equilibrium solutions of two-player zero-sum games
  can be found in polynomial time using the ellipsoid or interior-point
  algorithms and in smoothed polynomial time
  using the simplex algorithm.

However, 
  no polynomial-time algorithm has been found for
  computing discrete fixed points or approximate fixed points,
 rendering the
  equilibrium proofs based on fixed point theorems
  non-constructive in the view of polynomial-time computability.

The difficulty of discrete fixed point computation is partially
  justified in the query model.
In 1989, 
  Hirsch, Papadimitriou, and Vavasis \cite{HPV}
  proved an exponential lower bound on the number
  of function evaluations necessary to find a discrete fixed point,
  even in two dimensions, assuming algorithms only have a black-box access
  to the fixed point function.
Their bound has recently been improved \cite{XX05} and
  extended to the randomized query model \cite{ChenTengSTOC07}
  and to the quantum query model \cite{FISV, ChenTengSTOC07}.

Motivated by the pivoting structure used in the Lemke-Howson algorithm,
  Papadimitriou introduced the complexity class \textbf{PPAD} \cite{PAP91}.
\textbf{PPAD} is an abbreviation for
  {\em Polynomial Parity Argument in a Directed graph}.
He introduced several search problems concerning the computation
  of discrete fixed points.
For example, he defined the problem {\sc Sperner} to be the search problem
  of finding a Sperner simplex
  given a polynomial-sized circuit for assigning colors to a particular
   simplicial decomposition of a hypercube.
Extending the model of \cite{HPV}, he also defined
  a search problem for computing approximate Brouwer fixed points.
He proved that even in three dimensions,
  these fixed point problems are complete for the \textbf{PPAD} class.
Recently, Chen and Deng \cite{XX06}
  proved that the problem of finding a discrete fixed point in two dimensions
  is also complete for \textbf{PPAD}.

In \cite{PAP91}, Papadimitriou also proved that
  {\sc Bimatrix}, the problem of finding a Nash equilibrium
  in a two-player game with rational payoffs is member of \textbf{PPAD}.
His proof can be extended to show that finding
  a (properly defined) approximate equilibrium in a non-cooperative game among
  three or more players is also in \textbf{PPAD}.
Thus, if  these problems are
\textbf{PPAD}-complete, then the problem of finding an equilibrium
  is polynomial-time equivalent
  to the search problem for finding a discrete fixed point.

It is conceivable that Nash equilibria might be
  easier to compute than discrete fixed points.
In fact, by taking advantage of the special structure of Nash's normal
  form games,
  Lipton, Markarkis, and Mehta \cite{Lipton} developed a
  sub-exponential time algorithm for finding
  an approximate Nash equilibrium.
In their notion
  of an $\epsilon$-approximate Nash equilibrium, for a positive parameter $\epsilon$, 
  each players' strategy is at most
  an additive $\epsilon$ worse than the best response
  to other players' strategies.
They proved that if all payoffs are in $[0,1]$,
  then an $\epsilon$-approximate Nash equilibrium
  can be found in $n^{O(\log n/\epsilon^2)}$ time.

In a complexity-theoretic breakthrough, Daskalakis, Goldberg and
  Papadimitriou  \cite{DAS05}
  proved that the problem of computing a Nash equilibrium
  in a game among four or more players
  is complete for \textbf{PPAD}.
To cope with the fact that equilibria may not be rational,
  they considered an approximation version of equilibria
  by allowing exponentially small errors.
The complexity result was soon extended to the three-player game
  independently by Chen and Deng \cite{CHE05} and
  Daskalakis and Papadimitriou \cite{DAS06}, with different proofs.
The reduction of \cite{DAS05} has two steps:
First,  it reduces a \textbf{PPAD}-complete
  discrete fixed point problem,
  named {\sc 3-Dimensional Brouwer},
  to the problem of finding a
  Nash equilibrium in a degree-three graphical game
  \cite{KEA01}.
Then, it reduces the graphical game to a four-player game,
  using a result of Goldberg and Papadimitriou~\cite{GOL05}.
This reduction cleverly encodes fixed points by Nash equilibria.

The results of \cite{DAS05,CHE05,DAS06} characterize 
  the complexity of computing $k$-player Nash 
  equilibria for $k\geq 3$.
They also show that
   the fixed point approach is
   necessary in proving Nash's Equilibrium Theorem,
   at least for games among three or more players.
However, these latest complexity advances on the three/four-player
  games have fallen short on the two-player game.

\subsection{Computing Two-Player Nash Equilibria and Smoothed Complexity}


There have been amazing parallels between
    discoveries concerning the two-player zero-sum game
  and the general two-player game.
First, von Neumann proved the existence of
   an equilibrium for the zero-sum game,
   then Nash did the same for the general game.
Both classes of games have rational equilibria when payoffs are rational.
Second, more than a decade after von Neumann's Minimax Theorem,
  Dantzig developed the simplex algorithm, which can
   find a solution of a two-player zero-sum game in a finite number of steps.
A decade or so after Nash's work,
   Lemke and Howson developed their finite-step algorithm for
   {\sc Bimatrix}.
Then, about a quarter century after
  their respective developments, both the simplex algorithm \cite{KleeMinty}
  and the Lemke-Howson algorithm \cite{SavanivonStengel}
  were shown to have exponential worst-case complexity.

A half century after von Neumann's Minimax Theorem,
  Khachiyan proved that the ellipsoid algorithm can solve
  a linear program and hence can find a solution of a two-player zero-sum game
  with rational payoffs in polynomial time.
Shortly after that, Borgwardt \cite{Borg82} showed that
  the simplex algorithm has polynomial average-case complexity.
Then, Spielman and Teng \cite{SpielmanTengSimplex} proved that
  the smoothed complexity of the simplex algorithm is polynomial.
If history is of any guide, then a half century after Nash's Equilibrium
  Theorem, one should be quite optimistic to
  prove the following two natural conjectures.

\begin{itemize}
\item {\em Polynomial 2-{\sc Nash} Conjecture}:
There exists a (weakly)
  polynomial-time algorithm for {\sc Bimatrix}.

\item {\em Smoothed Lemke-Howson Conjecture}:
The smoothed complexity of the Lemke-Howson algorithm for {\sc Bimatrix}
  is polynomial.
\end{itemize}

An upbeat attitude toward the first conjecture has been encouraged
  by the following two facts.
First, unlike three-player games, every rational bimatrix game
  has a rational equilibrium.
Second, a key technical step involving coloring the graphical games
  in the \textbf{PPAD}-hardness proofs for three/four-player games
  fails to extend to two-player games \cite{DAS05,CHE05,DAS06}.
The Smoothed Lemke-Howson Conjecture was asked by 
  a number of~people \cite{list}.
Indeed,  whether the smoothed analysis of the simplex algorithm can be
  extended to the Lemke-Howson algorithm~\cite{LemkeHowson}
  has been the question most frequently raised during talks
  on smoothed analysis.
The conjecture is a special case  of the following conjecture
  posted by Spielman and Teng \cite{SpielmanTengSurvey} in a survey
  of smoothed analysis of algorithms.
\begin{itemize}
\item {\em  Smoothed 2-{\sc Nash} Conjecture}: 
The smoothed complexity of {\sc Bimatrix} is polynomial.
\end{itemize}

The Smoothed 2-{\sc Nash} Conjecture was inspired by the result
  of B\'ar\'any, Vempala and Vetta
  \cite{BaranyVempalaVetta} that an equilibrium of a random two-player game
  can be found in polynomial time.



\subsection{Our Contributions}

Despite much effort in the last half century, no
 significant progress has been made in characterizing the algorithmic
 complexity of finding a Nash equilibrium in a two-player game.
Thus, {\sc Bimatrix}, the most studied computational
  problem about Nash equilibria,
  stood out as the last open problem
  in equilibrium computation for normal form games.
Papadimitriou \cite{PapadimitriouInternet} named it,
  along with {\sc Factoring},
  as one of the two ``most concrete open problems'' at the
  boundary of \textbf{P}.
In fact, ever since Khachiyan's discovery \cite{KHA79},
  {\sc Bimatrix}
  has been on the frontier of natural problems
  possibly solvable in polynomial time.
Now, it is also on the frontier of the hard problems, assuming
  \textbf{PPAD} is not contained in \textbf{P}.

In this paper, we settle the computational complexity of the two-player
  Nash equilibrium.
We prove:

\begin{theo}
   {\sc Bimatrix} is {\em \textbf{PPAD}}-complete.
\end{theo}

Our result demonstrates that, even
  in this simplest form of non-cooperative games,
  equilibrium computation is polynomial-time equivalent
  to discrete fixed point computation.
In particular, we show that from each
  discrete Brouwer function $f$, we can build
  a two-player game $\calG$ and a polynomial-time
  map $\Pi$ from the Nash equilibria of
  $\calG$ to the fixed points of $f$.
Our proof complements Nash's proof that
  for each two-player game $\calG$, there is a
  Brouwer function $f$ and a map $\Phi$ from the fixed points of $f$ to
  the equilibrium points of $\calG$.

The success in proving the \textbf{PPAD} completeness of {\sc Bimatrix}
  inspires us to attempt to disprove the
  Smoothed 2-{\sc Nash} Conjecture.
A connection between the smoothed complexity and approximation
  complexity of Nash equilibria
  (\cite{SpielmanTengSurvey}, Proposition 9.12) then leads us to
  prove the following result.
\begin{theo}
For any $c>0$, the problem of computing
  an $n^{-c}$-approximate Nash equilibrium
  of a two-player game is {\rm \textbf{PPAD}}-complete.
\end{theo}

This result enables us to establish the following fundamental theorem
  about the approximation of Nash equilibria.
It also enables us answer the question about
  the smoothed complexity of the Lemke-Howson algorithm
  and disprove the Smoothed 2-{\sc Nash} Conjecture
  assuming \textbf{PPAD} is not contained in \textbf{RP}.
\begin{theo}
  {\sc Bimatrix} does not have a fully polynomial-time approximation scheme
  unless {\em \textbf{PPAD} is contained in \textbf{P}}.
\end{theo}

\begin{theo}
{\sc Bimatrix} is not in smoothed polynomial time
  unless {\rm \textbf{PPAD}} is contained in {\rm \textbf{RP}}.
\end{theo}

Consequently, it is unlikely that the $n^{O(\log
    n/\epsilon^2)}$-time algorithm of Lipton, Markakis, and Mehta
  \cite{Lipton}, the fastest algorithm known today for finding an
  $\epsilon $-approximate Nash equilibrium, can be improved to
  poly$(n,1/\epsilon)$. Also, it is unlikely that the average-case
  polynomial time result of \cite{BaranyVempalaVetta} can be extended
  to the smoothed model.

Our advances in the computation, approximation,
  and smoothed analysis of two-player Nash equilibria
  are built on several novel techniques that might be interesting on their own.
We introduce a new method for encoding
  boolean and arithmetic variables using the probability vectors
  of the mixed strategies.
We then develop a set of perturbation
  techniques to simulate the boolean and arithmetic
  operations needed for fixed point computation
  using the equilibrium conditions of two-player games.
These innovations enable us to bypass the graphical game model and
  derive a direct reduction from fixed point computation
  to {\sc Bimatrix}.
To study the approximation and smoothed complexity
  of the equilibrium problem,
  we introduce a new discrete fixed point
  problem on a high-dimensional grid graph with a constant side-length.
We then show that it can host the embedding of the proof structure of
  any $\text{{\rm \textbf{PPAD}}}$ problem.
This embedding result not only enriches the family of \textbf{PPAD}-complete
  discrete fixed point problems,  but also provides
  a much needed trade-off between precision and dimension.
We prove a key geometric
  lemma for finding a high-dimensional discrete fixed point, a new concept defined on
  a simplex inside a unit hypercube.
This geometric lemma enables us to
  overcome the curse of dimensionality in reasoning about fixed points
  in high dimensions.

\subsection{Implications and Impact}

Because the two-player Nash equilibrium enjoys
  several structural properties that Nash equilibria with three or more players
  do not have,  our result enables us to answer some
  other long-standing open questions in mathematical economics
  and operations research.
In particular, we have derived the following two important corollaries.
\begin{coro}
Arrow-Debreu market equilibria are {\em \textbf{PPAD}}-hard
  to compute.
\end{coro}

\begin{coro}
The P-matrix Linear Complementary Problem
   is computationally harder than convex programming, unless
   {\em \textbf{PPAD}} is contained in  {\em \textbf{P}},
 where a P-matrix is a square matrix with positive principle minors.
\end{coro}

To prove the first corollary, we use a recent
  discovery of Ye~\cite{YEwine2005} (see also \cite{CSVY})
  on the connection between two-player Nash equilibria
  and Arrow-Debreu equilibria in two-group Leontief
  exchange economies.
The second corollary concerns the linear complementary problem,
 in which we are given
  a rational $n$-by-$n$ matrix $\MM$ and a rational $n$-place
  vector $\qq$, and are asked to find vectors $\xx$ and $\yy$ such that
  $\yy = \MM\xx + \qq$, $\xx, \yy \geq \00$, and $\xx^T\yy = 0$.
Our result complements Megiddo's
  observation \cite{MegiddoLCP} that if it is \textbf{NP}-hard to solve the P-Matrix linear
  complementarity problem, then \textbf{NP} = \textbf{coNP}.

By applying a recent reduction of Abbott, Kane, and Valiant
  \cite{AbbottKaneValiant}, our result also implies the following
  corollary.

\begin{coro}
{\sc Win-Lose Bimatrix} is {\rm \textbf{PPAD}}-complete, where, in
  a win-lose bimatrix game, each payoff entry is either 0 or 1.
\end{coro}

We further refine our reduction to show that the Nash equilibria
  in sparse two-player games are hard to compute and hard to approximate
  in fully polynomial time.

We have also discovered several new structural properties about Nash equilibria.
In particular, we prove an equivalence result about various
  notions of approximate Nash equilibria.
We exploit these equivalences in the study of the complexity of
  finding an approximate Nash equilibrium and in the
  smoothed analysis of {\sc Bimatrix}.
Using them, we can also extend our
  result about approximate Nash equilibria as follows.
\begin{theo}
For any $c>0$,
  the problem of finding the first $(1+c)\log n$ bits
  of an exact Nash equilibrium in a two-player game, even
  when the payoffs are integers of polynomial magnitude,
  is polynomial-time equivalent to {\sc Bimatrix}.
\end{theo}

Recently,
   Chen, Teng, and Valiant \cite{ChenTengValiant}
    extended our approximation complexity
   result to win-lose two-player games;
 Huang and Teng \cite{HuangTeng} extended both the smoothed
    complexity and the approximation results to the computation of Arrow-Debreu
    equilibria.
Using the connection between Nash equilibria and Arrow-Debreu equilibria,
  our complexity result on sparse games
  can be extended to market equilibria in economies with sparse
  exchange structures \cite{ChenHuangTengWine}.

\subsection{Paper Organization}

In Section \ref{sec:Equilibrium},
  we review concepts in equilibrium theory.
We also prove an important equivalence between various notions
  of approximate Nash equilibria.
In Section \ref{sec:Complexity}, we recall the
  complexity class \textbf{PPAD}, the smoothed analysis framework,
  and the concept of polynomial-time reduction
  among search problems.
In Section \ref{sec:search}, we introduce two concepts:
  high-dimensional discrete Brouwer fixed points and generalized
  circuits, followed by the definitions of two search problems based on these concepts.
In Section \ref{sec:Main}, we state our main results and also provide
  an outline of our proofs.
In Section \ref{GCtoTG}, we show that one can simulate generalized circuits
  with two-player Nash equilibria.
In Section \ref{BROUWER}, we prove a \textbf{PPAD}-completeness result for
  a large family of high-dimensional fixed point search problems.
In Section \ref{Final}, we complete our proof by showing that
  discrete fixed points can be modeled by generalized circuits.
In Section \ref{sec:Open}, we discuss some extensions of our work and present
   several open questions and conjectures motivated by this research.
In particular, we will show that sparse {\sc Bimatrix}
  does not have a fully polynomial-time approximation
  scheme unless \textbf{PPAD} is contained in  \textbf{P}.
Finally, in Section \ref{sec:ACK}, we thank many wonderful
  people who helped us in this work.


This paper combines the papers
``Settling the Complexity of 2-Player Nash-Equilibrium'',
  by Xi Chen and Xiaotie Deng,
  and  ``Computing Nash Equilibria:
  Approximation and Smoothed Complexity'',
  by the three of us.
The extended abstracts of both papers appeared in the {\em
  Proceedings of the 47th Annual Symposium on
  Foundations of Computer Science, IEEE.}
The result that {\sc Bimatrix} is \textbf{PPAD}-complete
  is from the first paper.
We also include the main result from the paper
``Sparse Games are Hard'', by the three of us,
 presented at the {\em the 2nd International Workshop
  on Internet and Network Economics}.

\subsection{Notation}

We will use bold lower-case
  Roman letters such as $\xx $, $\aa$, $\bb_{j}$ to denote vectors.
Whenever a vector, say $\aa\in\Reals{n}$ is
  present, its components will be denoted by
  lower-case Roman letters with subscripts, such as $a_1,...,a_n$.
Matrices are denoted by bold upper-case Roman letters
  such as $\AA$ and scalars are usually denoted by lower-case roman
  letters, but sometimes by upper-case Roman letters such as $M$, $N$,
  and $K$.
The $(i,j)^{th}$ entry of a matrix $\AA$ is
  denoted by $a_{i,j}$.
Depending on the context, we may use $\aa_i$ to denote the $i^{th}$
  row or the $i^{th}$ column of $\AA$.

We now enumerate some other notations that are used in this paper.
\def\form#1#2{\left\langle#1 | #2 \right\rangle}
We will let $\ZZ_+^d$ to denote
    the set of $d$-dimensional vectors with positive
    integer entries; 
$\form{\aa}{\bb}$ to denote the dot-product of two vectors in 
  the same dimension;
  $\ee_i$ to denote the unit vector whose $i^{th}$ entry is equal to 1 and
    all other entries are 0.
Finally, for $a,b\in \mathbb{R}$, 
  by $a=b\pm \epsilon$, we mean $b-\epsilon\le a\le b+\epsilon$.

\section{Two-Player Nash Equilibria}\label{sec:Equilibrium}


A {\em two-player game} \cite{NashNonCooperative,Lemke,LemkeHowson}
  is a non-cooperative game between two players.
When the first player has $m$ choices of actions and the second player has $n$
  choices of actions, the game, in its normal form, can be specified
   by two $m \times n$ matrices
  $\AA =\left(a_{i,j}\right)$ and $\BB=$ $\left(b_{i,j}\right)$.
If the first player chooses action $i$ and the second player
  chooses action $j$, then their payoffs are $a_{i,j}$ and
  $b_{i,j}$, respectively.
Thus, a two-player game is also often referred to as a {\em bimatrix game}.
A mixed strategy of a player is a probability distribution
  over its choices. The Nash's Equilibrium Theorem
  \cite{NashNonCooperative,NAS50},
  when specialized to bimatrix games, asserts that
  every two-player game has an equilibrium point, i.e.,
  a profile of mixed strategies,
  such that neither player can
  gain by changing his or her strategy  unilaterally.
The zero-sum two-player game
  \cite{MOR47}
is a special case of the bimatrix game
  that satisfies $\BB  = -\AA$.

Let $\symP^n$ denote the set of all {\em probability
  vectors} in $\Reals{n}$, i.e., non-negative, $n$-place vectors
  whose entries sum to 1.
Then, a profile of mixed strategies can
  be expressed by two column vectors $(\xx^*
  \in\symP^{m},\yy^*\in\symP^{n})$.

Mathematically, a {\em Nash equilibrium} of a bimatrix game $(\AA,\BB)$
  is a pair $(\xx^*
  \in\symP^{m},\yy^*\in\symP^{n})$ such that
\begin{equation*} \label{defi:NashEquilibria}
(\xx^{*})^T \AA  \yy^{*} \geq \xx^{T} \AA  \yy^{*}\ \ \ \ \mbox{and}\ \ \ \ (\xx^{*})^T \BB \yy^{*}  \geq (\xx^{*})^T
\BB \yy, \quad \mbox{for all
  $\xx \in \symP^{m}$ and $\yy \in \symP^{n}$.}
\end{equation*}

Computationally, one might settle with an approximate
  Nash equilibrium.
There are several versions of approximate equilibrium points
  that have been defined in the literature.
The following are two most popular ones.

For a positive parameter $\epsilon$,
  an {\em $\epsilon$-approximate Nash equilibrium} of
  a bimatrix game $(\AA,\BB)$ is a pair
  $(\xx^* \in\symP^{m},\yy^*\in\symP^{n})$ such that
\begin{eqnarray*}
(\xx^{*})^T \AA  \yy^{*} \geq \xx^{T} \AA  \yy^{*}-\epsilon \ \ \mbox{and}\ \ (\xx^{*})^T \BB \yy^{*}  \geq
(\xx^{*})^T \BB \yy-\epsilon, \  \mbox{ for all $\xx \in \symP^{m}$ and $\yy \in \symP^{n}$.}
\end{eqnarray*}
An {\em  $\epsilon$-relatively-approximate Nash
  equilibrium} of $(\AA,\BB)$
  is a pair $(\xx^*,\yy^*)$ such that
\begin{equation*}
(\xx^{*})^{T} \AA  \yy^{*} \geq (1-\epsilon) \xx^{T} \AA
\yy^{*} \ \ \mbox{and} \ \ (\xx^{*})^{T} \BB \yy^{*}  \geq
(1-\epsilon) (\xx^{*})^{T} \BB \yy, \  \mbox{
  for all $\xx \in \symP^{m}$ and $\yy \in \symP^{n}$.}
\end{equation*}

Nash equilibria of a bimatrix game
  $(\AA,\BB)$ are invariant under positive scalings,
  meaning, the bimatrix game $(c_1\AA,c_2\BB)$
  has the same set of Nash equilibria as
  $(\AA,\BB)$, as long as $c_1, c_2>0$.
They  are also invariant under shifting:
  For any constants $c_1$ and $c_2$,
  the bimatrix game $(c_1+\AA,c_2+\BB)$
  has the same set of Nash equilibria as $(\AA,\BB)$.
It is  easy to verify that
    $\epsilon$-approximate Nash equilibria are also invariant
    under shifting.
However, each $\epsilon$-approximate Nash equilibrium
    $(\xx,\yy)$ of $(\AA,\BB)$ becomes
    a $(c\cdot\epsilon)$-approximate Nash equilibrium
    of the bimatrix game $(c\AA,c\BB)$ for $c>0$.
Meanwhile, $\epsilon$-relatively-approximate Nash equilibria
  are invariant under positive scaling, but may not
   be invariant under shifting.

The notion of the $\epsilon$-approximate Nash equilibrium
  is defined in the additive fashion.
To study its complexity, it is important to consider bimatrix games with
  normalized matrices in which the absolute value of each entry
  is bounded, for example, by 1.
Earlier work on this subject by Lipton, Markakis, and Mehta
   \cite{Lipton} used a similar normalization.
Let $\Reals{m\times n}_{[a:b]}$ denote the set of $m \times n$ matrices
      with real entries between $a$ and $b$.
In this paper, we say a bimatrix game $(\AA,\BB)$
  is {\em normalized} if  $\AA,\BB \in \Reals{m\times n}_{[-1,1]}$ and
 is {\em positively normalized} if $\AA,\BB \in \Reals{m\times n}_{[0,1]}$.

\begin{prop}\label{prop:relatively}
In a normalized two-player game $(\AA,\BB)$,
  every $\epsilon$-relatively-approximate Nash equilibrium
  is also an $\epsilon$-approximate Nash equilibrium.
\end{prop}


To define our main search problems of computing and
  approximating a two-player Nash equilibrium,
  we need to first define the input models.
The most general input model is the {\em  real model}
  in which a bimatrix game is specified by
  two real matrices $(\AA ,\BB)$.
In the {\em  rational model}, each entry of the payoff matrices
  is given by the ratio of two integers.
The {\em input size} is then the total number of bits 
  describing the payoff matrices.
Clearly, by multiplying the common denominators in
  a payoff matrix and using the fact that
  two-player Nash equilibria are invariant under positive
  scaling, we can transform a rational bimatrix game into
  an {\em integer bimatrix game}.
Moreover, the total number of bits in this game with integer payoffs
  is within a factor of poly$(m,n)$ of the input size of its
  rational counterpart.
In fact, Abbott, Kane, and Valiant \cite{AbbottKaneValiant}
   go one step further to show that
  from every bimatrix game with integer payoffs, one can
  construct a ``homomorphic'' bimatrix game
  with 0-1 payoffs who size is within a polynomial factor of
  the input size of the original game.

\def\support#1{\mbox{\rm Support}\left(#1\right)}

It is well known that each rational bimatrix game
  has a rational Nash equilibrium.
We may verify this fact as following.
Suppose $(\AA ,\BB )$ is a rational two-player game and
  $(\uu,\vv)$ is one of its Nash equilibria.
Let $\mbox{row-support} = \{i\ |\ u_{i} > 0 \}$ and
 $\mbox{column-support} = \{i\ |\ v_{i} > 0 \}$.
Let $\aa_i$ and $\bb_j$
  denote the $i^{th}$ row of $\AA$ and the $j^{th}$ column
  of $\BB$, respectively.
Then, by the condition of the Nash equilibrium,
  $(\uu,\vv)$ is a feasible solution
  to the following linear program:
\begin{eqnarray*}
& \sum_{i} x_{i} = 1 \mbox{ and } \sum_{i} y_{i} = 1 & \\
& x_{i} = 0, & \forall i \not\in \mbox{row-support}\\
& y_{i} = 0, & \forall i \not\in \mbox{column-support}\\
& x_{i} \geq  0, &  \forall i\in \mbox{row-support}\\
& y_{i} \geq  0, & \forall i \in \mbox{column-support}\\
& \aa_{i} \yy = \aa_{j}\yy, & \forall i, j\in \mbox{row-support}\\
& \xx^T\bb_{i} = \xx^T\bb_{j}, &  \forall i,j\in \mbox{column-support}\\
& \aa_{i} \yy \leq  \aa_{j}\yy, & \forall i\not\in \mbox{row-support}, j\in \mbox{row-support}\\
& \xx^T\bb_{i} \leq  \xx^T\bb_{j}, & \forall i\not\in \mbox{column-support}, j\in \mbox{column-support}.
\end{eqnarray*}
In fact, any solution to this linear program is a Nash
  equilibrium of $(\AA,\BB )$.
Therefore, $(\AA,\BB )$ has at least one rational
  equilibrium point such that the total number of bits describing 
  this equilibrium is within a polynomial factor of the
  input size of $(\AA ,\BB )$.
By enumerating all possible row supports and column supports
  and applying the linear program above,
  we can find a Nash equilibrium in the bimatrix game $(\AA,\BB)$.
This exhaustive-search algorithms takes $2^{m+n}{\mbox{\rm poly} (L)}$ time
  where $L$ is the input size of the game, and
  $m$ and $n$ are, respectively, the number of rows and
  the number of columns.

In this paper, we use {\sc Bimatrix} to denote
  the problem of finding a Nash equilibrium
  in a rational bimatrix game.
Without loss of generality, we make two assumptions
  about our search problem:
  all input bimatrix games are positively normalized
  in which both players have the same number of choices of actions.
Thus, two important parameters associated with each instance to {\sc Bimatrix}
  are:  $n$, the number of actions, and $L$,
  the total number of bits in the description of
  the game.
Thus, {\sc Bimatrix} is in \textbf{P} if there exists an
  algorithm for {\sc Bimatrix} with running time poly$(n,L)$.
As a matter of fact, for the two-player games that we will design
  in our complexity studies,
  $L$ is bounded by a polynomial in $n$.

We also consider two families of approximation problems for
  two-player Nash equilibria.
For a positive constant $c$,
\begin{itemize}
\item  let {\sc Exp$^c$-Bimatrix} denote the following search problem:
   Given a rational and positively normalized bimatrix game
   $(\AA ,\BB )$, compute a $2^{-cn}$-approximate Nash equilibrium
   of $(\AA ,\BB )$, if $\AA $ and $\BB $ are $n \times n$ matrices;

\item let {\sc Poly$^c$-Bimatrix} denote the following search problem:
   Given a rational and positively normalized bimatrix game
   $(\AA ,\BB )$, compute an $n^{-c}$-approximate Nash equilibrium
   of $(\AA ,\BB )$, if $\AA $ and $\BB $ are $n \times n$ matrices.
\end{itemize}

In our analysis, we will use an alternative notion of
  approximate Nash equilibria as introduced in \cite{DAS05},
  originally called {\em $\epsilon$-Nash equilibria}.
In order to avoid confusion with more commonly used $\epsilon$-approximate Nash equilibria,
  we will refer to this alternative approximation as
  the {\em $\epsilon$-well-supported Nash  equilibrium}.
For a bimatrix game $(\AA,\BB)$, let $\aa_i$ and $\bb_j$
  denote the $i^{th}$ row of $\AA$ and the $j^{th}$ column
  of $\BB$, respectively.
In a profile of mixed strategies $(\xx,\yy)$, the expected payoff of the
  first player when choosing the $i^{th}$ row is
  $\aa_i {\yy}$, and the expected payoff of the second player when choosing
  the $i^{th}$ column is $\xx^{T}{\bb_i}$.

For a positive parameter $\epsilon$,
  a pair of strategies $(\xx^* \in \symP^n,\yy^*\in\symP^n)$
  is an {\em $\epsilon$-well-supported Nash equilibrium} of
  $(\AA,\BB)$ if for all $j$ and $k$,
\begin{eqnarray*}
({\xx^*})^T \bb_j > (\xx^*)^T \bb_k +\epsilon\ \Rightarrow\ y^*_k=0 \ \ \ \ \mbox{and}\ \ \ \ \aa_j {\yy^*}
> \aa_k {\yy^*}+\epsilon\ \Rightarrow\ x^*_k=0.
\end{eqnarray*}
A Nash equilibrium is a $0$-well-supported Nash equilibrium as well as
  a $0$-approximate Nash equilibrium.
The following lemma, a key lemma in our complexity study of
  equilibrium approximation,
  shows that approximate Nash equilibria and well-supported
  Nash equilibria are polynomially related.
This polynomial relation allows us to focus our attention on
  pair-wise approximation conditions.
Thus, we can locally argue certain properties of
  the bimatrix game in our analysis.

\begin{lemm}[Polynomial Equivalence] \label{lem:equivalence}
In a bimatrix game $(\AA,\BB)$
  with $\AA, \BB \in \Reals{n\times n}_{[0:1]}$,
  for any $0\leq \epsilon \leq 1$,
\begin{enumerate}
\item each $\epsilon$-well-supported Nash equilibrium is also an $\epsilon$-approximate Nash
equilibrium; and
\item from any $\epsilon^2/8$-approximate Nash equilibrium $(\uu,\vv)$, one can find in polynomial
time an $\epsilon$-well-supported Nash equilibrium $(\xx,\yy)$.
\end{enumerate}
\end{lemm}
\begin{proof}
The first statement follows from the definitions.
Because $(\uu,\vv)$ is an $\epsilon^2/8$-approximate Nash
  equilibrium, we have
\begin{equation*}
 \forall\ \uu' \in \symP^n,\ (\uu')^T\AA\vv \leq  \uu^T\AA\vv + \epsilon^2/8,
 \ \ \mbox{and}\ \
 \forall\ \vv'\in \symP^n,\ \uu^T\BB\vv' \leq  \uu^T\BB\vv +
\epsilon^2/8.
\end{equation*}
Recall that $\aa_i$ denotes the $i^{th}$ row
  of $\AA$ and $\bb_i$ denotes the $i^{th}$ column of $\BB$.
We use $J_1$ to denote the set of indices $j:1\le j\le n$ such that
  $\aa_i \vv \ge \aa_j \vv + \epsilon/2$, for some $i\in [1:n]$.
Let $i^* $ be an index such that
  $\aa_{i^*}\vv=\max_{1\le i\le n} \aa_i\vv$.
Now by changing $u_j, j\in J_1$, to $0$ and changing
   $u_{i^*}$ to $u_{i^*} + \sum_{j\in J_1}u_j$
  we can increase the first-player's profit by at least $(\epsilon/2)\sum_{j\in J_1}u_j$,
  implying $\sum_{j\in J_1}u_j < \epsilon/4$.
Similarly, we define $J_2= \{\hspace{0.05cm}j:1\le j\le n: \exists\ i,\hspace{0.05cm}
  \uu^T{\bb_i} \ge \uu^T{\bb_j}+ \epsilon/2\hspace{0.05cm} \}$. Then we have
  $\sum_{j\in J_2} v_j< \epsilon/4$.

We now set all these $\{u_j\ |\  j\in J_1\}$ and $\{v_j\ | \ j\in J_2
  \}$ to zero,
  and uniformly increase the probabilities of other strategies to obtain a
  new pair of mixed strategies $(\xx,\yy)$.

Note for all $i \in [1: n]$, $|\hspace{0.06cm} \aa_i {\yy} -
  \aa_i {\vv}\hspace{0.04cm}|  \le \epsilon/4$,
  because we assume the value of
each entry in $\aa_i$ is between $0$ and $1$.
Therefore, for every pair $i,j:1\le i,j\le n$, the relative change between
  $\aa_i {\yy}-\aa_j {\yy}$ and $\aa_i {\vv}-\aa_j {\vv}$
is no more than $\epsilon/2$.
 Thus, any $j$ that is beaten by some $i$ by a gap of $\epsilon$
  is already set to zero in $(\xx,\yy)$.
\end{proof}

We conclude this section by pointing out that there
  are other natural notions of approximation for equilibrium points.
In addition to the rational representation of a rational
  equilibrium, one can use binary representations
  to define entries in an equilibrium.
As each entry $p$ in an equilibrium is a number
   between $0$ and $1$, we can specify it using its  binary representation
  $(0\mbox{\huge\bf .}c_{1}\cdots c_{P}\cdots)$,
  where $c_i \in \setof{0,1}$ and
$
  p = \lim_{i\rightarrow\infty} \sum_{i=1} c_i/ 2^i.
$
Some rational numbers may not have a finite binary representation.
Usually, we round off the numbers to store their finite approximations.
The first $P$ bits $c_1,...,c_{P}$ give us
   a {\em $P$-bit approximation} $\tilde{c}$ of $c$.

For a positive integer $P$, we will use $P$-{\sc Bit-Bimatrix} to
  denote the search problem of computing the first $P$ bits of
  the entries of a Nash equilibrium in a rational
  bimatrix game.
The following proposition relates $P$-{\sc Bit-Bimatrix} with
  {\sc Poly$^c$-Bimatrix}.

\begin{prop}\label{pro:bit}
Suppose $(\xx,\yy)$ is a Nash equilibrium of a positively normalized
  two-player game $(\AA,\BB)$ with $n$ rows and  $n$ columns.
For a positive integer $P$, let
   $(\tilde{\xx},\tilde{\yy})$ be the $P$-bit
   approximation of $(\xx,\yy)$.
Let  $\orig{\xx} = \tilde{\xx}/\pnorm{1}{\tilde{\xx}}$
  and $\orig{\yy} = \tilde{\yy}/\pnorm{1}{\tilde{\yy}}$.
Then, $(\orig{\xx},\orig{\yy})$
  is a $(3\hspace{0.04cm}n\hspace{0.04cm}2^{-P})$-approximate
  Nash equilibrium of $(\AA ,\BB)$.
\end{prop}
\begin{proof}\label{}
A similar proposition is stated and proved in \cite{ChenTengValiant}.

Let $a=2^{-P}$.
Suppose $(\orig{\xx},\orig{\yy})$ is not a $(3na)$-approximate
  Nash equilibrium.
Without loss of generality, assume there
  exists $\xx'\in\mathbb{P}^n$ such that $(\xx')^T \AA \orig{\yy}
  > \orig{\xx}^T \AA \orig{\yy}+3na$.
We have
\begin{eqnarray*}
(\xx')^T\AA\orig{\yy} &\le& (\xx')^T\AA\tilde{\yy}+na\ \le\ (\xx')^T\AA\yy+na
  \ \le\ \xx^T \AA\yy+ na\\ &\le& \tilde{\xx}^T\AA \yy + 2na\ \le\
  \tilde{\xx}^T \AA\tilde{\yy}+3na\ \le\ \orig{\xx}^T \AA\tilde{\yy}+3na
  \ \le\ \orig{\xx}^T\AA\orig{\yy}+3na,
\end{eqnarray*}
which contradicts our assumption.
To see the first inequality,
  note that since the game is positively normalized,
  every component in $(\xx')^T\AA$ is between
  $0$ and $1$.
The inequality follows from the fact that
$\orig{y}_i\ge  \tilde{y}_i$ for
  all $i \in [1:n]$, and $\pnorm{1}{\tilde{\yy}}\ge 1-na$.
The other inequalities can be proved similarly.
\end{proof}

\section{Complexity and Algorithm Analysis}\label{sec:Complexity}

In this section, we review the complexity class \textbf{PPAD} and
  the concept of polynomial-time reduction
  among search problems.
We then define the perturbation models in the smoothed analysis
   of {\sc Bimatrix} and show that
   if the smoothed complexity of {\sc Bimatrix} is polynomial,
   then we can compute an $\epsilon$-approximate Nash equilibrium
   of a bimatrix game in randomized poly$(n,1/\epsilon)$ time.

\subsection{PPAD and Polynomial-Time Reduction Among Search Problems}

\def\search#1{\mbox{{\sc Search}$^#1$}}

A binary relation $R\subset \setof{0,1}^*\times\setof{0,1}^*$
  is {\em polynomially balanced}
  if there exist constants $c$ and $k$ such that for all
  pairs $(x,y)\in R$, $\sizeof{\hspace{0.04cm} y\hspace{0.04cm}}
  \leq c\sizeof{\hspace{0.04cm}x\hspace{0.04cm}}^k$, where $\sizeof{\hspace{0.04cm}x\hspace{0.04cm}}$
  denotes the length of string $x$.
It is {\em polynomial-time computable} if
  for each pair $(x,y)$, one can decide whether or
  not $(x,y) \in R$ in time polynomial in
  $\sizeof{\hspace{0.04cm}x\hspace{0.04cm}}+
  \sizeof{\hspace{0.04cm}y\hspace{0.04cm}}$.
One can define the \textbf{NP} search problem $\search{R}$
  specified by $R$ as: Given $x\in \setof{0,1}^*$, 
  return a $y$ such satisfying $(x,y)\in$ $ R$, if such $y$ exists,
  otherwise, return a special string ``no''.

A relation $R$ is {\em total} if for every string
  $x \in \setof{0,1}^*$, there exists $y$ such
  that $(x,y)\in R$.
Following Megiddo and Papadimitriou \cite{MEG91},
  let \textbf{TFNP} denote the
  class of all \textbf{NP} search problems specified
  by total relations.
A search problem $\search{{R_1}}\in {\textbf{TFNP}}$ is
  {\em polynomial-time reducible} to problem $\search{{R_2}}
  \in {\textbf{TFNP}}$ if there exists a pair of
  polynomial-time computable functions $(f,g)$ such
  that for every $x$ of $R_1$, if $y$ satisfies that
  $(f(x),y)\in R_2$, then $(x,g(y))\in R_1$.
Search problems $\search{{R_1}}$ and $\search{{R_2}}$ are
  polynomial-time equivalent
  if $\search{{R_2}}$ is also reducible to $\search{{R_1}}$.

The complexity class \textbf{PPAD} \cite{PAP94} is a
  sub-class of \textbf{TFNP}, containing all search
  problems polynomial-time reducible to
  following problem called {\sc End-of-Line}:
\begin{defi}[{\sc End-of-Line}]
The input instance of {\sc End-of-Line} is a pair $({\cal M},0^n)$
   where ${\cal M}$ is a circuit of size polynomial in $n$ that
   defines a function $M$ satisfying\hspace{0.04cm}:
\begin{itemize}
\item for every $v\in \setof{0,1}^n$, $M(v)$ is an ordered pair $(u_1,u_2)$ where $u_1,u_2\in
\setof{0,1}^n \cup \setof{\text{``no''}}$.

\item $M(0^n)=(\text{``no''}, 1^n)$ and the first component of $M(1^n)$ is $0^n$.
\end{itemize}

This instance defines a directed graph $G_{M}=(V,E_{M})$ with
  $V=\setof{0,1}^n$ and $(u,v)\in E_M$, if and only if
  $v$ is the second component of $M(u)$ and
  $u$ is the first component of $M(v)$.

The output of this problem is an end vertex $G_{M}$ other than $0^n$,
  where a vertex of $V$ is an {\rm end vertex} if the summation of its
  in-degree and out-degree is equal to one.
\end{defi}

Note that in graph $G_{M}$,
   both the in-degree and the out-degree
   of each vertex  are at most 1.
Thus, edges of $G_{M}$ form a collection of directed paths and
  directed cycles.
Because $0^{n}$ has in-degree 0 and out-degree 1,
  it is an end vertex in $G_{M}$.
$G_{M}$  must have at least one directed path.
Hence, it has another end vertex and {\sc End-of-Line} is a member of \textbf{TFNP}.

In fact, $G_{M}$ has an odd number of end vertices other than $0^{n}$.
By evaluating the polynomial-sized circuit ${\cal M}$ on an input
  $v \in V$, we can access the predecessor and the successor of $v$.

Many important problems, such as the search versions of Brouwer's
  Fixed Point Theorem, Kakutani's Fixed Point Theorem, Smith's Theorem,
  and Borsuk-Ulam Theorem, have been shown to be in the class
  \textbf{PPAD}~\cite{PAP91}.

{\sc Bimatrix} is also in \textbf{PPAD} \cite{PAP91}.
As a corollary, for all $c>0$,
  {\sc Poly$^c$-Bimatrix} and {\sc Exp$^c$-Bimatrix}
  are  in \textbf{PPAD}.
However, it is not clear whether $P$-{\sc Bit-Bimatrix},
  for a positive integer $P$, is in \textbf{PPAD}.

\subsection{Smoothed Models of Bimatrix Games}

\def\SS{\mathbf{S}} \def\TT{\mathbf{T}}

In the smoothed analysis of the bimatrix game, we consider perturbed games
  in which each entry of the payoff
  matrices is subject to a small and independent random
  perturbation.
For a pair of $n\times n$ normalized  matrices $\overline{\AA}=
  (\overline{a}_{i,j})$ and $\overline{\BB}=(\overline{b}_{i,j})$,
 in the smoothed model, the input instance\footnote{
For the simplicity of presentation, in this subsection, we
  model the entries of payoff matrices and perturbations
  by real numbers.
Of course, to connect with the complexity result of
  the previous section, where entries of matrices are
  in finite representations,
  we are mindful that some readers may prefer that
  we state our result and write the proof more explicitly using the finite
  representations.
Using Equations (\ref{eqn:perb1}) and (\ref{eqn:perb2}) in the proof
  of Lemma \ref{lem:smoothed} (see Appendix \ref{App:SmoothedApproximation}),
  we can define a discrete version
  of the uniform and Gaussian perturbations and state and prove
  the same result.
} is then defined by $(\AA,\BB)$ where $a_{i,j}$ and
  $b_{i,j}$ are, respectively, independent perturbations of
  $\orig{a}_{i,j}$ and $\orig{b}_{i,j}$ with magnitude $\sigma$.

There might be several models of perturbations for
   $a_{i,j}$ and $b_{i,j}$ with magnitude $\sigma$ \cite{SpielmanTengSurvey}.
The two common perturbation models
   are the uniform perturbation and the Gaussian perturbation.

In the {\em uniform perturbation} with magnitude $\sigma$,
  $a_{i,j}$ and $b_{i,j}$ are chosen uniformly from the
  intervals $[\orig{a}_{i,j}-\sigma,\orig{a}_{i,j}+\sigma]$
  and $[\orig{b}_{i,j}-\sigma,\orig{b}_{i,j}+\sigma]$, respectively.
In the {\em Gaussian perturbation} with variance $\sigma^2$,
   $a_{i,j}$ and $b_{i,j}$ are, respectively, chosen with density
\begin{equation*}
\frac{1}{\sqrt{2 \pi} \sigma} e^{-|a_{i,j}-\orig{a}_{i,j}|^{2}/ 2
\sigma^{2}}\quad \mbox{and}
\quad \frac{1}{\sqrt{2 \pi} \sigma} e^{-|b_{i,j}-\orig{b}_{i,j}|^{2}/ 2 \sigma^{2}}.
\end{equation*}
We refer to these perturbations as {\em $\sigma $-uniform}
  and {\em $\sigma$-Gaussian perturbations}, respectively.

The smoothed complexity of
  an algorithm $J$ for {\sc Bimatrix} is defined as following:
Let $T_J(\AA,\BB)$ be the complexity of $J$ for
  finding a Nash equilibrium in a bimatrix game $(\AA,\BB)$.
Then, the {\em smoothed complexity} of $J$ under perturbations
  $N_{\sigma } ()$ of magnitude $\sigma$ is
\[
\smoothed{J}{n,\sigma} = \max_{\orig{\AA}, \orig{\BB}\in\Reals{n\times n}_{[-1,1]}} \expec{\AA\leftarrow N_{\sigma } (\orig{\AA }),\BB\leftarrow N_{\sigma } (\orig{\BB })}{T_J(\AA,\BB)},
\]
where we use
  $\AA\leftarrow N_{\sigma } (\orig{\AA }) $
  to denote that $\AA $  is a perturbation of $\orig{\AA}$
  according to $N_{\sigma } (\orig{\AA })$.

An algorithm $J$ has a {\em polynomial smoothed time complexity}
   \cite{SpielmanTengSurvey}
   if for all $0<\sigma < 1$ and
   for all positive integer $n$,
   there exist positive constants $c$, $k_1$ and $k_2$ such that
\[
\smoothed{J}{n,\sigma} \leq c\cdot n^{k_1} \sigma^{-k_2}.
\]
{\sc Bimatrix} is in {\em  smoothed polynomial time} if
  there exists an algorithm $J$ with polynomial smoothed time complexity
   for computing a two-player Nash equilibrium.


The following lemma shows that if the smoothed complexity
  of {\sc Bimatrix} is low, under uniform or Gaussian
  perturbations, then one can quickly find an
  approximate Nash equilibrium.

\begin{lemm}[Smoothed Nash vs Approximate Nash]\label{lem:smoothed}
If \mbox{\sc Bimatrix} is in smoothed polynomial time under
  uniform or Gaussian perturbations, then for all $\epsilon >0$, there exists
  a randomized algorithm to compute an $\epsilon $-approximate
  Nash equilibrium in a two-player game with expected
  time $O\hspace{0.04cm}\big(\hspace{0.04cm}\mbox{\rm poly}(m,
  n,1/\epsilon)\hspace{0.04cm}\big)$ or $O\hspace{0.04cm} \big(
  \hspace{0.04cm}\mbox{\rm poly}
  (m,n,\sqrt{\log \max (m,n)} /\epsilon)\hspace{0.04cm}\big)$, respectively.
\end{lemm}

\begin{proof}
Informally argued in \cite{SpielmanTengSurvey}.
See Appendix \ref{App:SmoothedApproximation} for a proof.
\end{proof}

\section{Two Search Problems}\label{sec:search}

In this section, we consider two search problems that are essential
  to our main results.
In the first problem, the objective is to find
  a high-dimensional discrete Brouwer fixed point.
To define the second problem, we introduce
  a concept of the generalized circuit.

\subsection{Discrete Brouwer Fixed Points}

The following is an oblivious fact:
Suppose we color the endpoints of
  an interval $[0,n]$ by two distinct colors, say
  red and blue, insert $n-1$ points evenly into this interval to
  subdivide it into $n$ unit subintervals, and color these new
  points arbitrarily by one of the two colors.
Then, there must be a {\em bichromatic subinterval}, i.e.,
  an unit subinterval whose two endpoints have distinct colors.

Our first search problem is built on a high-dimensional
  extension of this fact.
Instead of coloring points in a subdivision of an intervals, we
  color the vertices in a hypergrid.
If the dimension is $d$, we will use $d+1$ colors.

For $d$ $\in \ZZ_{+}^{1}$ and $\rr$ in $\ZZ_+^d$,
 let $A_{\rr}^d= \{\hspace{0.05cm} \qq\in \ZZ^d\ \big|
  \ 0\le q_i\le r_i-1, \forall\ i \in [1: d]
  \hspace{0.05cm} \} $ denote the vertices of the {\em
  hypergrid} with side lengths specified by $\rr $.
The {\em boundary} of $A_{\rr}^d$, $\partial(A_{\rr}^d)$,
  is the set of points $\qq \in A_{\rr}^d$ with
  $q_i \in \{\hspace{0.04cm}0,r_i-1\hspace{0.04cm}\}$
  for some $i$.
Let
  $\size{\textbf{r}}=\sum_{1\le i\le d}\hspace{0.06cm}
  \lceil\hspace{0.06cm} \log (r_i+1) \hspace{0.06cm}\rceil$.

In one dimension, the interval $[0,n]$ is the union
  of $n$ unit subintervals.
A hypergrid can be viewed as the union of a collection of unit hypercubes.
For a point $\pp\in \ZZ^d$, let $K_{\pp}=\{\hspace{0.05cm} \qq\in \ZZ^d\ \big|
  \ q_i \in \setof{ p_i, p_i+1 }, \forall
  \ i \in [1:d]\hspace{0.05cm}\}$ be the vertices of the  unit hypercube
  with $\pp$ as its corner closest to the origin.

We can color the vertices of a hypergrid with $(d+1)$ colors
  $\setof{1,2,...,d+1}$.
Like in one dimension, the coloring of the boundary vertices
  needs to meet certain requirements
  in the context of the discrete Brouwer fixed point problem.
A color assignment $\phi$ of $A_{\rr}^d$ is {\em valid}
if $\phi(\pp)$ satisfies the following condition:
   For $\pp \in \partial(A_{\rr}^d)$,
   if there exists an $i \in [1: d]$
   such that $p_i=0$
   then $\phi(\pp) = \max\{\hspace{0.06cm}i\ \big|
   \ p_{i}=0\hspace{0.06cm}\}$; otherwise $\phi(\pp) = d+1$.
In the later case, $\forall\hspace{0.08cm} i$, $p_i\neq 0$
     and $\exists i$, $p_{i} = r_i-1$.

The following theorem is a high-dimensional extension
  of the one-dimensional fact mentioned above.
It is also an extension of the two-dimensional
  Sperner's Lemma.

\begin{theo}[High-Dimensional Discrete Brouwer Fixed Points]\label{the:dis}
For $d$ $\in \ZZ_{+}^{1}$ and $\rr$ in $\ZZ_+^d$, for any valid coloring
  $\phi$ of $\AA_{\pp}^d$, there is a unit hypercube in $\AA_{\pp}^d$
  whose vertices have all $d+1$ colors.
\end{theo}

In other words, Theorem \ref{the:dis} asserts
  that there exists a $\pp \in A_{\rr}^d$ such that
  $\phi$ assigns all $(d+1)$ colors to $K_{\pp}$.
We call $K_{\pp}$ a {\em panchromatic cube}.
However, in $d$-dimensions, a panchromatic cube contains $2^d$
  vertices.
This exponential dependency in the dimension makes it inefficient
  to check whether a hypercube is panchromatic.
We introduce the following notion of discrete fixed points.

\begin{defi}[Panchromatic Simplex]
A subset $P\subset A_{\rr}^d$ is
  {\em accommodated} if $P\subset K_{\pp}$ for
  some point $\pp\in A_{\rr}^d$.
$P\subset A_{\rr}^d$ is a {\em panchromatic simplex}
  of a color assignment $\phi$ if it is accommodated and contains exactly
  $d+1$ points with $d+1$ distinct colors.
\end{defi}

\begin{coro}[Existence of Panchromatic Simplex]\label{the:dissimplex}
For $d$ $\in \ZZ_{+}^{1}$ and $\rr$ in $\ZZ_+^d$, for any valid coloring
  $\phi$ of $\AA_{\pp}^d$, there exists a panchromatic simplex in
  $\AA_{\pp}^d$.
\end{coro}

We can define a search problem based on Theorem \ref{the:dis}, or precisely,
  based on Corollary \ref{the:dissimplex}.
An input instance is a hypergrid together with
  a polynomial-sized circuit for coloring the vertices of the hypergrid.

\begin{defi}[Brouwer-Mapping Circuit and Color Assignment]
    \label{def:validBrouwer}
For $d\in\ZZ_{+}^{1}$ and $\rr$ $\in \mathbb{Z}_+^d$,
  a Boolean circuit $C$ with
  $\size{\rr}$ input bits and $2d$ output bits
  $\Delta_1^+, \Delta_1^-,...,\Delta_d^+,\Delta_d^-$
is a {\rm valid}  \emph{Brouwer-mapping circuit} \emph{(}with parameters $d$ and $\rr$\emph{)}
if the following is true.
\begin{itemize}
\item For every $\pp\in A_{\rr}^d$, the $2d$ output
  bits of $C$ evaluated at $\pp$ satisfy one of the following $(d+1)$ cases:
\begin{itemize}
\item Case $i$, $1\leq i\leq d$:
  $\Delta_i^+=1$ and all other $2d-1$ bits are $0$;
\item Case $(d+1)$: $\forall\hspace{0.08cm} i$, $\Delta_i^+=0$
  and $\Delta_i^-=1$.
\end{itemize}

\item For every $\pp \in \partial(A_{\rr}^d)$, if
  there exists an $i \in [1:d]$ such that $p_i=0$, letting
  $i_{\max} =\max\{\hspace{0.06cm}i\ \big|
  \ p_{i}=0\hspace{0.06cm}\}$, then the output bits
  satisfy Case $i_{\max}$,
otherwise \emph{(}\hspace{0.04cm}$\forall\hspace{0.08cm} i$,
  $p_i\neq 0$ and $\exists i$, $p_{i} = r_i-1$\hspace{0.04cm}\emph{)},
  the output bits satisfy Case $d+1$.
\end{itemize}

The circuit $C$ defines a valid color
  assignment $\mbox{Color}_C: A_{\rr}^d \rightarrow
  \setof{\hspace{0.05cm}1,2,...,d+1 }$ by setting
  $\color{C}{\pp} = i$, if the output bits of $C$
  evaluated at $\pp$ satisfy Case $i$.
\end{defi}

To define our high-dimensional Brouwer's fixed point problems, we need
  a notion of {\em well-behaved} functions
(\hspace{0.04cm}please note that this is not the function for the fixed point problem\hspace{0.04cm})
  to parameterize the shape
  of the search space.
An integer function $f(n)$ is called
  {\em well-behaved} if it is polynomial-time
  computable and there exists an integer
  constant $n_0$ such that $3\le f(n)\le n/2$
  for all $n\ge n_0$.
For example, $f_1(n)=3$, $f_2(n)=\lfloor n/2\rfloor$,
  $f_3(n)=\lfloor n/3\rfloor$, and
  $f_4(n)=\lfloor \log n\rfloor$ are all well-behaved.

\begin{defi}[{\sc Brouwer}$^f$]\label{BrouwerDef}
For each well-defined function $f$, the search problem
   {\sc Brouwer}$^f$ is defined as following:
Given an input instance of {\sc Brouwer}$^f$,
  $(C,0^n)$, where
  $C$ is a valid Brouwer-mapping
  circuit with parameters $d=\lceil n/f(n) \rceil$ and $\rr\in
  \mathbb{Z}_+^d$ where $\forall i\in [1:d]$, $r_i=2^{f(n)}$,
  find a panchromatic simplex of $C$.
\end{defi}

The {\em input size} of {\sc Brouwer}$^f$ is
  the sum of $n$ and the size of the circuit $C$.
{\sc Brouwer}$^{f_2}$ is a two-dimensional search problem
  over grid $[0:2^{\lfloor n/2\rfloor}-1]^2$
  and {\sc Brouwer}$^{f_3}$
  is a three-dimensional search problem
  over grid $[0:2^{\lfloor n/3\rfloor}-1]^3$, while
 {\sc Brouwer}$^{f_1}$ is a $\lceil n/3 \rceil$-dimensional search problem
  over grid $[0:7]^{\lceil n/3 \rceil}$.
Each of these three grids contains about $2^n$ hypercubes.
Both {\sc Brouwer}$^{f_2}$~\cite{XX06} and
  {\sc Brouwer}$^{f_3}$~\cite{DAS05} are known to be
  \textbf{PPAD}-complete.
In section \ref{BROUWER}, we will prove the
  following theorem, which states that the complexity of finding a
  panchromatic simplex is essentially independent of the shape or dimension
  of the search space.
In particular, it implies that  {\sc Brouwer}$^{f_1}$ is also
    \textbf{PPAD}-complete.

\begin{theo}[High-Dimensional Discrete Fixed Points]\label{mainFixed}
For each well-behaved function $f$,
  {\sc Brouwer}$^f$ is \textbf{\emph{PPAD}}-complete.
\end{theo}

\subsection{Generalized Circuits and Their Assignment Problem}

To effectively connect discrete Brouwer fixed points
  with two-player Nash equilibria,
  we use an intermediate structure called the generalized circuit.
This family of circuits, motivated by the reduction of
  \cite{DAS05,CHE05,DAS06}, extends the standard classes of
  Boolean or Arithmetic
  circuits in several aspects.

Syntactically, a {\em generalized circuit}
  $\calS=(V,\calT)$ is a pair, where $V$ is a
  set of nodes and $\calT$ is a collection of
  gates.
Every gate $T\in \calT$
  is a $5$-tuple $T=(G,v_1,v_2,v,\alpha)$ in which
\begin{itemize}
\item $G\in \setof{\hspace{0.05cm}
  G_\zeta, G_{\times \zeta}, G_=, G_{+}, G_{-},
  G_{<},G_\land,G_\lor,G_{\lnot}\hspace{0.05cm}}$
  is the type of the gate;

\item $v_1,v_2\in V\cup\{\hspace{0.04cm}nil
  \hspace{0.04cm}\}$ are the first
  and second input nodes of the gate;

\item $v\in V$ is the output node, and $\alpha\in \mathbb{R}
  \cup \{\hspace{0.04cm}nil\hspace{0.04cm}\}$.
\end{itemize}
The collection $\calT$ of gates must satisfy the
  following property: For every two gates $T=(G,v_1,
  v_2,v,\alpha)$ and $T'=(G',v_1',v_2',v',\alpha')$ in $\calT$,
  $v\not= v'$.

\begin{figure}[!t]
\centering
\includegraphics[width=5cm]{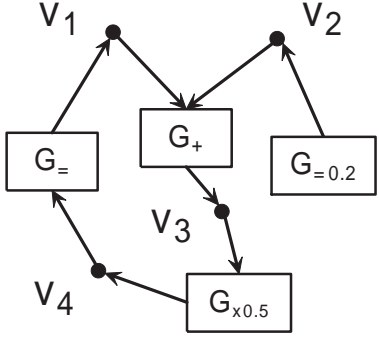}
\caption{An example of generalized circuits}\label{CIRCUIT}
\end{figure}

Suppose $T=(G,v_1,v_2,v,\alpha)$ in  $\calT$.
If $G=G_\zeta$, then the gate
  has no input node and $v_1=v_2=nil$.
If $G\in
  \{\hspace{0.04cm}G_{\times \zeta}, G_=,G_{\lnot}
  \hspace{0.04cm}\}$, then $v_1\in V$ and $v_2=nil$.
If $G\in \setof{\hspace{0.05cm}
  G_{+}, G_{-},  G_{<},G_\land,G_\lor\hspace{0.05cm}}$,
  then $v_1, v_2\in V$ and $v_1\not= v_2$.
Parameter $\alpha$ is only used in $G_\zeta$ and $G_{\times\zeta}$
  gates.
If $G=G_\zeta$, then $\alpha\in\mathbb{R}$ and
  $0\le \alpha\le 1/|V|$.
If $G=G_{\times\zeta}$, then $0\le \alpha \le 1$.
For other types of gates, $\alpha=\emph{nil}$.

The {\em input size} of a generalized circuit  is
  the sum of $|V|$ and the total number of bits
  needed to specify the $\alpha$ parameters in
  $\calS=(V,\calT)$.
As an important point which will become clear later,
   we make the following remark:
   In all generalized circuits that we will construct,
   the number of bits of each $\alpha$ parameter
   is upper bounded by poly$(|V|)$.

In addition to its more  expanded list of gate types,
  the generalized circuit differs crucially from the standard circuit
 in that it does not require the circuit to be acyclic.
In other words, in a generalized circuit, the directed graph
   defined by connecting input nodes of all gates to their output counterparts
  may have cycles.
We shall show later that the presence of cycles is necessary and
  sufficient to express fixed point computations
  with generalized circuits.

Semantically, we associate every node $v\in V$ with a real variable $\xx[v]$.
Each gate $T\in \calT$ requires that
   the variables of its input and output nodes
   satisfy certain constraints, either arithmetic
  or logical, depending on the type of the gate.
By setting $\epsilon = 0$, the constraints are defined in Figure~\ref{GG}.
The notation $=^{\hspace{0.06cm}\epsilon}_B$ in Figure~\ref{GG}
  will be defined shortly.
A generalized circuit defines
  a set of $K = |V|$ constraints, or a mathematical program,
  over the set of variables
  $\setof{\xx[v]\ |\ v \in V}$.

\begin{figure}[!t]
{ \rule{\textwidth}{1pt}\vspace{0.25cm}
\begin{tabular}{ll}
$G=G_\zeta$\hspace{0.06cm}: & \hspace{-0.3cm}$\calP[T, \epsilon]=\hspace{0.06cm}\Big[\hspace{0.14cm}
\xx[v]=\alpha\pm \epsilon\hspace{0.14cm} \Big]$\vspace{0.2cm}\\

$G=G_{\times \zeta}$\hspace{0.06cm}:\hspace{0.2cm} & \hspace{-0.3cm}$\calP[T,\epsilon]=\hspace{0.06cm}\Big[
\hspace{0.14cm}\xx[v]=\min\big( \hspace{0.04cm}\alpha\hspace{0.04cm} \xx[v_1],
1/K\hspace{0.04cm}\big)\pm \epsilon\hspace{0.14cm}\Big]$\vspace{0.2cm}\\

$G=G_=$\hspace{0.06cm}: & \hspace{-0.3cm}$\calP[T,\epsilon]= \hspace{0.06cm}\Big[\hspace{0.14cm}
\xx[v]=\min\big(\hspace{0.04cm}\xx[v_1],1/K\hspace{0.04cm}\big)\pm
\epsilon\hspace{0.14cm}\Big]$\vspace{0.2cm}\\

$G=G_+$\hspace{0.06cm}: & \hspace{-0.3cm}$\calP[T,\epsilon] =\hspace{0.06cm}\Big[\hspace{0.14cm} \xx[v]=\min
\big(\hspace{0.03cm}\xx[v_1]+\xx[v_2],1/K\hspace{0.04cm}
\big)\pm \epsilon\hspace{0.14cm}\Big]$\vspace{0.2cm}\\

$G=G_-$\hspace{0.06cm}: & \hspace{-0.3cm}$\calP[T,\epsilon]=\hspace{0.06cm}\Big[\hspace{0.04cm}
\min\big(\hspace{0.03cm}\xx[v_1]-\xx[v_2],1/K\hspace{0.03cm}\big)-\epsilon \le \xx[v] \le \max
\big(\hspace{0.03cm}\xx[v_1]-\xx[v_2],0\hspace{0.04cm}
\big)+\epsilon \hspace{0.14cm}\Big]$\vspace{0.2cm}\\

$G=G_<$\hspace{0.06cm}: & \hspace{-0.3cm}$\calP[T,\epsilon]=\hspace{0.06cm} \Big[\hspace{0.14cm}
\xx[v]=^{\hspace{0.06cm}\epsilon}_B 1 \text{\ \hspace{0.05cm}if\ \hspace{0.05cm}} \xx[v_1]<\xx[v_2]-\epsilon$;
$\xx[v]=^{\hspace{0.06cm}\epsilon}_B 0 \text{\ \hspace{0.05cm}if\ \hspace{0.05cm}}
\xx[v_1]>\xx[v_2]+\epsilon \hspace{0.14cm} \Big]$\vspace{0.25cm}\\

$G=G_{\lor}$\hspace{0.06cm}: & \hspace{-0.3cm}$\calP[T,\epsilon]=\hspace{0.05cm} \left[ \begin{array}{c}\xx[v]\BOO=
1\text{\ \hspace{0.05cm}if\ \hspace{0.05cm}}
\xx[v_1]\BOO= 1 \text{\ \hspace{0.05cm}or\ \hspace{0.05cm}} \xx[v_2]\BOO= 1\\[1.1ex]
\xx[v]\BOO= 0\text{\ \hspace{0.05cm}if\ \hspace{0.05cm}}\xx[v_1]\BOO= 0 \text{\ \hspace{0.05cm}and\ \hspace{0.05cm}}
\xx[v_2]\BOO= 0\end{array} \right]$\vspace{0.25cm}\\

$G=G_{\land}$\hspace{0.06cm}: & \hspace{-0.3cm}$\calP[T,\epsilon]=\hspace{0.05cm}\left[
\begin{array}{c}\xx[v]\BOO= 0\text{\ \hspace{0.05cm}if\ \hspace{0.05cm}}
\xx[v_1]\BOO= 0\text{\ \hspace{0.05cm}or\ \hspace{0.05cm}}\xx[v_2]\BOO= 0\\[1.1ex]
\xx[v]\BOO= 1\text{\ \hspace{0.05cm}if\ \hspace{0.05cm}}\xx[v_1]\BOO= 1 \text{\ \hspace{0.05cm}and\ \hspace{0.05cm}}
\xx[v_2]\BOO= 1\end{array} \right]$\vspace{0.25cm}\\

$G=G_{\lnot}$\hspace{0.06cm}: & \hspace{-0.3cm}$\calP[T,\epsilon] =\hspace{0.06cm}\Big[\hspace{0.14cm}\xx[v]\BOO=
0\text{\ \hspace{0.05cm}if\ \hspace{0.05cm}} \xx[v_1]\BOO= 1;\ \xx[v]\BOO= 1\text{\ \hspace{0.05cm}if\ \hspace{0.05cm}}
\xx[v_1]\BOO= 0\hspace{0.14cm}\Big]$\\
\end{tabular}}\vspace{0.25cm}

\rule{\textwidth}{1pt} \caption{Constraints $\calP[T,\epsilon]$,
  where $T=(G,v_1,v_2,v,\alpha)$} \label{GG}
\end{figure}

Suppose $\calS=(V,\calT)$ is a generalized circuit and
  $K=|V|$.
For every $\epsilon\ge 0$, an {\em$\epsilon$-approximate
  solution} to circuit $\calS$ is an assignment to
  the variables $\setof{\xx[v]\ |\ v \in V}$ such that
\begin{itemize}
\item the values of $\xx$ satisfy constraint
$\calP[\epsilon]=
\big[\hspace{0.08cm}0\le \xx[v]\le 1/K+\epsilon,
  \forall\ v\in V\hspace{0.08cm}\big];$ and

\item for each gate $T=(G,v_1,v_2,v,\alpha)\in \calT$,
  the values of  $\xx[v_1],
  \xx[v_2]$ and $\xx[v]$ satisfy the constraint $\calP[T,\epsilon]$,
  defined in Figure~\ref{GG}.
\end{itemize}




Among the nine types of gates, $G_\zeta, G_{\times \zeta},
  G_=, G_{+}$ and $G_{-}$ are arithmetic gates implementing
  arithmetic constraints like addition, subtraction and
  constant multiplication.
$G_<$ is a \emph{{brittle}} comparator; it only
  distinguishes values that are properly separated.
Finally, $G_\land,G_\lor$ and
  $G_{\lnot}$ are logic gates.
For an  assignment to variables $\setof{\xx[v]\ |\ v \in V}$,
   the value of $\xx[v]$
  represents boolean $1$ with precision
  $\epsilon$, denoted by $\xx[v] =^{\hspace{0.06cm}\epsilon}_B 1$,
  if $1/K-\epsilon \leq \xx[v] \leq 1/K+\epsilon$;
it represents boolean $0$ with precision $\epsilon$,
  denoted by $\xx[v]=^{\hspace{0.06cm}\epsilon}_B 0$,
 if $0\le \xx[v]\le \epsilon$.
We will use $\xx[v]=1/K\pm \epsilon$ to denote
the constraint that the value of $\xx[v]$ lies
in $[1/K-\epsilon,1/K+\epsilon]$.
The logic constraints implemented by the three
  logic gates are defined similarly as the classical ones.

From the reduction in Section~\ref{GCtoTG}, we can prove
  the following theorem.
A proof can be found in \textbf{Appendix~\ref{app:gcexist}}.
\begin{theo}\label{gcexist}
For any constant $c > 0$, every generalized circuit
  $\calS=(V,\calT)$ has a $1/|\hspace{0.04cm}V\hspace{0.04cm}|^c$-approximate
  solution.
\end{theo}

Let $c$ be a positive constant. We use {\sc Poly$^c$-Gcircuit}, and
  {\sc Exp$^c$-Gcircuit} to denote the problems of
  finding a $K^{-c}$-approximate
  solution and a $2^{-cK}$-approximate solution,
  respectively, of a given generalized circuit with $K$ nodes.

\section{Main Results and Proof Outline}\label{sec:Main}

As the main technical result of our paper, we prove the following theorem.

\begin{theo}[Main]\label{ppadb}
For any constant $c > 0$, {\sc Poly$^c$-Bimatrix}
  is \emph{\textbf{PPAD}}-complete.
\end{theo}

This theorem immediately implies the following statements about the complexity
  of computing and approximating two-player Nash equilibria.

\begin{theo}[Complexities of {\sc Bimatrix}]\label{thm:bimatrixMain}
{\sc Bimatrix} is \emph{\textbf{PPAD}}-complete. Moreover,
it does not have a fully-polynomial-time approximation
scheme, unless \emph{\textbf{PPAD}} is contained in \emph{\textbf{P}}.
\end{theo}

By Proposition \ref{prop:relatively}, {\sc Bimatrix} does not have
  a fully polynomial-time approximation scheme in the relative
  approximation of Nash equilibria.

Setting $\epsilon = 1/\mbox{poly} (n)$, by Theorem
  \ref{ppadb} and  Lemma \ref{lem:smoothed}, we obtain
  following theorem on the smoothed complexity of
  two-player Nash equilibria:

\begin{theo}[Smoothed Complexity of {\sc Bimatrix}]
{\sc Bimatrix} is not in smoothed polynomial time,
  under uniform or Gaussian perturbations, unless
  \emph{\textbf{PPAD} is contained in \textbf{RP}}.
\end{theo}

\begin{coro}[Smoothed Complexity of Lemke-Howson]\label{Theo:LemkeHowson}
If \emph{\textbf{PPAD} is not contained in \textbf{RP}},
  then the  smoothed complexity of the Lemke-Howson
  algorithm is not polynomial.
\end{coro}

By Proposition \ref{pro:bit}, we obtain the following corollary from
Theorem \ref{ppadb} about the complexity of {\sc Bit-Bimatrix}.

\begin{coro}[{\sc Bit-Bimatrix}]\label{cor:}
For any constant $c>1$, $(c\log n)$-{\sc Bit-Bimatrix}, the problem of
  finding the first $c\log n$ bits of a Nash equilibrium in a
  bimatrix game is polynomial-time equivalent to {\sc Bimatrix}.
\end{coro}

To prove Theorem~\ref{ppadb}, we will start with
  the discrete fixed point problem
  {\sc Brouwer}$^{f_1}$ (recall that
   $f_1(n)=3$ for all $n$).
As $f_1$ is a well-behaved function, Theorem \ref{mainFixed} implies that
  {\sc Brouwer}$^{f_1}$
  is a \textbf{PPAD}-complete problem.
We then apply the following three lemmas
  to reduce   {\sc Brouwer}$^{f_1}$ to {\sc Poly$^c$-Bimatrix}.

\begin{lemm}[{\sc FPC} to {\sc Gcircuit}]\label{fixedtogc}
{\sc Brouwer}$^{f_1}$ is polynomial-time reducible to {\sc Poly$^3$-Gcircuit}.
\end{lemm}

\begin{lemm}[{\sc Gcircuit} to {\sc Bimatrix}]\label{gctobimatrix}
{\sc Poly$^3$-Gcircuit} is polynomial-time reducible
  to {\sc Poly$^{12}$-Bimatrix}.
\end{lemm}


\begin{lemm}[Padding Bimatrix Games]\label{equivcc}
If {\sc Poly$^c$-Bimatrix} is \textbf{\emph{PPAD}}-complete
  for some constant $c>0$,
  then {\sc Poly$^{c'}$-Bimatrix} is
  \emph{\textbf{PPAD}}-complete for every constant $c'>0$.
\end{lemm}

We will prove Lemma \ref{fixedtogc} and Lemma \ref{gctobimatrix},
  respectively, in Section \ref{Final} and Section \ref{GCtoTG}.
A proof of Lemma \ref{equivcc} can be found in Appendix \ref{app:padding}.


\section{Simulating Generalized Circuits with Nash Equilibria}\label{GCtoTG}

In this section, we reduce {\sc Poly$^3$-Gcircuit}, the problem
of computing a $1/K^{3}$-approximate solution~of
  a generalized circuit of $K$ nodes, to  {\sc Poly$^{12}$-Bimatrix}.
As every two-player game has a Nash equilibrium,
  this reduction also implies that every generalized circuit with
  $K$ nodes has a $1/K^3$-approximate solution.

\subsection{Outline of the Reduction}

Suppose $\calS=(V,\calT)$ is a generalized circuit.
Let $K = |\hspace{0.03cm} V\hspace{0.03cm}|$ and $N=2K$.
Let $\calC$ be a one-to-one map from $V$ to
$\{\hspace{0.04cm}1,3, ...,2K-3,2K-1\hspace{0.04cm}\}$.
From every vector $\xx\in \mathbb{R}^N$,
   we define two maps $\overline{\xx},
  \hspace{0.06cm} \overline{\xx}_C:V
  \rightarrow \mathbb{R}$: For every node
  $v\in V$, supposing $\calC(v) = 2k-1$, we set  $\overline{\xx}[v]=
  x_{2k-1}$ and $\overline{\xx}_C[v]=x_{2k-1}
  +x_{2k}$.

In our reduction, we will build an
  $N\times N$ bimatrix game $\mathcal{G}^\calS=(\AA^\calS,\BB^\calS)$.
Our construction will take polynomial time and ensure the following properties
  for $\epsilon=1/K^3$.
\begin{itemize}
\item {\textbf{Property} $\textbf{A}_1$}: $|\hspace{0.04cm}
  a^{\calS}_{i,j}\hspace{0.04cm}|,|\hspace{0.04cm}b^{\calS}_{i,j}
\hspace{0.04cm} | \le N^3$, for all $i,j:1\le i,j\le N$ and

\item {\textbf{Property} $\textbf{A}_2$}: for every $\epsilon$-well-supported
  Nash equilibrium $(\xx,\yy)$ of game $\mathcal{G}^\calS$, $\overline{\xx}$
  is an $\epsilon$-approximate solution to $\calS$.
\end{itemize}
Then, we normalize $\calG^\calS$ to obtain $\overline{
  \mathcal{G}^\calS}=
  (\overline{\AA^\calS}, \overline{\BB^\calS})$ by setting
\begin{equation*}
\overline{a^{\calS}}_{i,j}=
  \frac{{a^{\calS}_{i,j}+N^3}}{2N^3}\ \ \ \mbox{and}\ \ \
\overline{b^{\calS}}_{i,j}=
  \frac{{b^{\calS}_{i,j}+N^3}}{2N^3},\ \ \ \mbox{for all }i,j:1\le i,j\le N.
\end{equation*}
\textbf{Property} $\textbf{A}_2$ implies that for every
  $\epsilon/(2N^3)$-well-supported equilibrium $(\xx,\yy)$ of
  $\overline{\cal{G}^\calS}$, $\overline{\xx}$ is
  an $\epsilon$-approximate solution to $\calS$.
By Lemma~\ref{lem:equivalence}, from every
  $2/N^{12}$-approximate Nash equilibrium of
  $\overline{\cal{G}^\calS}$, we can compute an
  $\epsilon$-approximate solution to $\calS$ in
  polynomial time.


\begin{figure}[!t]

\rule{ \textwidth}{1pt}\vspace{0.07cm}

$\LL[T]$ \textbf{and} $\RR[T]$\textbf{, where gate} $T=(G,v_1,v_2,v,\alpha)$

\vspace{-0.1cm}\rule{ \textwidth}{1pt}\vspace{0.22cm}
\begin{tabular}{ll}
 Set $\LL[T]=(L_{i,j})=\RR[T]=(R_{i,j})=0$, $k=\calC(v)$,
   $k_1=\calC(v_1)$ and $k_2=\calC(v_2)$ & \\ [1ex]

 \begin{tabular}{ll}
 \hspace{-0.2cm}$G_+$\hspace{0.05cm}:\hspace{-0.4cm} & \hspace{0.1cm}$L_{2k-1,2k-1}=L_{2k,2k}=
  R_{2k_1-1,2k-1}=R_{2k_2-1,2k-1}=R_{2k-1,2k}=1.$ \\ [1ex]

 \hspace{-0.2cm}$G_\zeta$\hspace{0.05cm}:\hspace{-0.4cm} & \hspace{0.1cm}$L_{2k-1,2k}=
 L_{2k,2k-1}=R_{2k-1,2k-1}=1, \hspace{0.06cm} R_{i,2k}=\alpha,
 \forall\hspace{0.06cm} i:1\le i\le 2K. $ \\ [1ex]

 \hspace{-0.2cm}$G_{\times \zeta}$\hspace{0.05cm}:\hspace{-0.4cm} & \hspace{0.1cm}$L_{2k-1,2k-1}=
 L_{2k,2k}=R_{2k-1,2k}=1,\hspace{0.06cm} R_{2k_1-1,2k-1}=\alpha.$ \\ [1ex]

 \hspace{-0.2cm}$G_=$\hspace{0.05cm}:\hspace{-0.4cm} & \hspace{0.1cm}$L_{2k-1,2k-1}=L_{2k,2k}=
 R_{2k_1-1,2k-1}=R_{2k-1,2k}=1.$ \\ [1ex]

 \hspace{-0.2cm}$G_-$\hspace{0.05cm}:\hspace{-0.4cm} & \hspace{0.1cm}$L_{2k-1,2k-1}=L_{2k,2k}=
 R_{2k_1-1,2k-1}=R_{2k_2-1,2k}=R_{2k-1,2k}=1.$ \\ [1ex]

 \hspace{-0.2cm}$G_<$\hspace{0.05cm}:\hspace{-0.4cm} & \hspace{0.1cm}$L_{2k-1,2k}=L_{2k,2k-1}=
 R_{2k_1-1,2k-1}=R_{2k_2-1,2k}=1.$ \\ [1ex]

 \hspace{-0.2cm}$G_\lor$\hspace{0.05cm}:\hspace{-0.4cm} & \hspace{0.1cm}$L_{2k-1,2k-1}=L_{2k,2k}=R_{2k_1-1,2k-1}=
 R_{2k_2-1,2k-1}=1,\hspace{0.06cm} R_{i,2k}=1/(2K),\forall\hspace{0.06cm} i:1\le i\le 2K.$ \\ [1ex]

 \hspace{-0.2cm}$G_\land$\hspace{0.05cm}:\hspace{-0.4cm} & \hspace{0.1cm}$L_{2k-1,2k-1}=L_{2k,2k}=R_{2k_1-1,2k-1}=
 R_{2k_2-1,2k-1}=1,\hspace{0.06cm} R_{i,2k}=3/(2K), \forall\hspace{0.06cm} i: 1\le i\le 2K.$ \\ [1ex]

 \hspace{-0.2cm}$G_{\lnot}$\hspace{0.05cm}:\hspace{-0.4cm} & \hspace{0.1cm}$L_{2k-1,2k}=L_{2k,2k-1}=
 R_{2k_1-1,2k-1}=R_{2k_1,2k}=1.$\vspace{0.2cm} \\
\end{tabular} & \\ [1ex]
\end{tabular}

\rule{ \textwidth}{1pt} \caption{Matrices $\LL[T]$ and $\RR[T]$}\label{PART2} \vspace{0.15cm}\end{figure}

In the remainder of this section, we assume $\epsilon = 1/K^3$.

\subsection{Construction of Game $\calG^\calS$}

To construct $\calG^\calS$, we transform
  a prototype
  game $\calG^*\hspace{-0.08cm}=(\AA^*,\BB^*)$, an $N\times N$
  zero-sum game to be defined in Section \ref{sub:proto},
  by adding $|\calT |$ carefully designed ``gadget'' games:
For each gate $T\in \calT$, we define a pair of $N\times N$ matrices
  $(\hspace{0.02cm} \LL[T], \RR[T]\hspace{0.02cm})$,
  according to Figure~\ref{PART2}.
Then, we set
$$\calG^\calS=(\AA^\calS,\BB^\calS),\ \text{where $\AA^{\calS}
  = \AA^*+\sum_{T\in\calT} \LL[T]$\hspace{0.05cm} and
  \hspace{0.05cm}$\BB^{\calS}=\BB^*+\sum_{T\in \calT} \RR[T]$}. $$

For each gate $T\in \calT$, $\LL[T]$ and $\RR[T]$ defined in
  Figure~\ref{PART2} satisfy the following property.

\begin{prope}\label{PROIMP}
Let $T=(G,v_1,v_2,v,\alpha)$, $\LL[T]=(L_{i,j})$ and $\RR[T]=(R_{i,j})$.
Suppose $\calC(v) = 2k-1$. Then,
\begin{eqnarray*}
i \not\in \setof{2k, 2k-1} &\Rightarrow &
  L_{i,j}=0, \quad \forall\ j\in [1:2K];\\
j\not\in \setof{2k,2k-1} &\Rightarrow& R_{i,j}=0, \quad
  \forall\ i\in [1:2K];\\
i\in \setof{2k,2k-1} &\Rightarrow& 0\le L_{i,j} \le 1, \quad
  \forall\ j\in [1: 2K];\\
j \in \setof{2k, 2k-1} & \Rightarrow & 0\le R_{i,j}\le 1, \quad
 \forall\ i\in [1:2K].
\end{eqnarray*}
\end{prope}
\subsection{The Prototype Game and Its Properties}\label{sub:proto}

The prototype $\calG^*=(\AA^*,\BB^*)$ is the bimatrix game called
  \emph{Generalized Matching Pennies} with parameter
  $M=2K^3$:
\begin{equation*}
\AA^*=\left(\begin{array}{ccccccc}
M & M & 0 & 0 & \cdots & 0 & 0 \\
M & M & 0 & 0 & \cdots & 0 & 0 \\
0 & 0 & M & M & \cdots & 0 & 0 \\
0 & 0 & M & M & \cdots & 0 & 0 \\
\vdots & \vdots & \vdots & \vdots & \ddots & \vdots & \vdots \\
0 & 0 & 0 & 0 & \cdots & M & M \\
0 & 0 & 0 & 0 & \cdots & M & M \\
\end{array}\right).
\end{equation*}
$\AA^*$ is a $K\times K$ block-diagonal matrix where each diagonal
  block is a $2\times 2$ matrix of all $M$'s,
  and $\BB^* = -\AA^*$.
All games we will consider below belong to the following class:

\begin{defi}[Class $\mathcal{L}$]
A bimatrix game $(\AA,\BB)$ is a member of $\mathcal{L}$ if
 the entries in
  $\AA-\AA^*$ and $\BB-\BB^*$ are in $[0:1]$.
\end{defi}

Note that every Nash equilibrium $(\xx,\yy)$ of
  $\calG^*$ enjoys the following nice property:
  For all $v\in V$, $\oxx_{C}[v]=\oyy_{C} [v]=1/K$.
We first prove an extension of
  this property for bimatrix games in  $\calL$.
Recall $\epsilon = 1/K^3$.

\begin{lemm}[Nearly Uniform Capacities]\label{MATCHING}
For every bimatrix game $(\AA,\BB)\in \mathcal{L}$,
if $(\xx,\yy)$ is a $1.0$-well-supported Nash equilibrium  of $(\AA,\BB)$,
  then
  $$1/K-\epsilon\hspace{0.06cm}\le\hspace{0.06cm} \oxxc[v],\hspace{0.06cm}
  \oyyc[v]\hspace{0.06cm}\le\hspace{0.06cm} 1/K+\epsilon,\ \mbox{
  for all }v\in V.$$
\end{lemm}
\begin{proof}
Recall that $\form{\aa}{\bb}$ denotes the inner product
  of two vectors $\aa$ and $\bb$ of the same length.
By the definition of class $\mathcal{L}$, for each $k$,
    the $2k-1^{st}$ and $2k^{th}$ entries
  of rows $\aa_{2k-1}$ and $\aa_{2k}$ in $\AA$ are in $[M, M+1]$
  and all other entries in these two rows are in $[0, 1]$.
Thus, for any probability vector $\yy\in \symP^n$ and
  for each node $v\in V$, supposing $\calC(v)=2k-1$, we have
\begin{align}
M\oyyc[v]\hspace{0.06cm}\le\hspace{0.06cm}
  \form{\aa_{2k-1}}{\yy},\hspace{0.06cm}
  \form{\aa_{2k}}{\yy} \hspace{0.06cm}\le\hspace{0.06cm}
  M\oyyc[v]+1. \label{eqn:set1}
\end{align}
Similarly,   the $(2l-1)^{th}$ and $2l^{th}$ entries
  of columns $\bb_{2l-1}$ and $\bb_{2l}$ in $\BB$ are in $[-M, -M+1]$
  and all other entries in these two columns are in $[0, 1]$.
Thus, for any probability vector $\xx\in \symP^n$ and for each
  node $v\in V$, supposing $\calC(v)=2l-1$, we have
\begin{align}
-M\oxxc[v] \hspace{0.06cm} \le \hspace{0.06cm} \form{\bb_{2l-1}}{\xx},\hspace{0.06cm}
  \form{\bb_{2l}}{\xx}  \hspace{0.06cm}\le\hspace{0.06cm}
  -M\oxxc[v]+1. \label{eqn:set2}
\end{align}

Now, suppose $(\xx,\yy)$ is a $t$-well-supported Nash equilibrium
  of $(\AA,\BB)$ for $t\leq 1$.
To warm up, we first prove that
  for each node $v\in V$, if $\oyyc[v]=0$ then $\oxxc[v]=0$.
Note that $\oyyc[v] = 0$ implies there exists
  $v'\in V$ with $\oyyc[v']\ge 1/K$.
Suppose $\calC(v) = 2l-1$ and $\calC(v')=2k-1$.
By Inequality (\ref{eqn:set1}),
\begin{equation*}
\form{\aa_{2k}}{\yy}- \max\Big(\form{\aa_{2l}}{\yy},
  \form{\aa_{2l-1}}{\yy}\Big)
  \ge M\oyyc[v']-\big(M\oyyc[v]+1\big)\ge M/K-1 >1
\end{equation*}
In other words, the payoff of the first player $P_1$ when choosing
  the $2k^{th}$ row is more than 1 plus the payoff
  of $P_1$ when choosing the $2l^{th}$ or the $(2l-1)^{th}$ row.
Because $(\xx,\yy)$ is a $t$-well-supported Nash equilibrium
  with $t\leq 1$, we have $\oxxc[v]=0$.

Next, we prove $ |\hspace{0.04cm} \oxxc[v]-1/K\hspace{0.04cm}|< \epsilon$ for all $v\in V$. To derive a contradiction,
we assume that this statement is not true.
Then, there exist $v,v'\in V$ such that
  $\oxxc[v]-\oxxc[v']>\epsilon$.
Suppose $\calC(v)=2l-1$ and $\calC(v')=2k-1$. By Inequality (\ref{eqn:set2}),
\begin{eqnarray*}
  \form{\bb_{2k}}{\xx}- \max\Big(\form{\bb_{2l}}{\xx},
    \form{\bb_{2l-1}}{\xx}\Big)
    \ge -M\oxxc[v']-\big(-M\oxxc[v]+1\big) >1,
\end{eqnarray*}
since $M=2K^3=2/\epsilon$. This would imply $\oyyc[v]=0$,
  and in turn imply $\oxxc[v]=0$, contradicting 
  our assumption that $\oxxc[v]>\oxxc[v']+\epsilon>0$.

We can similarly show $ |\hspace{0.04cm}
 \oyyc[v]-1/K\hspace{0.04cm} |< \epsilon$ for all $v\in V$.
\end{proof}

\subsection{Correctness of the Reduction}

We now prove that, for every $\epsilon$-well-supported
  equilibrium $(\xx,\yy)$ of $\calG^\calS$, $\oxx$ is
  an $\epsilon$-approximate solution to $\calS=(V,\calT)$.
It suffices to show, to be accomplished by the next two lemmas,
  that $\oxx$ satisfies the following
  collection of $1+|\hspace{0.03cm}\calT\hspace{0.03cm}|$
  constraints.
\begin{equation*}
\Big\{\hspace{0.08cm}\calP[\epsilon],\ \mbox{and\ }\calP[T,
  \epsilon],\ T\in \calT \hspace{0.08cm}\Big\}.
\end{equation*}


\begin{lemm}[Constraint \mbox{$\calP[\epsilon]$}]\label{inL}
Bimatrix game $\calG^\calS$ is in $\cal{L}$.
Thus,  for every $\epsilon$-well-supported Nash
  equilibrium $(\xx,\yy)$ of $\calG^\calS$, $\oxx$ satisfies
  constraint $\calP[\epsilon]= [\hspace{0.1cm}0\le \oxx[v]\le 1/K+\epsilon,
  \forall\hspace{0.06cm} v\in V\hspace{0.1cm}] $.
\end{lemm}
\begin{proof}
For each gate $T\in \calT$, $\LL[T]$ and $\RR[T])$ defined in
  Figure~\ref{PART2} satisfy Property \ref{PROIMP}.
By the definition the generalized circuit, gates in $\calT$
  have distinct output nodes, so $\calG^\calS\in \calL$.
The second statement of the lemma then follows from Lemma \ref{MATCHING}.
\end{proof}


\begin{lemm}[Constraints \mbox{$P[T,\epsilon]$}]\label{INSERT}
Let $(\xx,\yy)$ be an $\epsilon$-well-supported Nash
  equilibrium of $\calG^\calS$.
Then, for each gate $T\in \calT$,
 $\oxx$ satisfies constraint $\calP[T,\epsilon]$.
\end{lemm}
\begin{proof}
Recall $\calP[T,\epsilon]$ is a constraint defined in Figure~\ref{GG}.
By Lemma~\ref{inL}, $\xx$ and $\yy$ satisfy
  $$1/K-\epsilon\hspace{0.06cm}\le\hspace{0.06cm} \oxxc[v],\hspace{0.06cm}
  \oyyc[v]\hspace{0.06cm}\le\hspace{0.06cm} 1/K+\epsilon,\ \mbox{
  for all }v\in V.$$

Let $T=(G,v_1,v_2,v,\alpha)$ be a gate in $\calT$.
Suppose $\calC(v)=2k-1$. Let $\aa^*_i$ and $\lll_i$
  denote the $i^{th}$ row vectors of $\AA^*$ and
  $\LL[T]$, respectively; let $\bb^*_j$ and $\rr_j$ denote
  the $j^{th}$ column vectors of $\BB^*$ and $\RR[T]$,
  respectively.

From Property~\ref{PROIMP},
  $\LL [T]$ and $\RR [T]$ are the only two
  gadget matrices that modify
  the entries in rows $\aa^*_{2k-1},\aa^*_{2k}$ and
  in columns $\bb^*_{2k-1}$, $\bb^*_{2k}$,
  in the transformation from the prototype $\calG^*$ to $\calG^\calS$.
Thus, we have
\begin{eqnarray}
&\aa^\calS_{2k-1}=\aa^*_{2k-1}+\lll_{2k-1}, \ \ \aa^\calS_{2k}=\aa^*_{2k}+\lll_{2k};& \mbox{and}\label{eqx1}\\[0.5ex]
&\bb^\calS_{2k-1}=\bb^*_{2k-1}+\rr_{2k-1}, \ \ \bb^\calS_{2k}=\bb^*_{2k}+\rr_{2k}.&\label{eqx2}
\end{eqnarray}
Now, we prove
  $\oxx$ satisfies constraint
  $\calP[T,\epsilon]$.
Here we only consider the case when $G=G_+$.
In this case, we need to prove
  $\oxx[v]=\min (\hspace{0.015cm}\oxx[v_1]+\oxx[v_2],1/K\hspace{0.02cm}
 )\pm \epsilon.$
Proofs for other types of gates are similar and can
  be found in \textbf{Appendix~\ref{app:Gates}}.

Since $\aa^*_{2k-1}=\aa^*_{2k}$ and $\bb^*_{2k-1}=\bb^*_{2k}$,
  from (\ref{eqx1}), (\ref{eqx2}) and Figure~\ref{PART2}, we have
\begin{eqnarray}
&\form{\xx}{\bb^\calS_{2k-1}}-\form{\xx}{\bb^\calS_{2k}}=\oxx[v_1]+\oxx[v_2]-\oxx[v],\ \ \ \text{and}& \label{eqn:T1}\\ [0.5ex]
&\form{\aa^\calS_{2k-1}}{\yy}-\form{\aa^\calS_{2k}}{\yy}= \oyy[v]-\big(\hspace{0.04cm}\oyyc[v]-\oyy[v]
\hspace{0.04cm}\big).&  \label{eqn:T2}
\end{eqnarray}
In a proof by contradiction, we consider two cases.
First, we assume $\oxx[v]> \min(\oxx[v_1]+\oxx[v_2],1/K)+\epsilon$.
Since $\oxx[v]\le 1/K+\epsilon$, the
  assumption would imply  $\oxx[v]> \oxx[v_1]+\oxx[v_2] +\epsilon$.
By Equation (\ref{eqn:T1}), we have $\oyy[v]=y_{2k-1}=0$,
  because $(\xx,\yy)$ is an
  $\epsilon$-well-supported Nash equilibrium.
On the other hand, since $\oyyc[v]=1/K\pm\epsilon \gg \epsilon$,
by Equation (\ref{eqn:T2}),
 we have $\oxx[v]=x_{2k-1}=0$, contradicting our
assumption that $\oxx[v]> \oxx[v_1]+\oxx[v_2]+\epsilon>0$.

Next, we assume $\oxx[v]<\min(\oxx[v_1]+\oxx[v_2],1/K)-\epsilon
  \le \oxx[v_1]+\oxx[v_2]-\epsilon$.
Then, Equation (\ref{eqn:T1}) implies $\oyy[v]=\oyyc[v]$.
By Equation (\ref{eqn:T2}), we have $\oxx[v]=\oxxc[v]$
  and thus, $\oxx[v]\ge 1/K-\epsilon$, which
  contradicts our assumption that
  $\oxx[v]<\min(\oxx[v_1]+\oxx[v_2],1/K)-\epsilon\le 1/K-\epsilon$.
\end{proof}

We have now completed the proof of Lemma \ref{gctobimatrix}.
To prove our main technical Theorem \ref{ppadb},
  we only need to prove Theorem \ref{mainFixed}
  and Lemma \ref{fixedtogc}.

\section{PPAD-Completeness of {\sc Brouwer$^f$}}\label{BROUWER}

To prove Theorem \ref{mainFixed},
  we reduce a two-dimensional instance of
  {\sc Brouwer}$^{f_2}$, that is,
   a valid $3$-coloring  of a 2-dimensional grid,
   to {\sc Brouwer$^f$}, where recall
  l $f_2(n)=\lfloor n/2\rfloor$.
The basic idea of the reduction is to iteratively
  embed an instance of {\sc Brouwer}
  into a hypergrid one dimension higher to eventually
  ``fold'' or embed this two-dimensional input instance
  into the desired hypergrid.
We use the following concept to describe our
  embedding processes.
A triple $T=(C,d,\textbf{r})$ is a {\em coloring
  triple} if $\rr\in \mathbb{Z}^d$ with $r_i\ge 7$
  for all $1\le i\le d$ and $C$ is a valid
  Brouwer-mapping circuit with parameters $d$
  and $\rr $.
Let $\size{C}$ denote the number of gates
  plus the number of input and output
  variables in a circuit $C$.

Our embedding is carried out by a sequence of
  three polynomial-time transformations:
  $\LL^1(T,t,u)$, $\LL^2(T,u)$, and
  $\LL^3(T,t,a,b)$.
They embed a coloring triple $T$ into a
  larger $T'$ (\hspace{0.04cm}that is,
  the volume of the search
  space of $T'$ is greater than the one
  of $T$\hspace{0.04cm}) {\em such that from every
  panchromatic simplex of $T'$, one can
  find a panchromatic simplex of $T$ efficiently.}

To simplify our proof, in the context of this section,
  we slightly modify the definition of {\sc Brouwer}$^f$: In the original
  definition, each valid Brouwer-mapping circuit $C$
  defines a color assignment from the search space
  to $\{\hspace{0.04cm}1,2,3,...,d,d+1\hspace{0.04cm}\}$.
In this section, we replace the color $d+1$ by a special
  color ``red''.
In other words, if the output bits of $C$ evaluated at
  $\pp$ satisfy Case $i$ with $1\le i\le d$, then
  $\color{C}{\pp}=i$; otherwise,  the output bits
  satisfy Case $d+1$, and  $\color{C}{\pp}=$``red''.

We first prove a useful property of valid Brouwer-mapping circuits.

\begin{prope}[Boundary Continuity]\label{imp}
Let $C$ be a valid Brouwer-mapping circuit with
  parameters $d$ and $\rr$.
If points $\pp,\pp'\in
  \partial({A_{\rr}^d})$ satisfy $\pp'=\pp+\ee_t$
  for some $1\le t\le d$ and $1\le p_t \le r_t-2$,
  then $\color{C}{\pp}=\color{C}{\pp'}$.
\end{prope}

\begin{proof}
By the definition, if $C$ is a  valid
  Brouwer-mapping circuit $C$ with parameters $d$ and $\rr$,
  then for each $\pp \in \bdry{A_{\rr}^d}$,
  $\color{C}{\pp}$ has the following property:
If there exists an $i \in [1:d]$ such
that $p_i=0$, then $\color{C}{\pp} = \max\setof{\hspace{0.06cm}
  i\ |\ p_{i}=0\hspace{0.06cm}}$; otherwise, $\forall i$, $p_i \neq 0$
  and $\exists i$, $p_i = r_i-1$, we have $\color{C}{\pp} = \mbox{``red''}$.
Thus, if $1\le p_t \le r_t-2$, then $\color{C}{\pp}=\color{C}{\pp'}$.
\end{proof}

\subsection{Reductions Among Coloring Triples}

Both $\LL^1(T,t,u)$ and $\LL^2(T,u)$ are very simple
  operations:
\begin{itemize}
\item Given a coloring triple $T=(C,d,\rr)$ and
  two integers $1\le t\le d$,
  $u>r_t$, $\LL^1(T,t,u)$ pads dimension $t$ to
  size $u$, i.e., it builds a new coloring triple
  $T'=(C',d,\rr')$ with $r'_t=u$ and $r'_i=r_i$,
  for all $i:1\le i\not=t\le d$.

\item For integer $u\ge 7$, $\LL^2(T,u)$ adds a dimension
  to $T$ by constructing $T'=(C',d+1,\rr')$ such
  that $\rr'\in \mathbb{Z}^{d+1}$, $r'_{d+1}=u$
  and $r'_i=r_i$, for all $i \in [1; d]$.
\end{itemize}
These two transformations are described in Figure~\ref{consC1}
  and Figure~\ref{consC2}, respectively.
We prove their properties in the following two lemmas.

\begin{figure}[!t]

\rule{\textwidth}{1pt}\vspace{0.07cm}

\textbf{$\color{C'}{\pp}$ of a point $\pp\in A_{\rr'}^{d}$
  assigned by $(C',d,\rr')=\LL^1(T,t,u)$}


\vspace{-0.15cm}\rule{\textwidth}{1pt}\vspace{0.18cm}

\begin{tabular}{@{\hspace{0.1cm}}r@{\hspace{0.2cm}}p{\textwidth}}
1: & \textbf{if} $\pp\in \bdry{A_{\rr'}^d}$ \textbf{then} \\ [0.8ex]

2: & \ \ \ \ \textbf{if} there exists $i$ such that $p_i=0$
  \textbf{then} \\ [0.8ex]

3: & \ \ \ \ \ \ \ \ $\color{C'}{\pp}=i_{\max}=\max
  \{\hspace{0.06cm}i\ \big|\ p_i=0\hspace{0.06cm}\}$ \\ [0.8ex]

4: & \ \ \ \ \textbf{else} \\ [0.8ex]

5: & \ \ \ \ \ \ \ \ $\color{C'}{\pp}=\text{red}$ \\ [0.8ex]

6: & \textbf{else if} $p_t\le r_t$ \textbf{then} \\ [0.8ex]

7: & \ \ \ \ $\color{C'}{\pp}=\color{C}{\pp}$ \\ [0.8ex]

8: & \textbf{else} \\ [0.8ex]

9: & \ \ \ \ $\color{C'}{\pp}=\text{red}$ \\ [0.8ex]
\end{tabular}

\rule{\textwidth}{1pt} \caption{How $\LL^1(T,t,u)$ extends
  the coloring triple $T=(C,d,\rr)$} \label{consC1}
\end{figure}

\begin{lemm}[$\LL^1(T,t,u)$: Padding a Dimension]\label{lem1}
Given a coloring triple $T=(C,d,\rr)$ and
  two integers $1\le t\le d$ and
  $u>r_t$, we can construct a new coloring triple
  $T'=(C',d,\rr')$ that satisfies
  the following two conditions:
\begin{itemize}
\item[\textbf{\emph{A.}}]
For all $i:1\le i\not=t\le d$, $r'_i=r_i$, and $r'_t=u$. In addition, there exists a polynomial $g_1(n)$ such that
  $\size{C'}=\size{C}+O(g_1(\size{\rr'}))$
  and $T'$ can be computed in time polynomial
  in $\size{C'}$.
We write $T' = \LL^1(T,t,u)$;

\item[\textbf{\emph{B.}}] From each panchromatic simplex
   $P'$ of coloring triple $T'$, we can compute a
   panchromatic simplex $P$ of $T$ in polynomial time.
\end{itemize}
\end{lemm}

\begin{proof}
We define circuit $C'$ by its color assignment in
  Figure~\ref{consC1}.
Property \textbf{A} is true according to this definition.

To show Property \textbf{B}, let $P'\subset K_{\pp}$
  be a panchromatic simplex  of $T'$.
We first note that $p_t< r_t-1$, because had $p_t
  \ge r_t-1$, $K_{\pp}$ would not contain color $t$
  according to the color assignment.
Thus, it follows from $\color{C'}{\qq}=\color{C}{\qq}$
  for each $\qq\in A_{\rr}^d$ that $P'$ is
  also a panchromatic simplex of the coloring
  triple $T$.
\end{proof}

\begin{lemm}[$\LL^2(T,u)$: Adding a Dimension]\label{lem2}
Given a coloring triple $T=(C,d,\rr)$ and integer $u\ge 7$,
  we can construct a new coloring triple
  $T'=(C',d+1,\rr')$ that satisfies the following conditions:
\begin{itemize}
\item[\emph{\textbf{A.}}]
For all $i: 1\le i\le d$, $r'_i=r_i$, and  $r'_{d+1}=u$.
  Moreover, there exists a polynomial $g_2(n)$
  such that $\size{C'}=\size{C}+O(g_2(\size{\rr'}))$.
  $T'$ can be computed in time polynomial
  in $\size{C'}$.
We write $T' = \LL^2(T,u)$.

\item[\emph{\textbf{B.}}]
From each panchromatic simplex $P'$ of
  coloring triple $T'$, we can compute a
  panchromatic simplex $P$ of $T$ in polynomial time.
\end{itemize}
\end{lemm}

\begin{proof}
For each point $\pp\in A_{\rr'}^{d+1}$,
  we use $\hat{\pp}$ to denote the
  point $\zz\in A_{\rr}^d$ with $z_i=p_i$, $\forall
  \hspace{0.06cm} i \in [1: d]$.
The color assignment of circuit $C'$ is given in
  Figure~\ref{consC2}. Clearly, Property \textbf{A} is true.

To prove Property \textbf{B}, we let $P'\subset
  K_{\pp}$ be a panchromatic simplex of $T'$.
We note that $p_{d+1}=0$, for otherwise, $K_{\pp}$
  does not contain color $d+1$.
Note also that $\color{C'}{\qq}=d+1$ for every $\qq\in
  A_{\rr'}^{d+1}$ with $q_{d+1}=0$.
Thus, for every point $\qq\in P'$ with $\color{C'}{\qq}
  \not= d+1$, we have $q_{d+1}=1$.
So,  because $\color{C'}{\qq}=\color{C}{\hat{\qq}}$
  for every $\qq\in A_{\rr'}^{d+1}$ with $q_{d+1}=1$,
  $P=\{\hspace{0.04cm}\hat{\qq}\ \big|\
  \qq\in P' \mbox{\ and\ }\color{C'}{\qq}\not=d+1
  \hspace{0.04cm}\}$
  is a panchromatic simplex of $T$.
\end{proof}

\begin{figure}[!t]

\rule{\textwidth}{1pt}\vspace{0.07cm}

\textbf{$\color{C'}{\pp}$ of a point $\pp\in A_{\rr'}^{d+1}$
  assigned by $(C',d+1,\rr')=\LL^2(T,u)$}

\vspace{-0.15cm}\rule{\textwidth}{1pt}\vspace{0.18cm}

\begin{tabular}{@{\hspace{0.1cm}}r@{\hspace{0.2cm}}p{\textwidth}}
1: & \textbf{if} $\pp\in \bdry{A_{\rr'}^d}$ \textbf{then} \\ [0.8ex]

2: & \ \ \ \ \textbf{if} there exists $i$ such that
  $p_i=0$ \textbf{then} \\ [0.8ex]

3: & \ \ \ \ \ \ \ \ $\color{C'}{\pp}=i_{\max}=\max\{\hspace{0.06cm}i\
  \big|\ p_i=0\hspace{0.06cm}\}$ \\ [0.8ex]

4: & \ \ \ \ \textbf{else} \\ [0.8ex]

5: & \ \ \ \ \ \ \ \ $\color{C'}{\pp}=\text{red}$ \\ [0.8ex]

6: & \textbf{else if} $p_{d+1}=1$ \textbf{then} \\ [0.8ex]

7: & \ \ \ \ $\color{C'}{\pp}=\color{C}{\hat{\pp}}$, where
  $\hat{\pp}\in\mathbb{Z}^d$ satisfying $\hat{p}_i=p_i$ for all
  $1\le i\le d$ \\ [0.8ex]

8: & \textbf{else} \\ [0.8ex]

9: & \ \ \ \ $\color{C'}{\pp}=\text{red}$ \\ [0.8ex]
\end{tabular}

\rule{\textwidth}{1pt} \caption{How $\LL^2(T,u)$ extends the coloring triple $T=(C,d,\rr)$} \label{consC2}
\end{figure}

Transformation $\LL^3(T,t,a,b)$ is the one that
  does all the hard work.

\begin{lemm}[$\LL^3(T,t,a,b)$: Snake Embedding]\label{lem3}
Given a coloring triple $T=(C,d,\rr)$ and integer $1\le t\le d$,
 if $r_t=a(2b+1)+5$ for two integers $a,b\ge 1$,
   then we can construct a new triple $T'=(C',d+1,\rr')$ that
  satisfies the following conditions:
\begin{itemize}
\item[\emph{\textbf{A.}}]
For $i: 1\le i\not=t\le d$, $r'_i=r_i$ and $r'_t=a+5$ and
  $r'_{d+1}=4b+3$. Moreover, there exists a polynomial $g_3(n)$
  such that $\size{C'}=\size{C}+O(g_3(\size{\rr'}))$
  and $T'$ can be computed in time polynomial in $\size{C'}$.
We write  $T' = \LL^3(T,t,a,b)$.

\item[\emph{\textbf{B.}}] From each panchromatic simplex
   $P'$ of coloring triple $T'$, we can compute a panchromatic
   simplex $P$ of $T$
   in polynomial time.
\end{itemize}
\end{lemm}

\begin{proof}
Consider the domains
   $A_{\rr}^d\subset\mathbb{Z}^d$ and
   $A_{\rr'}^{d+1}\subset\mathbb{Z}^{d+1}$ of our coloring
   triples.
We form the reduction $\LL^3(T,t,a,b)$ in three steps. First, we define a $d$-dimensional set $W \subset
  A_{\rr'}^{d+1}$ that is
  large enough to contain $A_{\rr}^d$.
Second, we define a map $\psi$ from $W$ to $A_{\rr}^d$
  that (\hspace{0.04cm}implicitly\hspace{0.04cm})
  specifies an embedding of $A_{\rr}^d$ into $W$.
Finally, we build a circuit $C'$ for  $A_{\rr'}^{d+1}$
  and show that from each panchromatic simplex of
  $C'$, we can, in polynomial time, compute a
  panchromatic simplex of $C$.

A two dimensional view of $W\subset A_{\rr'}^{d+1}$
  is illustrated in Figure~\ref{setW}.
We use a snake-pattern to realize
  the longer $t^{th}$ dimension of $A_{\rr}^d$ in the
  two-dimensional space defined by the shorter $t^{th}$
  and $(d+1)^{th}$ dimensions of $A_{\rr'}^{d+1}$.
Formally, $W$ consists of points $\pp\in
  A_{\rr'}^{d+1}$ satisfying $1\le p_{d+1}\le 4b+1$ and
\begin{itemize}
\item[] if $p_{d+1}=1$, then $2\le p_t\le a+4$; 
\vspace{-0.15cm}

\item[]
  if $p_{d+1}=4b+1$, then $0\le p_t\le a+2$;\vspace{-0.15cm}

\item[] if $p_{d+1}=4(b-i)-1$ where $0\le i\le b-1$,
  then $2\le p_t\le a+2$;\vspace{-0.15cm}

\item[] if $p_{d+1}=4(b-i)-3$ where $0\le i\le b-2$,
  then $2\le p_t\le a+2$;\vspace{-0.15cm}

\item[] if $p_{d+1}=4(b-i)-2$ where $0\le i\le b-1$,
  then $p_t=2$;\vspace{-0.15cm}

\item[] if $p_{d+1}=4(b-i)$ where $0\le i\le b-1$, then $p_t=a+2$.
\end{itemize}

\begin{figure}[!t]
\centering
\includegraphics[width=7cm]{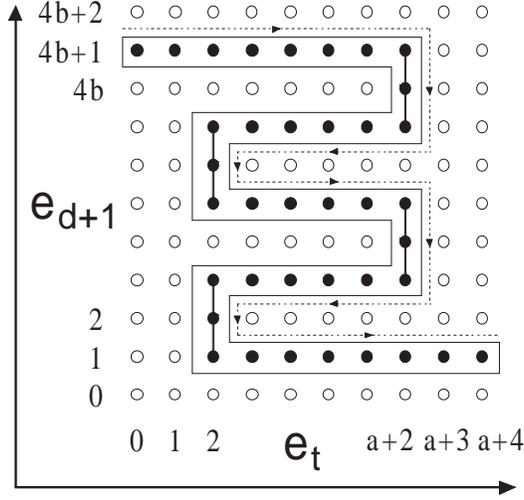}
\caption{The two dimensional view of set $W\subset A_{\rr'}^{d+1}$}\label{setW}
\end{figure}

To build $T'$, we embed the coloring triple
  $T$ into $W$.
The embedding is implicitly given
  by a natural surjective map $\psi$ from
  $W$ to $A_{\rr}^d$, a map that will play a
  vital role in our construction and analysis.
For each $\pp\in W$, we use $\pp[m]$ to denote
  the point $\qq$ in $\mathbb{Z}^d$ such that
  $q_t=m$ and $q_i=p_i$, for all $i:1\le i\not=t\le d$.
We define $\psi(\pp)$ according to the following
  cases:
\begin{itemize}
\item[] if $p_{d+1}=1$, then $\psi(\pp)=\pp[2ab+p_t]$
\vspace{-0.15cm}

\item[] if $p_{d+1}=4b+1$, then
  $\psi(\pp)=\pp[p_t]$;\vspace{-0.15cm}

\item[] if $p_{d+1}=4(b-i)-1$ where $0\le i\le b-1$,
  then $\psi(\pp)=\pp[(2i+2)a+4-p_t]$;\vspace{-0.15cm}

\item[] if $p_{d+1}=4(b-i)-3$ where $0\le i\le b-2$,
  then $\psi(\pp)=\pp[(2i+2)a+p_t]$;\vspace{-0.15cm}

\item[] if $p_{d+1}=4(b-i)-2$ where $0\le i\le b-1$,
  then $\psi(\pp)=\pp[(2i+2)a+2]$;\vspace{-0.15cm}

\item[] if $p_{d+1}=4(b-i)$ where $0\le i\le b-1$,
  then $\psi(\pp)=\pp[(2i+1)a+2]$.
\end{itemize}

Essentially, we map $W$ bijectively to $A_{\rr}^d$
  along its $t^{th}$ dimension
  with exception that when the snake pattern of $W$
  is making a turn, we stop the advance in $A_{\rr}^{d}$,
  and continue the advance after it completes the turn.
Let $\psi_i(\pp)$ denote the $i^{th}$ component of
  $\psi(\pp)$. Our embedding scheme guarantees the
  following important property of $\psi$.

\begin{prope}[Boundary Preserving]\label{prop:boundaryPreserving}
Let $\pp$ be a point in $W\cap \partial\hspace{0.04cm}({A_{\rr'}^{d+1}})$. If there exists $i$ such that $p_i=0$, then
  $\max\{\hspace{0.06cm}i\ |\ p_i=0\hspace{0.06cm}\}=
  \max\{\hspace{0.06cm}i\ |\ \psi_i(\pp)=0\hspace{0.06cm}\}.$
Otherwise, all entries of $\pp$ are non-zero and there
  exists $l$ such that $p_l=r'_l-1$, in which case,
  all entries of point $\psi(\pp)$ are
  nonzero and $\psi_l(\pp)=r_l-1$.
\end{prope}

\begin{figure}

\rule{\textwidth}{1pt}\vspace{0.07cm}

\textbf{$\color{C'}{\pp}$ of a point $\pp\in A_{\rr'}^{d+1}$
  assigned by $(C',d+1,\rr')= \LL^3(T,t,a,b)$}

\vspace{-0.15cm}\rule{\textwidth}{1pt}\vspace{0.1835cm}

\begin{tabular}{@{\hspace{0.1cm}}r@{\hspace{0.2cm}}p{\textwidth}}
1: & \textbf{if} $\pp\in W$ \textbf{then} \\ [0.8ex]

2: & \ \ \ \ $\color{C'}{\pp}=\color{C}{\psi(\pp)}$ \\ [0.8ex]

3: & \textbf{else if} $\pp\in \partial\hspace{0.05cm}
  {(A_{\rr'}^{d+1})}$ \textbf{then} \\ [0.8ex]

4: & \ \ \ \ \textbf{if} there exists $i$ such that $p_i=0$
  \textbf{then} \\ [0.8ex]

5: & \ \ \ \ \ \ \ \ $\color{C'}{\pp}=i_{\max}=\max
  \{\hspace{0.06cm}i\ \big|\ p_i=0\hspace{0.06cm}\}$ \\ [0.8ex]

6: & \ \ \ \ \textbf{else} \\ [0.8ex]

7: & \ \ \ \ \ \ \ \ $\color{C'}{\pp}=\text{red}$ \\ [0.8ex]

8: & \textbf{else if} $p_{d+1}=4i$ where $1\le i\le b$
  and $1\le p_t\le a+1$ \textbf{then} \\ [0.8ex]

9: & \ \ \ \ $\color{C'}{\pp}=d+1$ \\ [0.8ex]

10: & \textbf{else if} $p_{d+1}=4i+1$, $4i+2$ or $4i+3$
  where $0\le i\le b-1$ and $p_t=1$ \textbf{then} \\ [0.8ex]

11: & \ \ \ \ $\color{C'}{\pp}=d+1$ \\ [0.8ex]

12: & \textbf{else} \\ [0.8ex]

13: & \ \ \ \ $\color{C'}{\pp}=\text{red}$ \\ [0.8ex]
\end{tabular}

\rule{\textwidth}{1pt} \caption{How $\LL^3(T,t,a,b)$ extends
  the coloring triple $T=(C,d,\rr)$} \label{consC}
\end{figure}

The circuit $C'$ specifies a color assignment of $A_{\rr'}^{d+1}$
  according to Figure~\ref{consC}.
$C'$ is derived from circuit $C$ and map $\psi$. By Property \ref{prop:boundaryPreserving},
  we can verify that $C'$ is a valid Brouwer-mapping
  circuit with parameters $d+1$ and $\rr'$.

Property \textbf{A} follows directly from our
  construction.
In order to establish Property \textbf{B} of the lemma, we
  prove the following collection of statements to cover all
  possible cases of the given panchromatic simplex $P'$ of $T'$.
In the following statements, $P'$ is a panchromatic simplex of $T'$ in
  $A_{\rr'}^{d+1}$ and let $\pp^{*} \in A_{\rr'}^{d+1}$ be the point
  such that $P'\subset  K_{\pp^*}$.
We will also use the following notation: For each $\pp\in
  A_{\rr'}^{d+1}$, we will use $\pp[m_1,m_2]$ to denote the
  point $\qq\subset \mathbb{Z}^{d+1}$
  such that $q_t=m_1$, $q_{d+1}=m_2$ and $q_i=p_i$
  for all $i:1\le i\not=t\le d$.

\begin{stat}\label{st1}
If $p^*_t=0$, then $p^*_{d+1}=4b$ and furthermore, for every
  point $\pp\in P'$ such that
  $\color{C'}{\pp}\not=d+1$, $\color{C}{\psi(\pp[p_t,4b+1])}=\color{C'}{\pp}$.
\end{stat}
\begin{proof}
First, note that $p^*_{d+1} \neq  4b+1$, for otherwise, $K_{\pp^*}$ does not
  contain color $d+1$.
Second, if $p^*_{d+1}<4b$, then each point
   $\qq\in K_{\pp^*}$ is colored according one of the conditions in
  line 3, 8 or 10 of Figure~\ref{consC}.
Let $\qq^*\in K_{\pp^*}$ be the ``red'' point in $P'$. Then,
  $\qq^*$ must satisfy the condition in line 6 and hence
  there exists $l$ such that $q^*_l=r'_l-1$.
By our assumption, $p^*_t=0$. Thus, if $p^*_{d+1}<4b$,
  then  $l \not\in \setof{t,d+1}$, implying
   for each $\qq\in K_{\pp^*}$, $q_l>0$
(\hspace{0.04cm}as $q_l\ge q^*_l-1>0$\hspace{0.04cm}) and
  $\color{C'}{\qq}\not=l$.
Then, $K_{\pp^*}$ does not contain color $l$, contradicting the assumption
  of the statement.
Putting these two cases together, we have $p^*_{d+1}=4b$.

We now prove the second part of the statement. If $p_{d+1}=4b+1$, then we are done, because
  $\color{C}{\psi(\pp)}=\color{C'}{\pp}$ according to
  line 1 of Figure~\ref{consC}.
Let us assume $p_{d+1} = 4b$.
Since the statement
  assumes $\color{C'}{\pp}\not=d+1$,
  $\pp$ satisfies the condition in line 3
  and hence $\pp\in \partial\hspace{0.05cm}(
  {A_{\rr'}^{d+1}})$.
By Property~\ref{imp}, we have
  $\color{C'}{\pp[p_t,4b+1]}=\color{C'}{\pp}$,
  completing the proof of the statement.
\end{proof}

\begin{stat}\label{st2}
If $p^*_t=a+2$ or $a+3$, then $p^*_{d+1}=0$, and in addition,
   for each $\pp\in P'$ such that $\color{C'}{\pp}\not=d+1$,
   $\color{C}{\psi(\pp[p_t,1])}=\color{C'}{\pp}$.
\end{stat}
\begin{proof}
If $p^*_{d+1} > 0$, then $K_{\pp^*}$ does not contain color $d+1$. So
  $p^*_{d+1}=0$.
In this case, $p_{d+1}$ must be $1$, since $\color{C'}{\qq}=d+1$
  for any $\qq\in A_{\rr'}^{d+1}$ with $q_{d+1}=0$.
Thus, $\color{C}{\psi(\pp[p_t,1])}=
  \color{C'}{\pp[p_t,1]}=\color{C'}{\pp}$.
\end{proof}

\begin{stat}\label{st5}
If $p^*_{d+1}=4b$, then $0\le p^*_t\le a+1$. Moreover,
 for each $\pp\in P'$ such that $\color{C'}{\pp}\not= d+1$,
  $\color{C}{\psi(\pp[p_t,4b+1])}=\color{C'}{\pp}$.
\end{stat}
\begin{proof}
The first part of the statement is straightforward. Similar to the proof of Statement~\ref{st1},
  we can prove the second part for the case when $0\le p_t\le a+1$.
When $p_t=a+2$, we have $\psi(\pp)=\psi(\pp[p_t,4b+1])$. Thus,
$\color{C}{\psi(\pp[p_t,4b+1])}=\color{C}{\psi(\pp)}=\color{C'}{\pp}$.
\end{proof}

We can similarly prove the following statements.

\begin{stat}\label{st6}
If $p^*_{d+1}=4i+1$ or $4i+2$ for some $0\le i\le b-1$, then
  $p^*_t=1$. Moreover, for each
  $\pp\in P'$ such that $\color{C'}{\pp}\not=d+1$, $\color{C}{\psi(\pp[2,p_{d+1}])}$
$=\color{C'}{\pp}$.\vspace{0.1cm}
\end{stat}

\begin{stat}\label{st7}
  If $p^*_{d+1}=4i$ for some $1\le i\le b-1$, then $1\le p^*_t\le a+1$.
  In addition, for each
  $\pp\in P'$ such that $\color{C'}{\pp}\not=d+1$, if $2\le p_t\le a+1$, then
$\color{C}{\psi(\pp[p_t,4i+1])}=\color{C'}{\pp}$; if $p_t=1$, then
$\color{C}{\psi(\pp[2,4i+1])}=\color{C'}{\pp}$.\vspace{0.1cm}
\end{stat}

\begin{stat}\label{st8}
If  $p^*_{d+1}=4i-1$ for some $1\le i\le b$, then $1\le
  p^*_t\le a+1$. Moreover, for each
  $\pp\in P'$ such that $\color{C'}{\pp}\not=d+1$, if $2\le p_t\le a+1$, then
$\color{C}{\psi(\pp[p_t,4i-1])}=\color{C'}{\pp}$; if $p_t=1$, then
$\color{C}{\psi(\pp[2,4i-1])}=\color{C'}{\pp}$.\vspace{0.1cm}
\end{stat}

\begin{stat}\label{st4}
If $p^*_{d+1}=0$, then $1\le p^*_t\le a+3$. In addition,
  for each $\pp\in P'$ such that $\color{C'}{\pp}\not=d+1$, if $2\le p^*_t\le a+1$, then
$\color{C}{\psi(\pp[p_t,1])}=\color{C'}{\pp}$; if $p^*_t=1$, then
$\color{C}{\psi(\pp[2,1])}=\color{C'}{\pp}$.\vspace{0.1cm}
\end{stat}
In addition,
\begin{stat}\label{st3}
$p^*_{d+1}\neq 4b+1$.
\end{stat}
\begin{proof}
If $p^*_{d+1}=4b+1$ then $K_{\pp^*}$ does not
  contain color $d+1$.
\end{proof}

Now suppose that $P'$ is a panchromatic simplex of  $T'$.
Let $\pp^*\in \mathbb{Z}^{d+1}$ be the point such
  that $P'\subset K_{\pp^*}$.
Then, $P'$ and $\pp^{*}$ must satisfy the
  conditions of one of the statements above.
By that statement, we can transform every point $\pp\in P'$,
  (aside from the one that has color $d+1$)
  back to a point $\qq$ in $A_{\rr}^d$ to obtain
  a set $P$ from $P'$.
Since $P$ is  accommodated, it is a panchromatic simplex of $C$.
Thus, with all the statements above,
  we specify an efficient algorithm to
  compute a panchromatic simplex  $P$ of $T$ given  a
  panchromatic simplex  $P'$ of $T'$.
\end{proof}

\subsection{PPAD-Completeness of Problem {\sc Brouwer}$^f$}

We are now ready to prove the main result of this section.

\begin{proof}[Proof of Theorem~\ref{mainFixed}]
We reduce {\sc Brouwer}$^{f_2}$ to {\sc Brouwer}$^f$
  in order to prove that the latter is \textbf{PPAD}-complete.
Recall, $f_2(n)=\lfloor n/2\rfloor$. Suppose  $(C,0^{2n})$
  is an input instance of {\sc Brouwer}$^{f_2}$.
Let
\begin{equation*}
l=f(11n)\ge 3\hspace{0.06cm},\ \ \ \ \ m'=\left\lceil
  \frac{n}{l-2}\right\rceil\ \ \ \text{and} \ \ \ m=\left\lceil
\frac{11n}{l}\right\rceil.
\end{equation*}

We iteratively  construct a sequence of coloring triples
$\mathcal{T}=\{\hspace{0.05cm}T^0, T^1,...\hspace{0.09cm}T^{w-1},T^w\hspace{0.05cm}\}$
  for some  $w=O(m)$, starting with $T^0=\left(C,2,
  \left(2^n,2^n\right)\right)$ and ending with
  $T^w=\left(C^w,m,\rr^w\right)$ where $\rr^w\in \mathbb{Z}^m$  and
  $r^w_i=2^l$, for all $i\in [1: m]$.
At the $t^{th}$ iteration, we apply
  either $\textbf{L}^1,\textbf{L}^2$ or $\textbf{L}^3$
  with properly chosen parameters  to build $T^{t+1}$ from $T^t$.

\begin{figure}[!t]

\rule{\textwidth}{1pt}\vspace{0.07cm}

\textbf{The Construction of $T^{3m'-14}$ from $T^1$}

\vspace{-0.15cm}\rule{\textwidth}{1pt}\vspace{0.18cm}

\begin{tabular}{@{\hspace{0.1cm}}r@{\hspace{0.2cm}}p{\textwidth}}
1: & \textbf{for any} $t$ from $0$ to $m'-6$ \textbf{do} \\ [0.8ex]

2: & \ \ \ \ It can be proved inductively that $T^{3t+1}
  =(C^{3t+1},d^{3t+1},\rr^{3t+1})$ satisfies \\ [0.9ex]

 & \ \ \ \  $d^{3t+1}=t+2$, $r^{3t+1}_1=2^{(m'-t)(l-2)}$,
   $r^{3t+1}_2=2^n$ and $r^{3t+1}_i=2^l$ for all $3\le i\le t+2$ \\
[0.9ex]

3: & \ \ \ \ let $u=(2^{(m'-t-1)(l-2)}-5)(2^{l-1}-1)+5$ \\ [0.8 ex]

4: & \ \ \ \ [\hspace{0.12cm}$u\ge r^{3t+1}_1=2^{(m'-t)(l-2)}$
  under the assumption that $t\le m'-6$ and $l\ge
  3$\hspace{0.12cm}] \\ [0.8ex]

5: & \ \ \ \ $T^{3t+2}=\textbf{L}^1\hspace{0.03cm}
  (\hspace{0.03cm}T^{3t+1},\hspace{0.03cm}1,
  \hspace{0.03cm}u\hspace{0.03cm})$ \\ [0.8ex]

6: & \ \ \ \ $T^{3t+3}=\textbf{L}^3\hspace{0.03cm}(\hspace{0.03cm}
  T^{3t+2},\hspace{0.03cm}1,\hspace{0.03cm}
  2^{(m'-t-1)(l-2)},\hspace{0.03cm}2^{l-2}-1\hspace{0.03cm})$ \\ [0.8ex]

7: & \ \ \ \ $T^{3t+4}=\textbf{L}^1\hspace{0.03cm}(\hspace{0.03cm}
  T^{3t+3},\hspace{0.03cm}t+3,\hspace{0.03cm}
  2^l\hspace{0.03cm})$ \\ [0.95ex]
\end{tabular}

\rule{\textwidth}{1pt} \caption{The Construction
  of $T^{3m'-14}$ from $T^1$} \label{procedure1}
\end{figure}

Below we give the details of our construction. In the first step, we call $\textbf{L}^1 (T^0,1,
  2^{m'(l-2)} )$ to get $T^1= (C^1,2,
   (2^{m'(l-2)},2^n ) )$.
This step is possible because $m'(l-2)\ge n$. We then invoke the procedure in Figure~\ref{procedure1}. In each
for-loop, the first component of $\rr$
  decreases by a factor of $2^{l-2}$, while the
  dimension of
  the space increases by $1$.
After running the for-loop $(m'-5)$ times, we obtain a coloring triple
  $T^{3m'-14}= (C^{3m'-14},d^{3m'-14},\rr^{3m'-14})$ that
satisfies\footnote{{ Remark: the superscript of
 $C$, $d$, $r_{i}$, denotes the index of the
iterative step. It is not an exponent!}}
\begin{equation*}
d^{3m'-14}=m'-3\hspace{0.06cm}, \ r^{3m'-14}_1=2^{5(l-2)}\hspace{0.06cm}, \ r^{3m'-14}_2=2^n \  \text{ and } \
r^{3m'-14}_i=2^l \hspace{0.06cm},\ \forall\ i \in [3: m'-3].
\end{equation*}

Next, we call the procedure given in
  Figure~\ref{procedure2}.
Note that the while-loop must terminate in at most $8$
  iterations because we start with $r^{3m'-14}_1=2^{5(l-2)}$.
The procedure returns a coloring triple $T^{w'}= (C^{w'},d^{w'},\rr^{w'} )$ that satisfies
\begin{equation*}
w'\le 3m'+11\hspace{0.06cm}, \ d^{w'}\le m'+5\hspace{0.06cm},
 \ r^{w'}_1=2^l\hspace{0.06cm}, \
r^{w'}_2=2^n \ \text{ and } \ r^{w'}_i=2^l\hspace{0.06cm},\ \forall\ i
\in [3 : d^{w'}].
\end{equation*}

\begin{figure}

\rule{\textwidth}{1pt}\vspace{0.07cm}

\textbf{The Construction of $T^{w'}$ from $T^{3m'-14}$}

\vspace{-0.15cm}\rule{\textwidth}{1pt}\vspace{0.18cm}

\begin{tabular}{@{\hspace{0.1cm}}r@{\hspace{0.2cm}}p{\textwidth}}
1: & let $t=0$ \\ [0.6ex]

2: & \textbf{while} $T^{3(m'+t)-14}=(C^{3(m'+t)-14},m'+t-3,
\rr^{3(m'+t)-14})$ satisfies $r^{3(m'+t)-14}_1>2^l$ \textbf{do} \\
[0.4ex]

3: & \ \ \ \ let $k=  \lceil\hspace{0.05cm}(r^{3(m'+t)-14}_1-5)
  /(2^{l-1}-1)\hspace{0.05cm} \rceil+5$ \\ [0.8ex]

4: & \ \ \ \ $T^{3(m'+t)-13}=\textbf{L}^1\hspace{0.03cm}
  (\hspace{0.03cm}T^{3(m'+t)-14},\hspace{0.03cm}1,
  \hspace{0.03cm}(k-5)(2^{l-1}-1)+5\hspace{0.03cm})$ \\ [0.8ex]

5: & \ \ \ \ $T^{3(m'+t)-12}=\textbf{L}^3\hspace{0.03cm}(\hspace{0.03cm}
  T^{3(m'+t)-13},\hspace{0.03cm}1,\hspace{0.03cm}
  k,\hspace{0.03cm}2^{l-2}-1\hspace{0.03cm})$ \\ [0.8ex]

6: & \ \ \ \ $T^{3(m'+t)-11}=\textbf{L}^1\hspace{0.03cm}(\hspace{0.03cm}
  T^{3(m'+t)-12},\hspace{0.03cm}m'+t-2,\hspace{0.03cm}
  2^l\hspace{0.03cm})$, set $t=t+1$ \\ [0.8ex]

7: & let $w'=3(m'+t)-13$ and $T^{w'}=\textbf{L}^1(
  T^{3(m'+t)-14},1,2^l)$ \\ [1ex]
\end{tabular}

\rule{\textwidth}{1pt} \caption{The Construction of $T^{w'}$
  from $T^{3m'-14}$} \label{procedure2}
\end{figure}

We then repeat the whole process above
  on the second coordinate and obtain a coloring triple
  $T^{w''}= (C^{w''},d^{w''},\rr^{w''} )$ that satisfies
\begin{equation*}
w''\le 6m'+21\hspace{0.06cm}, \ d^{w''}\le 2m'+8  \ \text{ and } \
r^{w''}_i=2^l\hspace{0.06cm},
   \ \forall\ i \in [1\: d^{w''}].
\end{equation*}
The way in which we define $m$ and $m'$ guarantees
\begin{equation*}
d^{w''}\le 2m'+8\le 2\left(\frac{n}{l-2}+1\right)+8\le 2\left(\frac{n}{l/3}\right)+10=\frac{6n}{l}+10\le
\frac{11n}{l}\le m.
\end{equation*}

Finally, by applying $\textbf{L}^2$ on coloring triple $T^{w''}$
  for $m-d^{w''}$ times with parameter $u=2^l$,
  we obtain $T^w=\left(C^w,m,\rr^w\right)$
  with $\rr^w\in\mathbb{Z}^m$ and $r^w_i=2^l$,
  $\forall\ i \in [1: m]$.
It follows from our construction that $w=O(m)$.

To see why the sequence $\mathcal{T}$ gives a reduction
  from {\sc Brouwer}$^{f_2}$ to {\sc Brouwer}$^f$,
  let $T^i= (C^i,d^i,\rr^i )$
  (\hspace{0.04cm}again the superscript of $C$, $d$,
  $\rr$, denotes the index of the
  iteration\hspace{0.04cm}).
As sequence $\{\hspace{0.06cm}\size{\rr^i}\hspace{0.06cm}\}_{\hspace{0.03cm} 0\le i\le w}$ is nondecreasing and $w=O(m)
= O(n)$,
  by the Property \textbf{A} of Lemma~\ref{lem1},~\ref{lem2} and~\ref{lem3},
  there exists a polynomial $g(n)$ such that
\begin{equation*}
\size{C^w}=\size{C}+O\big(g\left(n\right)\big).
\end{equation*}
By these Properties \textbf{A} again,
  we can construct the whole sequence
  $\mathcal{T}$ and in particular,
  triple $T^{w}= (C^{w},m,\rr^{w} )$,
  in time polynomial in $\size{C}$.

Pair $(C^w,0^{11n})$ is an input instance of {\sc Brouwer}$^f$. 
Given any panchromatic simplex $P$ of $ (C^w,0^{11n})$ and
  using the algorithms in Properties \textbf{B} of
   Lemma~\ref{lem1},~\ref{lem2} and~\ref{lem3},
  we can compute a sequence of panchromatic simplices
  $P^{w}=P,P^{w-1}...,P^{0}$ iteratively in polynomial time,
  where $P^t$ is a panchromatic simplex of $T^t$
  and is computed from the panchromatic simplex $P^{t+1}$ of
  $T^{t+1}$.
In the end, we obtain $P^0$, which is a panchromatic
  simplex of $ (C,0^{2n} )$.
\end{proof}

\section{Computing Fixed Points with Generalized Circuits}\label{Final}

In this section, we show that fixed points can be
  modeled by generalized circuits.
In particular, we reduce the search of a panchromatic
  simplex in an instance of {\sc Brouwer}$^{f_1}$
  (\hspace{0.04cm}which will be simply referred to as
  {\sc Brouwer} in this section\hspace{0.04cm}) to
  {\sc Poly$^3$-Gcircuit}, the computation of a
  $1/K^3$-approximate solution of a generalized
  circuit of $K$ nodes.
Recall $f_1(n)=3$. Our reduction will use ideas from \cite{DAS05}. However, we will need to develop several new
techniques
  to meet the geometric and combinatorial challenges
  in the consideration of high-dimensional fixed points.

Suppose $U=(C,0^{3n})$ is an input instance of
  {\sc Brouwer}, which colors the hypergrid $B^n =
  \mathbb{Z}^n_{[0,7]}$ with colors from
  $\{\hspace{0.04cm}1, ..., n,n+1\hspace{0.04cm}\}$.
Let $m$ be the smallest integer such that
  $2^m\ge \size{C}>n$ and $K=2^{6m}$, where
  $\size{C}$~is the number of gates plus the number of
  input and output variables in a Boolean circuit $C$.
Please note that $m=O(\log |C|)$ and hence
   $2^{\Theta(m)}$ is polynomial in the input size of $U$.

We will construct a generalized circuit
  $\calS^U=(V,\calT^U)$ with $|V|=K$ in polynomial time.
Our construction ensures that,
\begin{itemize}
\item {Property $\textbf{R}$}: From every
$(1/K^3)$-approximate
  solution to $\calS^U$, we can compute a panchromatic
  simplex $P$ of circuit $C$ in polynomial time.
\end{itemize}

In the rest of this section, we assume $\epsilon = 1/K^3$.
\subsection{Overcome the Curse of Dimensionality}

We prove a key geometric lemma (Lemma \ref{lem:geo} below) for finding a
  panchromatic simplex in order to overcome the
  curse of dimensionality.
Our construction of $\calS^U$ will then build on this lemma.

For $a\in \mathbb{R}^+$, let $\pi(a)=\max\{\hspace{0.06cm}
  i\ |\ 0\le i\le 7\ \mbox{and}\ i<a\hspace{0.06cm}\}$ be
  the largest integer in $[0:7]$ that is smaller than $a$.
Let $E^n=\{\hspace{0.06cm}\zz^1,\zz^2,...,\zz^n,\zz^{n+1}
  \hspace{0.08cm}\}$ where $\zz^i=\ee_i/K^2$ and
  $\zz^{n+1}=-\sum_{1\le i\le n}\ee_i/K^2$.
For each $i \in [1:n+1]$, we encode the $i^{th}$ color
  in Color$_{C}$ by vector $\zz^i$.

For any $\pp\in \mathbb{R}_+^n$, let $\qq=\pi(\pp)$ be the
  integer point in $B^n=\mathbb{Z}_{[0,7]}^n$ with
  $q_i=\pi(p_i)$.
Let $\xxi(\pp) = \zz^t$, where $t=\color{C}{\pi(\pp)}$.

\begin{defi}[Well-Positioned Points]
A real number $a$ $\in \mathbb{R}^+$ is {\em poorly-positioned}
  if there is an integer $t \in [0:7]$ such that $|\hspace{0.05cm}a-t\hspace{0.05cm}|\le
  80K\epsilon=80/K^2$.
A point $\pp\in \mathbb{R}_+^n$ is {\em well-positioned} if
  none of its components is {\em
  poorly-positioned}, otherwise, it is {\em poorly-positioned}.
\end{defi}

Let $S=\{\hspace{0.04cm}\pp^{1},\pp^2,... ,\pp^{h}
  \hspace{0.04cm}\}$ be a set of $h$ points
  in $\mathbb{R}_{[0,8]}^n$.
We define
\begin{equation*}
I_B (S) =\big\{\hspace{0.08cm}k\ \Big|\ \text{$\pp^k$
  is poorly-positioned }\hspace{0.08cm}\big\}\ \
  \ \mbox{and}\ \ \ I_G (S) =\big\{\hspace{0.06cm}k\
  \Big|\ \text{$\pp^k$ is well-positioned}\hspace{0.06cm}\big\}.
\end{equation*}

\begin{lemm}[Key Geometry: Equiangle Averaging]\label{lem:geo}
Suppose $U=(C,0^{3n})$ is an instance of {\sc Brouwer}.
Let $S$ $= \{\hspace{0.04cm}\pp^i,1\le i\le n^3\hspace{0.04cm}\}$
  be $n^3$ points in $\mathbb{R}_{[0,8]}^n$ such that $\pp^i=\pp^{i-1}
  + \sum_{i=1}^n \ee_i/K$.
If there is a vector
  $\rr^k\in\mathbb{R}^n_{[0,1/K^2]}$ for each $k$ in $I_B (S)$, such that,
\begin{eqnarray*}
& \Big\|\hspace{0.07cm}\sum_{k\in I_G (S)} \xxi(\pp^k)\hspace{0.2cm}
  + \sum_{k\in I_B (S)}\rr^k\hspace{0.07cm}
  \Big\|_\infty = O(\epsilon),&
\end{eqnarray*}
then $Q=\{\hspace{0.03cm}\pi(\pp^k),k\in I_G (S)\hspace{0.03cm}\}$
  is a panchromatic simplex of $C$.
\end{lemm}
\begin{proof}
We first prove that $Q'=\{\hspace{0.06cm} \qq^k=\pi(\pp^k),
  1\le k\le n^3\hspace{0.06cm}\}$ is accommodated,
  and satisfies $|\hspace{0.03cm}Q'\hspace{0.03cm}|\le n+1$.
Let $\qq^k=\pi(\pp^k)$ for each $k \in [1 : n^3]$. As sequence $\{\pp^k\}_{1\le k\le n^3}$ is strictly
  increasing, $\{\qq^k\}_{1\le k\le n^3}$ is non-decreasing.
Since $n/K\ll 1$, there exists at most one $k_i$ for each $i \in [1: n]$, such that
  $q^{k_i}_i=q^{k_i-1}_i+1$, which implies that $Q'$ is accommodated.
Since $\{\qq^k\}$ is non-decreasing,
  $|\hspace{0.03cm}Q'\hspace{0.03cm}|\le n+1$.
Because $Q\subset Q'$, $Q$ is accommodated and
  $|\hspace{0.03cm}Q\hspace{0.03cm}|\le n+1$.

Next, we give an upper bound for $|\hspace{0.03cm}I_B(S)\hspace{0.03cm}|$. Because $1/K^2\ll 1/K\ll 1$, there is at
most one $k_i$ for each $i$, such that $p^{k_i}_i$ is poorly-positioned.
Since every poorly-positioned point has at least one
  poorly-positioned com\-ponent, $|\hspace{0.03cm}I_B (S)
  \hspace{0.03cm}|\le n$ and $|\hspace{0.03cm}I_G (S)\hspace{0.03cm} |\ge n^3-n$.

Let $W_i$ denote the number of
  points in $\{\hspace{0.04cm}\qq^{k}: k\in I_G (S)
  \hspace{0.04cm}\}$ that are colored $i$ by circuit $C$.
To prove $Q$ is a panchromatic simplex,
 it suffices to show that $W_i > 0$ for all $i \in [1: n+1]$.

Let $\rr^G=\sum_{k\in I_G (S)}\xxi(\pp^k)$ and $\rr^B=
  \sum_{k\in I_B (S)} \rr^k $.
Since $|\hspace{0.03cm}I_B (S)\hspace{0.03cm} |\le n$ and $\|\hspace{0.05cm}
  \rr^k\hspace{0.05cm}\|_\infty\le 1/K^2$,
\begin{eqnarray}
&&\hspace{-0.9cm}\|\hspace{0.05cm}\rr^B\hspace{0.05cm}\|_\infty \le  n/K^2, \quad \mbox{\rm and}\nonumber\\\label{cont}
&&\hspace{-0.9cm}\|\hspace{0.05cm}\rr^G \hspace{0.05cm}\|_\infty \le \|\hspace{0.05cm}\rr^B\hspace{0.05cm}\|_\infty +
O(\epsilon)\le n/K^2+O(\epsilon).
\end{eqnarray}

\noindent Assume by way of contradiction that one of $W_i$ is zero:
\begin{itemize}
\item If $W_{n+1}=0$, supposing $W_{i^*}=\max_{1\le i\le n}W_i$,
  then $W_{i^*}\ge n^2-1$, as $|\hspace{0.03cm}I_G (S)\hspace{0.03cm} |\ge n^3-n$.
But $r^G_{i^*}\ge
  (n^2-1)/K^2\gg n/K^2+O(\epsilon)$,
  which contradicts $(\ref{cont})$ above, since $\epsilon=1/K^3$.

\item If $W_t=0$ for $t\in [1:n]$, then we can assert $W_{n+1}\le n^2/2$,
  for otherwise, $|\hspace{0.03cm}r^G_{t}
  \hspace{0.03cm}|>n^2/(2K^2)\gg n/K^2+O(\epsilon)$,
  contradicting  $(\ref{cont})$.
Suppose $W_{i^*}=\max_{1\le
  i\le n+1}W_i$.
Then, $W_{i^*}\ge n^2-1$ and $i^*\not= n+1$.
So $r_{i^*}^G\ge (n^2-1-n/2)/K^2\gg
  n/K^2+O(\epsilon)$, contradicting $(\ref{cont})$.
\end{itemize}
As a result, $W_i>0$ for all $i \in [1: n+1]$, and we have completed the
  proof of the lemma.
\end{proof}

\subsection{Construction of the Generalized Circuit $\calS^\mathbf{U}$}

We will show how to implement Lemma~\ref{lem:geo}, using
  a generalized circuit.
Given an input $U=(C,0^{3n})$ of {\sc Brouwer},
  our objective is to design a generalized circuit
  $\calS^U=(V,\calT^U)$ with $|\hspace{0.03cm}V
  \hspace{0.03cm}|=K$, such that, from any
  $\epsilon$-approximate
  solution to $\calS^U$, one can find a panchromatic
  simplex of $C$ in polynomial time.
Recall that $\epsilon=1/K^3$.

More precisely, we will design the generalized circuit $\calS^U$
  that encodes $n^3$ points in $\mathbb{R}_{[0,8]}^n$,  simulates
  the $\pi$ function, and simulates the boolean circuit $C$.
Our $\calS^U$ has the property that in each of its
  $\epsilon$-approximate solution, the
  sum of the $n^3$ vectors as given in Lemma \ref{lem:geo}
  is close to zero, i.e., $O(\epsilon)$.
Then $Q$, as defined in Lemma~\ref{lem:geo}, is a panchromatic
  simplex of $C$, which can be computed from
  the approximate solution of $\calS^{U}$ in polynomial time.

Let us define some notations that will be useful.
Suppose $\calS=(V,\calT)$ is a generalized circuit with
  $|\hspace{0.03cm}V\hspace{0.03cm}|=K$.
A node $v\in V$ is said to be \emph{unused} in $\calS$ if
  none of the gates $T\in \calT$
  uses $v$ as its output node.
Now, suppose $T \not\in \calT $ is a gate such that
  the output node of $T$ is unused in $\calS$.
We will use $\text{\sc Insert}(\calS,T)$ to denote the
  insertion of $T$ into $\calS$.
After calling $\text{\sc Insert}(\calS,T)$,
  $\calS$ becomes $(V,\calT\cup \{T\})$.

\begin{figure}[!t]

\rule{\textwidth}{1pt}\vspace{0.07cm}

{\sc ExtractBits}$\hspace{0.04cm}(\calS,v,v^1,v^2,v^3)$

\vspace{-0.15cm}\rule{ \textwidth}{1pt}\vspace{0.16cm}
\begin{tabular}{@{\hspace{0.0cm}}r@{\hspace{0.2cm}}p{\textwidth}}
1: & pick unused nodes $v_1,v_2,v_3,v_4\in V$ \\ [0.8ex]

2: & {\sc Insert}$(\calS,(G_=,v,nil,v_1,nil))$\\ [0.8ex]

3: & \textbf{for} $j$ from $1$ to $3$ \textbf{do} \\ [0.8ex]

4: & \ \ \ \ pick unused $v_{j1},v_{j2}\in V$\\ [0.8ex]

5: & \ \ \ \ {\sc Insert}$\hspace{0.04cm}(\calS,(G_\zeta,nil,nil,
  v_{j1},2^{-(6m+j)} ))$, {\sc Insert}$\hspace{0.04cm}(\calS,(G_<,v_{j1},v_j,
  v^j,nil ))$\\ [0.8ex]

6: & \ \ \ \ {\sc Insert}$\hspace{0.04cm}(\calS,(G_{\times\zeta},
  v^j,nil,v_{j2},2^{-j} ))$, {\sc Insert}$\hspace{0.04cm}(\calS,(G_-,v_j,v_{j2},
  v_{j+1},nil ))$ \\ [1ex]
\end{tabular}

\rule{ \textwidth}{1pt} \caption{Function {\sc ExtractBits}} \label{EXTRACT}
\end{figure}

To encode these $n^{3}$ points, let
  $\{\hspace{0.01cm}v^k_i\hspace{0.01cm}\}_{
  1\le k\le n^3,1\le i\le n}$
  be $n^4$ distinguished nodes in $V$.
We start with $\calS^U=(V,\emptyset)$ and insert a
  number of gates into it so that, in every
  $\epsilon$-approximate solution $\xx$, values of these
  nodes encode $n^3$ points $S = \{\hspace{0.06cm}
  \pp^k: {1\le k\le n^3}\hspace{0.04cm}\}$ in
  $\Reals{n}_{[0,8]}$ that \emph{approximately}
  satisfy all the conditions of Lemma \ref{lem:geo}.
In our encoding, we let $p^k_i=8K\xx[v^k_i]$.

We define two functions $\text{\sc ExtractBits}$ and
  {\sc Color\-ingSimulation}.
They are the building blocks in our construction.
$\text{\sc ExtractBits}$ implements the $\pi$ function,
  and is given in Figure~\ref{EXTRACT}.

\begin{lemm}[Encoding Binary]\label{small}
Suppose $\calS=(V,\calT)$ is a generalized circuit with
  $|\hspace{0.03cm}V\hspace{0.03cm}|=K$.
For each $v\in V$ and three unused nodes $v^1,v^2, v^3
  \in V$, we let $\calS'$ be the generalized circuit
  obtained after calling $\text{\sc ExtractBits} (\calS,v,v^1,v^2,v^3)$.
Then, in every $\epsilon$-approximate solution $\xx$ of $\calS'$,
  if $a=8K\xx[v]$ is well-positioned, then $\xx[v^i]=^{\hspace{0.06cm}\epsilon}_B b_i$,
  where $b_1b_2b_3$ is the binary representation of
  integer $\pi(a)\in [0:7]$.
\end{lemm}

\begin{proof}
First we consider the case when $\pi(a)=7$. As $a\ge 7+80K\epsilon$, we have $\xx[v]\ge
  1/(2K)+1/(4K)+1/(8K)+10\epsilon$.
By Figure~\ref{EXTRACT}, $\xx[v_1]\ge
  \xx[v]-2\epsilon$, $\xx[v^1]=^{\hspace{0.06cm}\epsilon}_B
1$ in the first loop and
\begin{eqnarray*}
\xx[v_2 ]&\ge& \xx[v_1]-\xx[v_{12}]-\epsilon \ \ge\
   \xx[v]-2\epsilon-(2^{-1}\xx[v^1]+\epsilon)-\epsilon\\ &\ge&
\xx[v]-2^{-1}(1/K+\epsilon)-4\epsilon\ \ge \ 1/(4K)+1/(8K)+5\epsilon.
\end{eqnarray*}
Since $\xx[v_{21}]\le 1/(4K)+\epsilon$ and $\xx[v_2]-\xx[v_{21}]
  >\epsilon$, we have $\xx[v^2]=^{\hspace{0.06cm}\epsilon}_B 1$ and
\begin{equation*}
\xx[v_3]\ \ge\ \xx[v_2 ]-\xx[v_{22}]-\epsilon\ >\ 1/(8K)+2\epsilon.
\end{equation*}
As a result, $\xx[v_3]-\xx[v_{31}]>\epsilon$ and $\xx[v^3]
  =^{\hspace{0.06cm}\epsilon}_B 1$.

Next, we consider the general case that $t<\pi(a)<t+1$ for
  $0\le t\le 6$. Let $b_1b_2b_3$ be the binary
representation of $t$. As $a$ is well-positioned, we have
\begin{equation*}
b_1/(2K)+b_2/(4K)+b_3/(8K)+10\epsilon\ \le\ \xx[v]
  \ \le\ b_1/(2K)+b_2/(4K)+(b_3+1)/(8K)-10\epsilon.
\end{equation*}
With similar arguments, after the first loop one can
  show that $\xx[v^1]=^{\hspace{0.06cm}\epsilon}_B b_1$ and
\begin{equation*}
b_2/(4K)+b_3/(8K)+5\epsilon\ \le\ \xx[v_2]\ \le\ b_2/(4K)+(b_3+1)/(8K)-5\epsilon.
\end{equation*}
After the second loop, we have
  $\xx[v^2]=^{\hspace{0.06cm}\epsilon}_B b_2$ and
\begin{equation*}
b_3/(8K)+2\epsilon\ \le\ \xx[v_3]\ \le\ (b_3+1)/(8K)-2\epsilon.
\end{equation*}
Thus, $\xx[v^3]=^{\hspace{0.06cm}\epsilon}_B b_3$.
\end{proof}

Next, we introduce {\sc ColoringSimulation}. Suppose $\calS=(V,\calT)$ is a generalized circuit with
  $|\hspace{0.03cm}V\hspace{0.03cm}|=K$.
Let $\{\hspace{0.04cm}v_i\hspace{0.04cm}\}_{i\in[1:n]}$
  be $n$ nodes in $V$, and $\{\hspace{0.04cm}v_i^+,
  v_i^-\hspace{0.04cm}\}_{i\in[1:n]}\subset V$ be $2n$ \emph{unused} nodes.~We use $\pp\in \mathbb{R}^n_+ $
  to denote the point encoded by nodes $\{\hspace{0.01cm}v_i
  \hspace{0.01cm}\}_{i\in[1:n]}$, that is, $p_i=8K\xx[v_i]$.
Imagine that $\pp $ is a point in $S = \{\hspace{0.04cm}\pp^{k}:
  1\leq i\leq n^{3}\hspace{0.04cm}\}$.
$\text{\sc ColoringSimulation}(\calS ,\setof{v_i}_{i\in[1:n]},
  $ $\{v_i^+,v_i^-\}_{i\in[1:n]})$ simulates circuit
  $C$ on input $\pi (\pp )$, by inserting the following gates
  into $\calS$:\vspace{0.1cm}
\begin{enumerate}
\item Pick $3n$ \emph{unused} nodes $\{\hspace{0.04cm}
  v_{i,j}\hspace{0.04cm} \}_{i\in [1:n],j\in[1:3]}$ in $V$.
\\ Call $\text{\sc ExtractBits}\hspace{0.04cm}(\calS,v_t,v_{t,1},v_{t,2},v_{t,3})$, for
each $ 1\le t\le n$;\vspace{-0.1cm}

\item View the values of $\{\hspace{0.02cm}v_{i,j}\hspace{0.02cm}\} $
  as $3n$ input bits of $C$. \\
Insert the corresponding logic gates from
  $\setof{ G_\lor,G_\land,G_\lnot }$
  into $\calS$ to simulate the evaluation of $C$, one for
  each gate in $C$, and place the $2n$ output bits in
  $\{\hspace{0.02cm}v_i^+,v_i^-\hspace{0.02cm}\}$.\vspace{0.1cm}
\end{enumerate}
We obtain the following lemma for
  {\sc ColoringSimulation}\hspace{0.04cm}$(\calS ,
  \setof{v_i}_{i\in[1:n]}, \{v_i^+,v_i^-\}_{i\in[1:n]})$
  as a direct consequence of Lemma \ref{small}.
\begin{lemm}[Point Coloring]\label{lem:point}
Let $\calS'$ be the generalized circuit obtained after
  calling the above $\text{\sc ColoringSimulation}$,
and $\xx$ be an $\epsilon$-approximate solution to $\calS'$. We let $\pp\in \mathbb{R}_+^n$ denote the point with
$p_i=8K\xx[v_i]$
  for all $i\in [1:n]$, and $\qq=\pi(\pp)$.
We use $\{\hspace{0.04cm}\Delta^+_i[\qq],\Delta^-_i[\qq]\hspace{0.04cm}\}_{
  i\in [1:n]}$ to denote the $2n$ output bits
  of $C$ evaluated at $\qq$.
If $\pp$ is a well-positioned point, then $\xx[v_i^+]
  =^{\hspace{0.06cm}\epsilon}_B
  \Delta_i^+[\qq]$ and $\xx[v_i^-]=^{\hspace{0.06cm}
  \epsilon}_B \Delta_i^-[\qq]$ for all $i \in [1:n]$.
\end{lemm}

Note that if the point $\pp$ in the lemma above is
  not well-positioned, then the values of $\{ v_i^+,v_i^- \}$ could be
  arbitrary.
However, according to the definition of generalized circuits,
  $\xx$ must satisfy $$0\le
  \xx[v^+_i],\xx[v^-_i]\le 1/K+\epsilon,\ \ \forall\ i \in [1:n].$$

Finally, we build the promised generalized circuit $\calS^U$ with a four-step construction. At the beginning, $\calS^U=(V, \emptyset)$
  and $|\hspace{0.03cm}V\hspace{0.03cm}|=K$.\vspace{0.25cm}

\noindent \textbf{Part 1}:  [\hspace{0.08cm}Equiangle Sampling Segment\hspace{0.08cm}]

\noindent Let $\{\hspace{0.01cm}v^k_i\hspace{0.01cm}\}_{
  1\le k\le n^3,1\le i\le n}$ be
  $n^4$ nodes in $V$.
We insert $G_\zeta$ gates, with
  properly chosen parameters,
  and $G_+$ gates into $\calS^U$ to ensure that every
  $\epsilon$-approximate solution $\xx$ of
  $\calS^U$ satisfies
\begin{equation}
\xx[v^k_i]=\min \Big(\hspace{0.04cm}\xx[v^1_i]+(k-1)/(8K^2),
  1/K \hspace{0.04cm} \Big)\pm O(\epsilon),\label{eq:ooo}
\end{equation}
for all $ k\in[1:n^3]$ and $i\in[1:n]$.\vspace{0.25cm}

\noindent \textbf{Part 2}: [\hspace{0.08cm}Point Coloring\hspace{0.08cm}]

\noindent Pick $2n^4$ unused nodes $\{\hspace{0.04cm}
  v_i^{k+}, v_i^{k-} \hspace{0.04cm}\}_{i\in
  [1:n], k\in[1:n^3]}$ from $ V $.
For every $k\in [1:n^3]$, we call
  $$\mbox{\sc ColoringSimulation} \hspace{0.05cm}\big(\calS^U ,\{v^{k}_i\}, \{v_i^{k+},
v_i^{k-}\}_{i\in[1:n]}\big).$$

\noindent \textbf{Part 3}: [\hspace{0.08cm}Summing up the Coloring
Vectors\hspace{0.08cm}]

\noindent Pick $2n$ unused nodes $\{ v_i^+,
  v_i^- \}_{i\in [1:n]}\subset V$.
  Insert properly-valued $G_{\times\zeta}$
  gates and $G_+$ gates to ensure in the
  resulting generalized circuit $\calS^U$,
  each $\epsilon$-approximate solution $\xx$ satisfies
\begin{equation*}
\xx[v_i^+]=\sum_{1\le k\le n^3} \Big(\hspace{0.03cm} \frac{1}{K}\hspace{0.08cm}\xx[v_i^{k+}]\hspace{0.03cm} \Big)\pm
O(n^3\epsilon)\ \ \ \ \text{and}\ \ \ \ \hspace{0.04cm}\xx[v_i^-]=\sum_{1\le k\le n^3}
\Big(\hspace{0.03cm}\frac{1}{K}\hspace{0.08cm} \xx[v_i^{k-}]\hspace{0.03cm} \Big)\pm O(n^3\epsilon).
\end{equation*}

\noindent \textbf{Part 4}: [\hspace{0.08cm}Closing the Loop\hspace{0.08cm}]

\noindent For each $i\in [1:n]$, pick unused nodes $v_i',v_i'' \in V $ and insert the following gates:
\begin{eqnarray*}
&\text{\sc Insert}\left(\calS^U,(G_+,v^1_i,v_i^+,v_i',nil )\right),
 \ \text{\sc Insert}\left(\calS^U,(G_-,v_i',v_i^-,v_i'',nil )
\right),&\\[0.5ex]
&\mbox{and}\ \ \text{\sc Insert}\left(\calS^U,(G_=,v_i'',nil,v^1_i,nil )\right).&
\end{eqnarray*}

\subsection{Analysis of the Reduction} \label{app:analysis}

We now prove the correctness of our construction.

Let $\xx$ be an $\epsilon$-approximate solution to $\calS^U$. Let $S=\{\hspace{0.04cm}\pp^k, \mbox{with }
p_i^k=8K\xx[v_i^k],
  1\le k\le n^3 \hspace{0.04cm} \}$ be the set of $n^3$
  points that we want to produce from $\xx$.
Let $I_G = I_G(S)$ and $I_B = I_B(S)$. For each $t\in I_G$, let $c_t\in [1:n+1]$ be the color of
  point $\qq^t=\pi(\pp^t)$ assigned by $C$, and for each $i\in[1:n+1]$,
  let $ W_i= $ $\sizeof{\hspace{0.04cm}
\left\{\hspace{0.04cm}t\in I_G \ | \ c_t=i\hspace{0.04cm}\right\}\hspace{0.04cm}}$.

It suffices to prove, as $ Q=\{\hspace{0.05cm}\pi(\pp^k),k\in I_G  \hspace{0.05cm}\}$ can be computed in polynomial time, that $Q$
  is a panchromatic simplex of $C$.
The line of the proof is very similar to the
  one for Lemma~\ref{lem:geo}.
First, we use the constraints introduced by
  the gates in \textbf{Part 1} to prove the  following two lemmas:

\begin{lemm}[Not Too Many Poorly-Positioned Points]\label{001}
$|\hspace{0.04cm}I_B\hspace{0.04cm}|\le n$, and hence $|\hspace{0.04cm}I_G\hspace{0.04cm}|\ge n^3-n$.
\end{lemm}
\begin{proof}
For each $t\in I_B$, according to the definition
  of poorly-positioned points, there exists an
  integer $1\le l\le n$ such that $p^t_l$ is
  a poorly-positioned number.
We will prove that, for every integer $1\le l\le n$,
  there exists at most one $t\in [1: n^3]$ such that
  $p^t_l=8K\xx[v_l^t]$ is  poorly-positioned, which implies
  $|\hspace{0.04cm}I_B\hspace{0.04cm}|\le n$ immediately.

Assume $p^t_l$ and $p^{t'}_l$ are both poorly-positioned,
  for a pair of integers $1\le t<t'\le n^3$.
Then, from the definition, there exists a pair
  of integers $0\le k,k'\le 7$,
\begin{equation}
\big|\hspace{0.04cm}\xx[v^t_l]-k/(8K)\hspace{0.04cm}
  \big|\le 10\epsilon \quad \mbox{and} \quad
\big|\hspace{0.04cm}\xx[v^{t'}_l]-k'/(8K)\hspace{0.04cm}
  \big|\le 10\epsilon. \label{eq:ggrrdd}
\end{equation}
Because (\ref{eq:ggrrdd}) implies that $\xx[v^t_l]
  <1/K-\epsilon\le \xx_C[v_l^t]$ and $\xx[v^{t'}_l]<
  1/K-\epsilon\le \xx_C[v_l^{t'}]$,
  by Equation (\ref{eq:ooo}) of \textbf{Part 1}, we have
\begin{equation}\label{eq:gggg}
\xx[v_l^t]=\xx[v_l^1]+(t-1)/(8K^2)\pm O(\epsilon) \quad \mbox{and}
  \quad \xx[v_l^{t'}]=\xx[v_l^1]+(t'-1)/(8K^2)\pm O(\epsilon).
\end{equation}
Hence, $\xx[v_l^t]<\xx[v_l^{t'}]$, $k\le k'$ and
\begin{equation}\label{eq:dd}
\xx[v_l^{t'}]-\xx[v_l^t]=(t'-t)/(8K^2)\pm O(\epsilon)
\end{equation}
Note that
  when $k=k'$, Equation (\ref{eq:ggrrdd}) implies
  that $\xx[v_l^{t'}]-\xx[v_l^t]\le 20\epsilon$,
  while when $k<k'$, it implies
   that $\xx[v_l^{t'}]-\xx[v_l^t]\ge (k'-k)/(8K)-20
   \epsilon\ge 1/(8K)-20\epsilon$.
 In both cases, we derived an inequality that
  contradicts (\ref{eq:dd}).
Thus, only one of $p^t_l$ or $p^{t'}_l$ can be poorly-positioned.
\end{proof}

\begin{lemm}[Accommodated]\label{002}
$Q=\{\hspace{0.03cm}\pi(\pp^k),k\in I_G\hspace{0.03cm}\}$ is accommodated and $|\hspace{0.03cm}Q\hspace{0.03cm}|\le
n+1$.
\end{lemm}
\begin{proof}
To show $Q$ is accommodated, it is sufficient to prove
\begin{equation}\label{eq:state}
q^t_l\le q_l^{t'}\le q^t_l+1,\quad\quad\mbox{for all
  $l \in [1: n]$ and $t,t'\in I_G$ such that $t<t'$.}
\end{equation}
For the sake of contradiction,
  we assume that (\ref{eq:state}) is not true.
We need to consider the following two cases.

First, assume $q_l^t>q_l^{t'}$ for some $t,t'\in I_G$ with $t<t'$. Since $q_l^{t'}<q_l^t\le 7$, we have $p_l^{t'}<7$
and thus, $\xx[v_l^{t'}]<7/(8K)$.~As a result, the first component of the min operator in (\ref{eq:ooo})
  is the smallest for both $t$ and $t'$, implying
  that $\xx[v_l^t]<\xx[v_l^{t'}]$ and $p_l^t<p_l^{t'}$.
This contradicts the assumption that $q_l^t>q_l^{t'}$.

Second, assume $q_l^{t'}-q_l^t\ge 2$ for some $t,t'\in I_G$ with $t<t'$. From the definition of $\pi$, we have
  $p_l^{t'}-p_l^t>1$ and thus, $\xx[v_l^{t'}]-\xx[v_l^t]>1/(8K)$.
But from (\ref{eq:ooo}), we have
$$\xx[v_l^{t'}]-\xx[v_l^t]\le (t'-t)/(8K^2)+O(\epsilon)< n^3/(8K^2)+O(\epsilon)\ll 1/(8K).$$
As a result, (\ref{eq:state}) is true.

Next, we prove $|\hspace{0.03cm}Q\hspace{0.03cm}|\le n+1$. Note that the definition of $Q$ together with
(\ref{eq:state}) implies
  that there exist integers $t_1<t_2<...<t_{|Q|}\in I_G$
  such that $\qq^{t_i}$ is strictly dominated by
  $\qq^{t_{i+1}}$, that is, $\qq^{t_i}\not=\qq^{t_{i+1}}$ and
   $q_j^{t_i} \leq q_j^{t_{i+1}}$ for all $j \in [1:n]$.

On the one hand, for every $1\le l\le |\hspace{0.03cm}Q\hspace{0.03cm}|-1$,
  there exists an integer $1\le k_l\le n$
such that $q_{k_l}^{t_{l+1}}=q_{k_l}^{t_l}+1$. On the other hand, for every $1\le k\le n$, (\ref{eq:state})
  implies that there is at most
  one $1\le l\le |\hspace{0.03cm}Q\hspace{0.03cm}|-1$ such that
  $q_k^{t_{l+1}}=q_k^{t_l}+1$.
Therefore, $|\hspace{0.03cm}Q\hspace{0.03cm}|\le n+1$.
\end{proof}

The construction in \textbf{Part 2 }and Lemma~\ref{lem:point} guarantees that:

\begin{lemm}[Correct Encoding of Colors]\label{FGHJKL}
For each $1\le k\le n^3$, let $\rr^k$ denote the vector that
  satisfies $r^k_i=\xx[v_i^{k+}]-\xx[v_i^{k-}]$, $\forall i \in [1:n]$.
For each $t\in I_G$, $\rr^t=K\zz^{c_t}\pm 2\epsilon$; for each $t\in I_B$,
$\|\hspace{0.02cm}\rr^t\hspace{0.02cm}\|_\infty\le 1/K+ 2\epsilon$.
\end{lemm}

Recall that \textbf{Part 3} sums up these $n^3$ vectors $\setof{\rr^k}$. Let $\rr$ denote the vector that satisfies
  $r_i=\xx[v_i^+]-\xx[v_i^-]$, for all $i\in [1: n]$.
Ideally, with (\hspace{0.04cm}the constraints of\hspace{0.04cm})
  the gates inserted in \textbf{Part 4}, we wish to establish
  $\pnorm{\infty}{\hspace{0.02cm}\rr
  \hspace{0.024cm}} = O(\epsilon)$.
However, whether or not this condition holds depends on the values of
  $\{v_i^1\}_{1\le i\le n}$ in $\xx$, as the gate
$(G_-,a,b,c,nil)$ requires $c=max(a-b,0)$ within a difference of $\epsilon$.
For example, in the case when $\xx[v^1_i]=0$, the
  magnitude of $\xx[v_i^-]$ could be
  much larger than that of  $\xx[v_i^+]$.
We are able to establish the following lemma which is
  sufficient to carry out the correctness proof of our reduction.

\begin{lemm}[Well-Conditioned Solution] \label{nearzero}
For all $i\in [1: n]$,
\begin{enumerate}
\item  if $\xx[v_i^1]>4\epsilon$,
  then $r_i=\xx[v_i^+]-\xx[v_i^-]>-4\epsilon$; and
\item if   $\xx[v_i^1]<1/K-2n^3/K^2$, then $r_i=\xx[v_i^+]-\xx[v_i^-]<4\epsilon$.
\end{enumerate}
\end{lemm}
\begin{proof}
In order to set up a proof-by-contradiction
  of the first if-statement, we assume there exists
  some $i$ such that $\xx[v_i^1]>4\epsilon$
  and $\xx[v_i^+]-\xx[v_i^-]\le -4\epsilon$.

By the first gate $(G_+,v^1_i,v_i^+,v_i',nil)$ inserted
  in \textbf{Part 4}, we have
\begin{equation}\label{eq:eee}
\xx[v_i']=\min(\xx[v_i^1]+\xx[v_i^+],1/K)\pm \epsilon
  \le \xx[v_i^1]+\xx[v_i^+]+\epsilon \le
  \xx[v_i^1]+\xx[v_i^-]-3\epsilon.
\end{equation}
By the second gate $(G_-,v_i',v_i^-,v_i'',nil)$, we have
\begin{equation}\label{eq:uuu}
\xx[v_i'']\le \max(\xx[v_i']-\xx[v_i^-],0)+\epsilon \le
  \max(\xx[v_i^1]-3\epsilon,0)+\epsilon \le \xx[v_i^1]-2\epsilon,
\end{equation}
where the last inequality follows from the assumption
  that $\xx[v_i^1]>4\epsilon$.
Since $\xx[v_i^1]\le 1/K+\epsilon$,
    we have $\xx[v_i'']\le \xx[v_i^1]-2\epsilon\le 1/K
    -\epsilon<1/K$.
So, by the last gate $(G_=,v_i'',nil,v^1_i,nil)$, we have $\xx[v^1_i]=\min(\xx[v_i''],1/K)\pm \epsilon =\xx[v_i'']\pm
\epsilon,$ which contradicts (\ref{eq:uuu}).

Similarly, to prove the second if-statement,
   we assume there exists some $1\le i\le n$ such that
   $\xx[v_i^1]<1/K-2n^3/K^2$ and $\xx[v_i^+]-\xx[v_i^-]\ge 4\epsilon$ in order
   to derive a contradiction.

By \textbf{Part 3}, $\xx[v_i^+]\le n^3/K^2+O(n^3\epsilon)$. Together with the assumption, we have $\xx[v_i^1]+
\xx[v_i^+] $ $\leq 1/K -
  n^3/K^2+O(n^3\epsilon) < 1/K$.
Thus, by the first gate $G_+$, we have
\begin{equation*}
\xx[v_i']=\min(\xx[v_i^1]+\xx[v_i^+],1/K)\pm
  \epsilon=\xx[v_i^1]+\xx[v_i^+]\pm \epsilon \ge
\xx[v_i^1]+\xx[v_i^-]+3\epsilon
\end{equation*}
and $\xx[v_i']\le \xx[v_i^1]+\xx[v_i^+]+\epsilon\le 1/K-n^3/K^2+O(n^3\epsilon)$.
  By the second gate $G_-$,
\begin{equation}\label{eq:ggggg}
  \xx[v_i'']\ge \min(\xx[v_i']-\xx[v_i^-],1/K)-
    \epsilon=\xx[v_i']-\xx[v_i^-]- \epsilon\ge
\xx[v_i^1]+2\epsilon.
\end{equation}
We also have $\xx[v_i'']\le \max(\xx[v_i']-\xx[v_i^-],0)+\epsilon\le \xx[v_i']+\epsilon
  <1/K$.
However, the last gate $G_=$ implies $\xx[v^1_i]=\min(\xx[v_i''],1/K)\pm \epsilon=\xx[v_i'']\pm \epsilon$, which
contradicts (\ref{eq:ggggg}).
\end{proof}

Now, we show that $Q$ is a panchromatic simplex of $C$.
By Lemma~\ref{002}, it suffices to prove that $W_i>0$,
for all $i \in [1: n+1]$.

By {\bf Part 3} of the construction and Lemma~\ref{FGHJKL},
\begin{eqnarray*}
\rr&=& \frac{1}{K}\sum_{1\le i\le n^3} \rr^i\pm O(n^3\epsilon)\ \
   =\ \
    \frac{1}{K}\sum_{i\in I_G} \rr^i+\frac{1}{K}\sum_{i\in
        I_B}\rr^i\pm O(n^3\epsilon)\nonumber\\
   &=&  \sum_{i\in I_G}
 \zz^{c_i} +\frac{1}{K}\sum_{i\in I_B}\rr^i\pm O(n^3\epsilon)\nonumber\ \ =
\sum_{1\le i\le n+1} W_i\hspace{0.08cm}\zz^i +\frac{1}{K}\sum_{i\in I_B}\rr^i\pm O(n^3\epsilon) \\
&=&\rr^G + \rr^B \pm O(n^3\epsilon).
\end{eqnarray*}
where $\rr^G = \sum_{1\le i\le n+1} W_i\zz^i $
  and $\rr^B =\sum_{i\in I_B}\rr^i /K$.
Since $|\hspace{0.03cm}I_B\hspace{0.03cm}|\le n$ and
  $\|\hspace{0.03cm}\rr^i\hspace{0.03cm}\|_\infty \le 1/K+\epsilon$
  for each $i\in I_B$,
  we have $\|\hspace{0.03cm}\rr^B\hspace{0.03cm}\|_\infty=O(n/K^2)$.

As $|\hspace{0.03cm}I_G\hspace{0.03cm}| \geq n^3 -n$,
  we have $\sum_{1\leq i\leq n+1} W_i \geq n^3-n$.
The next lemma shows that, if one of $W_i$ is
  equal to zero, then $\|\hspace{0.03cm}\rr^G\hspace{0.03cm}\|_\infty
  \gg \|\hspace{0.03cm}\rr^B\hspace{0.03cm}\|_\infty$.

\begin{lemm} \label{lem:colorGap}
If one of $W_i$ is equal to zero, then $\|\hspace{0.03cm}\rr^G\hspace{0.03cm}\|_\infty\ge n^2/(3K^2)$, and thus
$\|\hspace{0.03cm}\rr\hspace{0.03cm}\|_\infty \gg 4\epsilon$.
\end{lemm}
\begin{proof}
We divide the proof into two cases. First, assume $W_{n+1}=0$. Let $l \in
[1: n]$ be the integer such
  that $W_l=\max_{1\le i\le n} W_i$, then we have $W_l > $ $n^2-1$.
Thus, $r'_l=W_l/K\ge (n^2-1)/K>n^2/(3K^2)$.

Second, assume $W_t=0$ for some $1\le t\le n$. We have the following two cases\hspace{0.04cm}:
\begin{itemize}
\item $W_{n+1}\ge n^2/2$: $r'_t=-W_{n+1}/K\le -n^2/(2K^2)<-n^2/(3K^2)$.
\item $W_{n+1}<n^2/2$:
Let $l$ be the integer such that $W_l=\max_{1\le i\le n+1}W_i$, then
  $l\not=t,n+1$ and $W_l> n^2-1$.
Then, $r'_l=(W_l-W_{n+1})/K>(n^2/2-1)/K^2>n^2/(3K^2)$.
\end{itemize}
\end{proof}

Therefore, if $Q$ is not a panchromatic simplex,
  then one of the $W_i$'s is equal to zero, and hence
  $\|\hspace{0.03cm}\rr\hspace{0.03cm}\|_\infty \gg 4\epsilon$.
Had \textbf{Part 4} of our construction
  guaranteed that $\|\hspace{0.03cm}\rr\hspace{0.03cm}
  \|_\infty$ $=O(\epsilon)$, we would have completed the proof.
As it is not always the case, we prove the
  following lemma so that we can use
  Lemma \ref{nearzero} to complete the proof.

\begin{lemm}[Well-Conditioned]\label{lemm:wellconditionness}
For all $i \in [1: n]$, $4\epsilon < \xx[v_i^1] < 1/K-2n^3/K^2$.
\end{lemm}
\begin{proof}

In this proof, we will use the following
   boundary condition of circuit $C$:
   For each $\qq\in B^n$ and $1\le k\not=l\le n$,
\begin{itemize}
\item[\ \ \ \ ] \textbf{B.1:} if $q_k=0$, then $\color{C}{\qq}\not=n+1$;
\item[\ \ \ \ ] \textbf{B.2:} if $q_k=0$ and $q_l>0$, then $\color{C}{\qq}\not=l$;
\item[\ \ \ \ ] \textbf{B.3:} if $q_k=7$, then $\color{C}{\qq}\not=k$; and
\item[\ \ \ \ ] \textbf{B.4:} if $q_k=7$ and $\color{C}{\qq}=l\not=k$, then $q_l=0$.
\end{itemize}
These conditions follow directly from the definition of valid circuits.
Recall $1/K =2^{-6m}$, $\epsilon = 2^{-18m}=1/K^3$ and $2^m > n$.

First, if there exists an integer $ k \in [1: n]$
  such that $\xx[v_k^1]\le 4\epsilon$, then $q^t_k=0$ for all
  $t\in I_G$.
By B.1, $W_{n+1}=0$. Let $l$ be the integer such that $W_l=\max_{1\le i\le n} W_i$. As
$\sum_{i=1}^{n+1} W_i = |\hspace{0.04cm}I_G\hspace{0.04cm}|\ge n^3-n$,
  we have $W_l \ge n^2-1$.
So, $r_l \ge W_l/K^2 - O(n/K^2) -O(n^3\epsilon) \gg
  4\epsilon$.
Now consider the following two cases\hspace{0.04cm}:
\begin{itemize}
\item If $\xx[v_l^1]<1/K-2n^3/K^2$, then we get a
  contradiction in Lemma~\ref{nearzero}.
\item If $\xx[v_l^1]\ge 1/K-2n^3/K^2$, then for all
  $t\in I_G$, $$ p^t_l=8K \Big(\min\big(\xx[v_l^1]+(t-1)/(8K^2),1/K\big)\pm
  O(\epsilon)\Big)>1  $$ and hence $q^t_l>0$.
By B.2, we have $W_l=0$,
  contradicting the assumption.
\end{itemize}

Second, if there exists an integer $ k \in [1: n]$ such
  that $\xx[v_k^1]\ge 1/K-2n^3/K^2$, then for all
  $t\in I_G$, we have $q^t_k=7$.
By B.3, $W_k=0$. If $W_{n+1}\ge n^2/2$, then $$r_k \le -W_{n+1}/K^2 + O(n/K^2) +
  O(n^3\epsilon) \ll -4\epsilon,$$ which contradicts 
  the assumption that $\xx[v_k^1]\ge 1/K-2n^3/K^2>$
  $4\epsilon$ (\hspace{0.04cm}see Lemma \ref{nearzero}.1\hspace{0.04cm}).
Below, we assume $W_{n+1}< n^2/2$.

Let $l$ be the integer such that $W_l=\max_{1\le i\le n+1} W_i$. Since $W_k=0$, we have $W_l\ge n^2-1$ and
  $l\not=k$. As $W_{n+1}< n^2/2$, $W_l-W_{n+1}>n^2/2-1$ and thus,
  $$r_l \ge (W_l-W_{n+1})/K^2 - O(n/K^2) -
  O(n^3\epsilon) \gg 4\epsilon.$$
We now consider the following two cases\hspace{0.04cm}:
\begin{itemize}
\item If
  $\xx[v_l^1]<1/K-2n^3/K^2$, then we get a contradiction in
  Lemma~\ref{nearzero}.2;
\item If $\xx[v_l^1]\ge 1/K$ $-2n^3/K^2$, then $p^t_l>1$ and thus
  $q^t_l>0$ for all $t\in I_G$.
By B.4, we have $W_l=0$
  which contradicts the assumption.\vspace{-0.7cm}
\end{itemize}
\end{proof}

\section{Extensions and Open Problems}\label{sec:Open}

\subsection{Sparse Games are Hard}\label{}

As fixed points and Nash equilibria are fundamental to many
  other search and optimization problems, our results and techniques
  may have a broader scope of applications and implications.
So far, our complexity results on the computation and approximation
  of Nash equilibria have been extended to Arrow-Debreu equilibria
  \cite{HuangTeng}.
They can also be naturally extended to both $r$-player
  games \cite{NashNonCooperative} and $r$-graphical games
  \cite{KEA01}, for every fixed $r\geq 3$.
Since the announcement of our work,
  it has been shown that the Nash equilibrium
  is \textbf{PPAD}-hard to approximate in fully polynomial
  time even for bimatrix games with some special payoff structures,
  such as bimatrix games in which
  all payoff entries are either 0 or 1 \cite{ChenTengValiant},
  or in which most of the payoff entries are 0.
In the latter case, we can strengthen our gadgets to
  prove the following theorem:

\begin{theo}[{\sc Sparse Bimatrix}]
Nash equilibria remains \emph{\textbf{PPAD}}-hard to approximate in fully
  polynomial time for sparse bimatrix games in which each row and column of
  the two payoff matrices contains at most $10$ nonzero entries.
\end{theo}

The reduction needed in proving this theorem is similar
  to the one used in proving Theorem \ref{ppadb}.
The key difference is that we first reduce {\sc Brouwer}$^{f_1}$
  to a sparse generalized circuit,
  where a generalized circuit is \emph{sparse}
  if each node is used by at most two gates as their input nodes.
We then refine our gadget games for $G_\zeta$, $G_{\land}$ or $G_{\lor}$,
  to guarantee that the resulting bimatrix game is sparse.
Details of the proof can be found in~\cite{ChenDengTengSparse}.

\subsection{Open Questions and Conjectures}\label{}
There remains a complexity gap in the approximation of
  two-player Nash equilibria:
 Lipton, Markakis and Mehta \cite{Lipton} show
  that an \vspace{-0.02cm} $\epsilon$-approximate Nash equilibrium
  can be computed in $n^{O(\log n/\epsilon^2)}$-time, while this paper
  shows that, for $\epsilon$ of order $1/\mbox{poly} (n)$,
 no algorithm can find an $\epsilon$-approximate Nash
 equilibrium in $\mbox{poly} (n,1/\epsilon)$-time,
 unless $\text{\textbf{PPAD}}$ is contained in $\text{\textbf{P}}$.
However, our hardness result does not cover the
  case when $\epsilon$ is a constant between $0$ and $1$, or of order
  $1/\mbox{polylog} (n)$.
Naturally, it~is unlikely that finding an
  $\epsilon$-approximate Nash equilibrium is \textbf{PPAD}-complete when
  $\epsilon$ is an absolute constant, for otherwise, all search
  problems in \textbf{PPAD} would be solvable in $n^{O(\log n)}$-time,
  due to the result of \cite{Lipton}.\vspace{0.03cm}

Thinking optimistically, we would like to see the following
    conjectures turn out to be true.

\begin{conj}[PTAS for {\sc Bimatrix}]\label{conj:PTAS}
There is an $O(n^{k+\epsilon^{-c}})$-time algorithm for
  finding an $\epsilon$-approximate Nash equilibrium in a
  two-player game, for some constants $c$ and $k$.
\end{conj}

\begin{conj}[Smoothed {\sc Bimatrix}]\label{smoothedTwoNashC}
There is an algorithm for {\sc Bimatrix} with smoothed complexity
  $O(n^{k+\sigma^{-c}})$ under perturbations
  with magnitude $\sigma$, for some constants $c$ and $k$.
\end{conj}

For a sufficiently large $\epsilon$, such as $\epsilon\geq 1/2$,
  an $\epsilon$-approximate Nash equilibrium can be found in polynomial
  time \cite{constant1,constant2}, by examining small-support strategies.
However, new techniques are needed to prove Conjecture \ref{conj:PTAS}
  \cite{DederNazerzadehSaberi}.
Lemma 2.2 implies that Conjecture~\ref{conj:PTAS}
  is true for $\epsilon$-well-supported
  Nash equilibrium if and only if it is true for
  $\epsilon$-approximate Nash equilibrium.
For each bimatrix game $(\AA ,\BB )$ such that rank$(\AA +\BB )$ is a
  constant, Kannan and Theobald \cite{KannanTheobald}
  found a fully-polynomial-time algorithm
  to approximate Nash equilibria.

For Conjecture \ref{smoothedTwoNashC},
   one might be able to  prove a weaker version of this conjecture by extending the analysis of
  \cite{BaranyVempalaVetta} to show that
  there is an algorithm for {\sc Bimatrix}
   with smoothed complexity $n^{O(\log n/\sigma^2)}$.
We also conjecture that Corollary \ref{Theo:LemkeHowson}
  remains true without any complexity assumption on \textbf{PPAD}.
A positive answer would extend the result
 of Savani and von Stengel \cite{SavanivonStengel} to smoothed bimatrix games.
Another interesting question is whether the average-case complexity of the
  Lemke-Howson algorithm is polynomial.

Of course, the fact that two-player Nash equilibria and Arrow-Debreu
  equilibria are \textbf{PPAD}-hard to compute in the smoothed model
  does not necessarily imply that game and market problems
  are hard to solve in practice.
In addition to possible noise and imprecision in inputs,
  practical problems might have other special structure
  that makes equilibrium computation or approximation more tractable.
The game and market problems and their hardness results might
  provide an opportunity and a family of concrete problems for discovering new
  input models that can help us to rigorously evaluate the performance of
  practical equilibrium algorithms and heuristics.

Theorem \ref{thm:bimatrixMain} implies that for any $r>2$, the computation of
 an $r$-player Nash equilibrium can be reduced in polynomial time
 to the computation of a two-player Nash equilibrium.
However, the implied reduction is not very natural:
The $r$-player Nash equilibrium problem is first reduced to
  {\sc End-of-Line}, then to {\sc Brouwer}, and then to {\sc
  Bimatrix}.
It remains an interesting question to find a more direct reduction
  from $r$-player Nash equilibria to two-player Nash equilibria.

The following complexity question about Nash equilibria is due to
 Vijay Vazirani:
 Are the counting versions of all {\rm\textbf{PPAD}} complete-problems
   as hard as the counting version of {\sc Bimatrix}?
Gilboa and Zemel  \cite{GilboaZemel} showed
  that deciding whether a bimatrix game has a unique Nash equilibrium
  is \textbf{NP}-hard.
Their technique was extended in \cite{ConitzerSandholm} to prove
  that counting the number of Nash equilibria is \textbf{\#P}-hard.
Because the reduction between search
  problems only requires a many-to-one map between
  solutions,  the number of solutions is not necessarily preserved.
More restricted reductions are needed to solve Vazirani's question.

Finally, even though our results in this paper as well as the results of
  \cite{DAS05,CHE05,DAS06} provide strong evidence
  that equilibrium computation might be hard for \textbf{P},
  the hardness of the \textbf{PPAD} complexity class
  itself is largely unknown \cite{JohnsonColumn}.
On one hand, Megiddo \cite{MegiddoLCP} proved that
  if {\sc Bimatrix} is \textbf{NP}-hard, then
  \textbf{NP} = \textbf{coNP}.
On the other hand, there are oracles that separate
  \textbf{PPAD} from \textbf{P}, and various
  discrete fixed point problems such as
  the computational version of Sperner's Lemma,
  requires an exponential number of functional evaluations
  in the query model, deterministic  \cite{HPV,XX05} or randomized
  \cite{ChenTengSTOC07}, and in the
  quantum query  model \cite{FISV,ChenTengSTOC07}.
It is  desirable to find stronger evidences
  that \textbf{PPAD} is not contained in \textbf{P}.
Does the existence of one-way functions imply that
    \textbf{PPAD} is not contained in \textbf{P}?
Does ``{\sc Factoring} is not in \textbf{P}'' imply that
     \textbf{PPAD} is not contained in \textbf{P}?
Characterizing the hardness of the \textbf{PPAD} class
   is a great and a challenging problem.

\section{Acknowledgments}\label{sec:ACK}

We would like to thank Kyle Burke, Li-Sha Huang,
   Jon Kelner, Rajmohan Rajaraman,
   Dan Spielman, Ravi Sundaram,  Paul Valiant, and
   Vijay Vazirani for helpful comments and suggestions.
We would like to thank everyone who asked about the smoothed
  complexity of the Lemke-Howson algorithm, especially John Reif
  for being the first player to ask us this question.

Xi Chen's work was supported by the Chinese National Key
  Foundation Plan (\hspace{0.03cm}2003CB31\-7807, 2004CB318108\hspace{0.03cm}), the
  National Natural Science Foundation of China Grant 60553001
  and the National Basic Research Program of China
  Grant (\hspace{0.03cm}2007CB807900, 2007CB807901\hspace{0.03cm}).
Part of his work was done while visiting the City University of
  Hong Kong.
Xiaotie Deng's work was supported by City
  University of Hong Kong for his research.
Shang-Hua Teng's work was supported by the NSF grants CCR-0311430 and ITR
  CCR-0325630.
Part of his work was done while visiting
  Tsinghua University and Microsoft Beijing Research  Lab.
Several critical ideas on the approximation and smoothed complexities
  of two-player Nash equilibrium
  were shaped when the authors were attending ISAAC2005 at
  Sanya, Hainan, China.

\bibliographystyle{plain}
\bibliography{BIMATRIX}

\begin{thebibliography}{10}

\bibitem{list}
John Reif, Nicole Immorlica, Steve Vavasis, Christos Papadimitriou, Mohammad
  Mahdian, Ding-Zhu Du, Santosh Vempala, Aram Harrow, Adam Kalai, Imre
  B\'ar\'any, Adrian Vetta, Jonathan Kelner and a number of other people asked
  whether the smoothed complexity of the Lemke-Howson algorithm or Nash
  Equilibria is polynomial, 2001--2005.

\bibitem{AbbottKaneValiant}
T.~Abbott, D.~Kane, and P.~Valiant.
\newblock On the complexity of two-player win-lose games.
\newblock In {\em FOCS '05: Proceedings of the 46th Annual IEEE Symposium on
  Foundations of Computer Science}, pages 113--122, 2005.

\bibitem{AD}
K.J. Arrow and G.~Debreu.
\newblock Existence of an equilibrium for a competitive economy.
\newblock {\em Econometrica}, 22:265--290, 1954.

\bibitem{BaranyVempalaVetta}
I.~B\'ar\'any, S.~Vempala, and A.~Vetta.
\newblock Nash equilibria in random games.
\newblock In {\em FOCS '05: Proceedings of the 46th Annual IEEE Symposium on
  Foundations of Computer Science}, pages 123--131, 2005.

\bibitem{BSS}
L.~Blum, M.~Shub, and S.~Smale.
\newblock On a theory of computation over the real numbers; {NP} completeness,
  recursive functions and universal machines.
\newblock {\em Bulletin of the AMS}, 21(1):1--46, July 1989.

\bibitem{Borg82}
K.-H. Borgwardt.
\newblock The average number of steps required by the simplex method is
  polynomial.
\newblock {\em Zeitschrift f\"ur Operations Research}, 26:157--177, 1982.

\bibitem{BRO10}
L.E.J. Brouwer.
\newblock {\"{U}ber Abbildung von Mannigfaltigkeiten}.
\newblock {\em Mathematische Annalen}, 71:97--115, 1910.

\bibitem{XX05}
X.~Chen and X.~Deng.
\newblock On algorithms for discrete and approximate {B}rouwer fixed points.
\newblock In {\em STOC '05: Proceedings of the 37th Annual ACM Symposium on
  Theory of computing}, pages 323--330, 2005.

\bibitem{XX06}
X.~Chen and X.~Deng.
\newblock On the complexity of {2D} discrete fixed point problem.
\newblock In {\em ICALP '06: Proceed\-ings of the 33rd International Colloquium
  on Automata, Languages and Programming}, pages 489--500, 2006.

\bibitem{CHE05}
X.~Chen and X.~Deng.
\newblock 3-{N}ash is {PPAD}-complete.
\newblock In {\em Electronic Colloquium in Computational Complexity}, TR05-134,
  2005.

\bibitem{ChenDengTengSparse}
X.~Chen, X.~Deng, and S.-H. Teng.
\newblock Sparse games are hard.
\newblock In {\em Proceedings of the 2nd Workshop on Internet and Network
  Economics}, pages 262--273, 2006.

\bibitem{ChenHuangTengWine}
X.~Chen, L.-S. Huang, and S.-H. Teng.
\newblock Market equilibria with hybrid linear-{L}eontief utilities.
\newblock In {\em Proceedings of the 2nd Workshop on Internet and Network
  Economics}, pages 274--285, 2006.

\bibitem{ChenTengSTOC07}
X.~Chen and S.-H. Teng.
\newblock Paths beyond local search: A nearly tight bound for randomized
  fixed-point computation.
\newblock {\em {\rm arXiv}}, 2007.
\newblock http://arxiv.org/abs/cs.GT/0702088.

\bibitem{ChenTengValiant}
X.~Chen, S.-H. Teng, and P.A. Valiant.
\newblock The approximation complexity of win-lose games.
\newblock In {\em SODA '07: Proceedings of the 18th Annual ACM-SIAM Symposium
  on Discrete Algorithms}, 2007.

\bibitem{CSVY}
B.~Codenotti, A.~Saberi, K.~Varadarajan, and Y.~Ye.
\newblock Leontief economies encode nonzero sum two-player games.
\newblock In {\em SODA '06: Proceedings of the 17th Annual ACM-SIAM Symposium
  on Discrete Algorithms}, pages 659--667, 2006.

\bibitem{NSFWorkshop1999}
A.~Condon, H.~Edelsbrunner, E.~Emerson, L.~Fortnow, S.~Haber, R.~Karp,
  D.~Leivant, R.~Lipton, N.~Lynch, I.~Parberry, C.~Papadimitriou, M.~Rabin,
  A.~Rosenberg, J.~Royer, J.~Savage, A.~Selman, C.~Smith, E.~Tardos, and
  J.~Vitter.
\newblock Challenges for theory of computing: Report of an {NSF}-sponsored
  workshop on research in theoretical computer science.
\newblock {\em SIGACT News}, 30(2):62--76, 1999.

\bibitem{ConitzerSandholm}
V.~Conitzer and T.~Sandholm.
\newblock Complexity results about nash equilibria.
\newblock In {\em In Proceedings of the International Joint Conference on
  Artificial Intelligence (IJCAI)}, 2003.

\bibitem{DAS05}
C.~Daskalakis, P.W. Goldberg, and C.H. Papadimitriou.
\newblock The complexity of computing a {N}ash equilibrium.
\newblock In {\em STOC '06: Proceedings of the 38th Annual ACM Symposium on
  Theory of Computing}, pages 71--78, 2006.

\bibitem{constant1}
C.~Daskalakis, A.~Mehta, and C.H. Papadimitriou.
\newblock A note on approximate {Nash} equilibria.
\newblock In {\em Proceedings of the 2nd Workshop on Internet and Network
  Economics}, pages 297--306, 2006.

\bibitem{DAS06}
C.~Daskalakis and C.H. Papadimitriou.
\newblock Three-player games are hard.
\newblock In {\em Electronic Colloquium in Computational Complexity}, TR05-139,
  2005.

\bibitem{DPS}
X.~Deng, C.~Papadimitriou, and S.~Safra.
\newblock On the complexity of price equilibria.
\newblock {\em Journal of Computer and System Sciences}, 67(2):311--324, 2003.

\bibitem{DederNazerzadehSaberi}
T.~Feder, H.~Nazerzadeh, and A.~Saberi.
\newblock Approximating nash equilibria using small-support strategies.
\newblock Stanford, 2006.

\bibitem{FISV}
K.~Friedl, G.~Ivanyos, M.~Santha, and F.~Verhoeven.
\newblock On the black-box complexity of {S}perner's lemma.
\newblock In {\em Proceedings of the 15th International Symposium on
  Fundamentals of Computation Theory}, pages 245--257, 2005.

\bibitem{GilboaZemel}
I.~Gilboa and E.~Zemel.
\newblock Nash and correlated equilibria: Some complexity considerations.
\newblock {\em Games and Economic Behavior}, 1(1).

\bibitem{GOL05}
P.W. Goldberg and C.H. Papadimitriou.
\newblock Reducibility among equilibrium problems.
\newblock In {\em STOC '06: Proceedings of the 38th Annual ACM Symposium on
  Theory of Computing}, pages 61--70, 2006.

\bibitem{HPV}
M.D. Hirsch, C.H. Papadimitriou, and S.~Vavasis.
\newblock {Exponential lower bounds for finding Brouwer fixed points}.
\newblock {\em Journal of Complexity}, 5:379--416, 1989.

\bibitem{NashSocial}
C.~A. Holt and A.~E. Roth.
\newblock The {Nash} equilibrium: A perspective.
\newblock {\em PNAS}, 101(12):3999--4002, March 2004.

\bibitem{HuangTeng}
L.-S. Huang and S.-H. Teng.
\newblock On the approximation and smoothed complexity of {L}eontief market
  equilibria.
\newblock In {\em Electronic Colloquium in Computational Complexity}, TR06-031,
  2006.

\bibitem{JohnsonColumn}
D.~Johnson.
\newblock The {NP}-completeness column: Finding needles in haystacks.
\newblock {\em ACM Transactions on Algorithms}, (to appear), April 2007.

\bibitem{KAKU}
S.~Kakutani.
\newblock A generalization of {Brouwer's} fixed point theorem.
\newblock {\em Duke Mathematical Journal}, 8:457--459, 1941.

\bibitem{KannanTheobald}
R.~Kannan and T.~Theobald.
\newblock Games of fixed rank: A hierarchy of bimatrix games.
\newblock In {\em SODA '07: Proceedings of the 18th Annual ACM-SIAM Symposium
  on Discrete Algorithms}, 2007.

\bibitem{Karmarkar}
N.~Karmarkar.
\newblock A new polynomial time algorithm for linear programming.
\newblock {\em Combinatorica}, 4:373--395, 1984.

\bibitem{KEA01}
M.~Kearns, M.~Littman, and S.~Singh.
\newblock Graphical models for game theory.
\newblock In {\em Proceedings of the Conference on Uncertainty in Artificial
  Intelligence}, pages 253--260, 2001.

\bibitem{KHA79}
L.G. Khachian.
\newblock A polynomial algorithm in linear programming.
\newblock {\em Doklady Akademia Nauk}, SSSR 244:1093--1096, {\it{English
  translation in Soviet Math. Dokl. 20, 191--194}}, 1979.

\bibitem{KleeMinty}
V.~Klee and G.J. Minty.
\newblock How good is the simplex algorithm?
\newblock In O.~Shisha, editor, {\em Inequalities -- {III}}, pages 159--175.
  Academic Press, 1972.

\bibitem{constant2}
S.~Kontogiannis, P.~Panagopoulou, and P.~Spirakis.
\newblock Polynomial algorithms for approximating {Nash} equilibria of bimatrix
  games.
\newblock In {\em Proceedings of the 2nd Workshop on Internet and Network
  Economics}, pages 286--296, 2006.

\bibitem{Lemke}
C.E. Lemke.
\newblock Bimatrix equilibrium points and mathematical programming.
\newblock {\em Management Science}, 11:681--689, 1965.

\bibitem{LemkeHowson}
C.E. Lemke and J.T. {Howson, Jr.}
\newblock Equilibrium points of bimatrix games.
\newblock {\em Journal of the Society for Industrial and Applied Mathematics},
  12:413--423, 1964.

\bibitem{Leonard}
R.J. Leonard.
\newblock Reading {Cournot}, reading {Nash}: The creation and stabilisation of
  the {Nash} equilibrium.
\newblock {\em Economic Journal}, 104(424):492--511, 1994.

\bibitem{Lipton}
R.J. Lipton, E.~Markakis, and A.~Mehta.
\newblock Playing large games using simple strategies.
\newblock In {\em Proceedings of the 4th ACM conference on Electronic
  commerce}, pages 36--41, 2004.

\bibitem{MegiddoLCP}
N.~Megiddo.
\newblock A note on the complexity of {P-matrix LCP} and computing an
  equilibrium.
\newblock {\em Research Report RJ6439}, IBM Almaden Research Center, San Jose,
  1988.

\bibitem{MEG91}
N.~Megiddo and C.H. Papadimitriou.
\newblock {On total functions, existence theorems and computational
  complexity}.
\newblock {\em Theoretical Computer Science}, 81:317--324, 1991.

\bibitem{MOR47}
O.~Morgenstern and J.~von Neumann.
\newblock {\em {Theory of Games and Economic Behavior}}.
\newblock Princeton University Press, 1947.

\bibitem{NAS50}
J.~Nash.
\newblock {Equilibrium point in n-person games}.
\newblock {\em Porceedings of the National Academy of the USA}, 36(1):48--49,
  1950.

\bibitem{NashNonCooperative}
J.~Nash.
\newblock Noncooperative games.
\newblock {\em Annals of Mathematics}, 54:289--295, 1951.

\bibitem{PAP91}
C.H. Papadimitriou.
\newblock {On inefficient proofs of existence and complexity classes}.
\newblock In {\em Proceedings of the 4th Czechoslovakian Symposium on
  Combinatorics}, 1991.

\bibitem{PAP94}
C.H. Papadimitriou.
\newblock {On the complexity of the parity argument and other inefficient
  proofs of existence}.
\newblock {\em Journal of Computer and System Sciences}, pages 498--532, 1994.

\bibitem{PapadimitriouInternet}
C.H. Papadimitriou.
\newblock Algorithms, games, and the internet.
\newblock In {\em STOC '01: Proceedings of the 33rd Annual ACM Symposium on
  Theory of Computing}, pages 749--753, 2001.

\bibitem{Sandholm2000}
T.~Sandholm.
\newblock Issues in computational vickrey auctions.
\newblock {\em International Journal of Electronic Commerce}, 4(3):107 -- 129,
  March 2000.

\bibitem{SavanivonStengel}
R.~Savani and B.~von Stengel.
\newblock Exponentially many steps for finding a {Nash} equilibrium in a
  bimatrix game.
\newblock In {\em FOCS '04: Proceedings of the 45th Annual IEEE Symposium on
  Foundations of Computer Science}, pages 258--267, 2004.

\bibitem{Scarf}
H.~Scarf.
\newblock The approximation of fixed points of a continuous mapping.
\newblock {\em SIAM Journal on Applied Mathematics}, 15:997--1007, 1967.

\bibitem{Scarfprice}
H.~Scarf.
\newblock On the computation of equilibrium prices.
\newblock In W.~Fellner, editor, {\em Ten Economic Studies in the Tradition of
  Irving Fisher}. New York: John Wiley \& Sons, 1967.

\bibitem{SPE28}
E.~Sperner.
\newblock {Neuer Beweis f\"ur die Invarianz der Dimensionszahl und des
  Gebietes}.
\newblock {\em Abhandlungen aus dem Mathematischen Seminar Universit\"at
  Hamburg}, 6:265--272, 1928.

\bibitem{SpielmanTengSimplex}
D.A. Spielman and S.-H. Teng.
\newblock Smoothed analysis of algorithms: Why the simplex algorithm usually
  takes polynomial time.
\newblock {\em Journal of the ACM}, 51(3):385--463, 2004, also in {\it STOC
  '01: Proceedings of the 33rd Annual ACM Symposium on the Theory of
  Computing}.

\bibitem{SpielmanTengSurvey}
D.A. Spielman and S.-H. Teng.
\newblock Smoothed analysis of algorithms and heuristics: Progress and open
  questions.
\newblock In {L. Pardo, A. Pinkus, E. S\"{u}li and M.J. Todd}, editor, {\em
  {Foundations of Computational Mathematics}}, pages 274--342. Cambridge
  University Press, 2006.

\bibitem{vN1928}
J.~von Neumann.
\newblock Zur theorie der gesellschaftsspiele.
\newblock {\em Mathematische Annalen}, 100:295--320, 1928.

\bibitem{Wilson}
R.~Wilson.
\newblock Computing equilibria of n-person games.
\newblock {\em SIAM Journal on Applied Mathematics}, 21:80--87, 1971.

\bibitem{YEwine2005}
Y.~Ye.
\newblock Exchange market equilibria with {L}eontief's utility: Freedom of
  pricing leads to rationality.
\newblock In {\em Proceedings of the 1st Workshop on Internet and Network
  Economics}, pages 14--23, 2005.

\end{thebibliography}
\appendix

\section{Perturbation and Probabilistic Approximation}
  \label{App:SmoothedApproximation}

In this section, we prove Lemma \ref{lem:smoothed}.
To help explain the probabilistic reduction from
  the approximation of bimatrix games to
  the solution of perturbed bimatrix games,
we first define the notion of many-way polynomial
  reductions among \textbf{TFNP} problems.

\begin{defi}[Many-way Reduction]\label{manyWayREDUCTION}
Let $\calF$ be a set of polynomial-time computable
   functions and $g$ be a polynomial-time computable
   function.
A search problem $\search{{R_1}}\in \emph{\textbf{TFNP}}$ is
  {\em ($\calF,g$)-reducible} to $\search{{R_2}}\in \emph{\textbf{TFNP}}$
  if, for all $y\in\setof{0,1}^*$,
   $(f(x),y)\in R_2$ implies $(x,g(y))\in R_1$
  for every input $x$ of $R_1$ and for every function $f\in \calF$.
\end{defi}
\begin{proof} ({\bf of Lemma \ref{lem:smoothed}})
We will only give a proof of the lemma under uniform perturbations.
With a slightly more complex argument to handle
   the low probability case when the absolute value of the
   perturbation is too large, we can similarly prove
   the lemma under Gaussian perturbations.

Suppose $J$ is an algorithm with polynomial smoothed complexity for
 {\sc Bimatrix}.
Let $T_J(\AA,\BB)$ be the complexity of $J$ for solving
  the bimatrix game defined by  $(\AA,\BB)$.
Let $N_{\sigma }()$ denotes the uniform perturbation with magnitude
  $\sigma $.
Then there exists constants  $c$, $k_{1}$ and $k_{2}$ such that
  for all $0 < \sigma  < 1$,

\[
\max_{\orig{\AA}, \orig{\BB}\in\Reals{n\times n}_{[-1,1]},
   } \expec{\AA\leftarrow
N_{\sigma } (\orig{\AA }),\BB\leftarrow N_{\sigma } (\orig{\BB
})}{T_J(\AA,\BB)} \leq c\cdot n^{k_{1}} \sigma^{-k_{2}}.
\]

For each pair of perturbation matrices $\SS, \TT
  \in \Reals{n\times n}_{[-\sigma,\sigma]}$,
  we can define a function $f_{(\SS,\TT)}: \Reals{n\times n}\times
  \Reals{n\times n} \rightarrow \Reals{n\times n}\times
  \Reals{n\times n}$ as
$f_{(\SS,\TT)} ((\orig{\AA} ,\orig{\BB})) = (\orig{\AA}+\SS,\orig{\BB}
  +\TT).
$
Let $\calF_{\sigma }$ be the set of all such functions, i.e.,
\[
\calF_{\sigma } = \setof{f_{(\SS,\TT)} | \SS, \TT
  \in \Reals{n\times n}_{[-\sigma,\sigma]}}.
\]

Let $g$ be the identity function from $\Reals{n}\times\Reals{n}$ to $\Reals{n}\times\Reals{n}$.

We now show that the problem of computing an $\epsilon $-approximate
  Nash equilibrium is $(\calF_{\epsilon /2},g)$-reducible to the
  problem of finding a Nash equilibrium of perturbed instances.
More specifically, we prove that
  for every bimatrix game $(\orig{\AA },\orig{\BB })$ and  for every
  $f_{(\SS,\TT)} \in \calF_{\epsilon /2}$,
  an Nash equilibrium  $(\xx, \yy)$ of $f_{(\SS,\TT)} ((\orig{\AA},\orig{\BB}))$
  is an $\epsilon$-approximate Nash equilibrium of $(\orig{\AA},\orig{\BB})$.

Let $\AA = \orig{\AA } + \SS$ and $\BB  = \orig{\BB }+ \TT$.
Then,
\begin{eqnarray}
& |{\xx^T \AA \yy - \xx^T \orig{\AA} \yy }| &  =
  |\xx^T \SS \yy| \leq \epsilon /2 \label{eqn:perb1}\\
&|\xx^T \BB \yy - \xx^T \orig{\BB} \yy| & =  |\xx^T \TT \yy|  \leq \epsilon /2.
  \label{eqn:perb2}
\end{eqnarray}

Thus, for each Nash equilibrium $(\xx, \yy)$ of $(\AA,\BB)$,
  for any $(\xx',\yy')$,
\begin{equation*}
(\xx')^T\orig{\AA}\yy - \xx^T \orig{\AA} \yy  \leq
  \left((\xx')^T\AA\yy - \xx^T \AA \yy\right) + \epsilon  \leq \epsilon.
\end{equation*}
Similarly, $\xx^T\orig{\BB}\yy' - \xx^T \orig{\BB} \yy  \leq \epsilon$.
Therefore, $(\xx, \yy)$ is an $\epsilon$-Nash equilibrium of
  game $(\orig{\AA}, \orig{\BB})$.

Now given the algorithm $J$ with polynomial smoothed time-complexity for
  {\sc Bimatrix},
  we can apply the following randomized algorithm (with the help of
  a $(\calF_{\epsilon /2},g)$-many-way reduction) to find an
  $\epsilon$-approximate Nash equilibrium of game $(\orig{\AA},\orig{\BB})$:

\begin{quotation}
\noindent {\bf Algorithm} {\tt  NashApproximationByPerturbations}$(\orig{\AA},\orig{\BB})$
\begin{itemize}
\item [1.] Randomly choose a pair of perturbation matrices $\SS,\TT$
           of magnitude $\sigma $ and set $\AA =\orig{\AA} +\SS$ and $\BB = \orig{\BB} + \TT$.

\item [2.] Apply algorithm $J$ to find a Nash equilibrium $(\xx,\yy)$ of $(\AA,\BB)$.

\item [3.] Return $(\xx,\yy)$.
\end{itemize}
\end{quotation}
The expected time complexity of {\tt  NashApproximationByPerturbation} is
   bounded from above by the smoothed complexity of $J$ when the
   magnitude perturbations is $\epsilon /2 $ and hence is at most
$2^{k_{2}} c\cdot n^{k_{1}} \epsilon ^{-k_{2}}.$
\end{proof}

\section{Padding Generalized Circuits: Proof of Theorem~\ref{gcexist}}\label{app:gcexist}

Suppose $\calS=(V,\calT)$ is a generalized circuit.
Let $K = |V|$.

First,  $\calS $ has a $1/K^3$-approximate
  solution because
1) {\sc Poly$^3$-Gcircuit} is reducible to {\sc
  Poly$^{12}$-Bimatrix}
  (\hspace{0.02cm}Section~\ref{GCtoTG}\hspace{0.02cm}); and
2) every two-player game has a Nash equilibrium.
Thus, the theorem is true for $c\leq 3$.

To prove the theorem for the case when $c>3$,
  we reduce {\sc Poly$^c$-Gcircuit} to {\sc Poly$^3$-Gcircuit}.
Suppose $c=2b+1$, where $b>1$.
We construct a new circuit $\calS'=(V',\calT')$ by inserting
  some dummy nodes into $\calS$ as following:
\begin{itemize}
\item $V\subset V'$, $|\hspace{0.04cm}V\hspace{0.04cm}|=K^b>K$ and
  $|\hspace{0.03cm}\calT'\hspace{0.03cm}| =
  |\hspace{0.03cm}\calT\hspace{0.03cm}|$\hspace{0.04cm};
\item For each gate $T=(G,v_1,v_2,v,\alpha)\in \calT$, if $G\notin \{\hspace{0.02cm}G_\zeta,
  G_{\times \zeta}\hspace{0.02cm}\}$ (\hspace{0.04cm}and thus,
  $\alpha=nil$\hspace{0.04cm}), then
  $T\in \calT'$; otherwise, gate $(G,v_1,v_2,v,K^{1-b}\alpha)\in \calT'$.
\end{itemize}
Let $\xx'$ be a $1/|\hspace{0.03cm}V'\hspace{0.03cm}|^3$-approximate solution of
  $\calS'$.
Note that $|V'|^{3} =1/K^{3b}$.
We construct an assignment $\xx:V\rightarrow \mathbb{R}$
  by setting $\xx[v]=K^{b-1}\xx'[v]$ for every $v\in V$.
One can easily check that $\xx$ is a $1/K^{2b+1}$-approximate
  solution to the original circuit $\calS$.
We then apply $1/K^{2b+1}=1/K^c$.

\section{Padding Bimatrix Games: Proof of Lemma~\ref{equivcc}}\label{app:padding}

Let $c$ be the constant such that {\sc Poly$^c$-Bimatrix}
  is known to be \textbf{PPAD}-complete.
If $c<2$, then finding an $n^{-2}$-approximate Nash
  equilibrium is harder, and thus is also complete in
  \textbf{PPAD}.
With this, without loss of generality, we
   assume that $c\ge 2$.
To prove the lemma, we only need to show that for every
  constant $c'$ such that $0<c'<c$, {\sc Poly$^c$-Bimatrix}
  is polynomial-time reducible to {\sc Poly$^{c'}$-Bimatrix}.

Suppose $\calG=(\AA,\BB)$ is an $n\times n$
  positively normalized two-player game.
We transform it into a new $n\times n$ game
  $\calG'=(\AA',\BB')$ as follows:
\begin{equation*}
a'_{i,j}=a_{i,j}+\Big(1-\max_{1\le k\le n}a_{k,j}\Big)\ \ \ \mbox{and}\ \ \ b'_{i,j}=b_{i,j}+\Big(1-\max_{1\le k\le
n}a_{i,k}\Big),\ \ \forall\ i,j:1\le i,j\le n.
\end{equation*}
One can verify that any $\epsilon$-approximate
  Nash equilibrium of $\calG'$ is also an
  $\epsilon$-approximate Nash equilibrium of $\calG$.
Besides, every column of $\AA'$ and every row
  of $\BB'$ has at least one entry with value $1$.

Next, we construct an $n''\times n''$ game $\calG''=
  (\AA'',\BB'')$ where $n''=n^\frac{2c}{c'}>n$ as follows:
$\AA''$ and $\BB''$ are both $2\times 2$ block
  matrices with $\AA''_{1,1}=\AA'$, $\BB''_{1,1}=\BB'$,
  $\AA''_{1,2}=\BB''_{2,1}=1$
  and $\AA''_{2,1}=\AA''_{2,2}=\BB''_{1,2}=\BB''_{2,2}=0$.
Now let $(\xx'',\yy'')$ be any $1/{n''}^{c'}=1/n^{2c}$-approximate
  Nash equilibrium of game $\calG''=(\AA'',\BB'')$.
By the definition of $\epsilon$-approximate Nash equilibria,
  one can show that $0\le \sum_{n<i\le n''}
  x''_i,\sum_{n<i\le n''} y''_i\le n^{1-2c}\ll 1/2$,
  since we assumed that $c\ge 2$.
Let $a=\sum_{1\le i\le n}x''_i$
and $b=\sum_{1\le i\le n} y''_i$.
We construct a
  profile of mixed strategies $(\xx',\yy')$ of $\calG'$ as follows:
  $x'_i=x''_i/a$ and $y'_i=y''_i/b$ for all
  $i \in [1:n]$.
Since $a,b>1/2$, one can show that
  $(\xx',\yy')$ is a $2/n^{2c}$-approximate
  Nash equilibrium of $\calG'$, which is
  also a $1/n^c$-approximate Nash equilibrium
  of the original game $\calG$.

\section{Gadget Gates: Complete the Proof of Lemma~\ref{INSERT}}\label{app:Gates}

\begin{proof}[Proof for \mbox{${G_\zeta}$} Gates]
From (\ref{eqn:set1}), (\ref{eqn:set2}) and Figure~\ref{PART2}, we have
\begin{eqnarray*}
&\form{\xx}{\bb^\calS_{2k-1}}-\form{\xx}{\bb^\calS_{2k}}=\oxx[v]-\alpha,\ \ \ \text{and}&\\ [0.5ex]
&\form{\aa^\calS_{2k-1}}{\yy}-\form{\aa^\calS_{2k}}{\yy}= \big(\hspace{0.04cm}\oyyc[v]-\oyy[v]
\hspace{0.04cm}\big)-\oyy[v].&
\end{eqnarray*}
If $\oxx[v]>\alpha+\epsilon$, then from the first equation,
  we have $\oyy[v]=\oyyc[v]$.
But the second equation implies $\oxx[v]=0$, which
  contradicts our assumption that $\oxx[v]>0$.

If $\oxx[v]<\alpha-\epsilon$, then from the first equation, we have
  $\oyy[v]=0$. But the second equation implies that
  $\oxx[v]=\oxxc[v]\ge 1/K-\epsilon$, which contradicts  the
  assumption that $\oxx[v]<\alpha-\epsilon$ and
  $\alpha\le 1/K$.
\end{proof}

\begin{proof}[Proof for \mbox{${G_{\times\zeta}}$} Gates]
From (\ref{eqn:set1}), (\ref{eqn:set2}) and Figure~\ref{PART2}, we have
\begin{eqnarray*}
&\form{\xx}{\bb^\calS_{2k-1}}-\form{\xx}{\bb^\calS_{2k}}=\alpha\hspace{0.05cm}\oxx[v_1]-\oxx[v],\ \ \ \text{and}&\\ [0.5ex]
&\form{\aa^\calS_{2k-1}}{\yy}-\form{\aa^\calS_{2k}}{\yy}= \oyy[v]-\big(\hspace{0.04cm}\oyyc[v]-\oyy[v]
\hspace{0.04cm}\big).&
\end{eqnarray*}
If $\oxx[v]>\min\hspace{0.04cm}(\hspace{0.04cm}\alpha\hspace{0.04cm}
  \oxx[v_1],1/K\hspace{0.04cm})+\epsilon$, then
  $\oxx[v]>\alpha\hspace{0.04cm}\oxx[v_1]+\epsilon$, since
  $\oxx[v]\le \oxxc[v]\le 1/K+\epsilon$.
By the first equation, we have $\oyy[v]=0$ and
  the second one implies that $\oxx[v]=0$, which
  contradicts the assumption that
  $\oxx[v]>\min\hspace{0.04cm}(\hspace{0.04cm}\alpha\hspace{0.04cm}
  \oxx[v_1],1/K\hspace{0.04cm})+\epsilon>0$.

If $\oxx[v]<\min\hspace{0.04cm}(\hspace{0.04cm}\alpha\hspace{0.04cm}
  \oxx[v_1],1/K\hspace{0.04cm})-\epsilon\le \alpha\hspace{0.04cm}\oxx[v_1]-\epsilon$,
  then the first equation shows $\oyy[v]=\oyyc[v]$ and thus
  by the second equation, we
  have $\oxx[v]=\oxxc[v]\ge 1/K-\epsilon$, which
  contradicts the assumption that $\oxx[v]<\min\hspace{0.04cm}
  (\hspace{0.04cm}\alpha\hspace{0.04cm} \oxx[v_1], 1/K\hspace{0.04cm})-\epsilon\le 1/K-\epsilon$.
\end{proof}

\begin{proof}[Proof for \mbox{${G_=}$} Gates]
$G_=$ is a special case of $G_{\times\zeta}$, with parameter $\alpha=1$.
\end{proof}

\begin{proof}[Proof for \mbox{${G_-}$} Gates]
From (\ref{eqn:set1}), (\ref{eqn:set2}) and Figure~\ref{PART2}, we have
\begin{eqnarray*}
&\form{\xx}{\bb^\calS_{2k-1}}-\form{\xx}{\bb^\calS_{2k}}= \oxx[v_1]-\oxx[v_2]-\oxx[v],\ \ \ \text{and}&\\ [0.5ex]
&\form{\aa^\calS_{2k-1}}{\yy}-\form{\aa^\calS_{2k}}{\yy}= \oyy[v]-\big(\hspace{0.04cm}\oyyc[v]-\oyy[v]
\hspace{0.04cm}\big).&
\end{eqnarray*}
If $\oxx[v]>\max\hspace{0.03cm} (\hspace{0.03cm}
  \oxx[v_1]-\oxx[v_2],0\hspace{0.03cm}) +\epsilon\ge
  \oxx[v_1]-\oxx[v_2]+\epsilon$, then the first equation implies
  $\oyy[v]=0$. By the second equation, we have
  $\oxx[v]=0$ which contradicts the assumption that
  $\oxx[v]>\max\hspace{0.03cm} (\hspace{0.03cm}\oxx[v_1]-
  \oxx[v_2],0 \hspace{0.03cm}) +\epsilon>0$.

If $\oxx[v]<\min\hspace{0.03cm}(\hspace{0.01cm}\oxx[v_1]
  -\oxx[v_2],1/K\hspace{0.01cm})-\epsilon\le
  \oxx[v_1]-\oxx[v_2]-\epsilon$, then by the first equation,
  we have $\oyy[v]=\oyyc[v]$. By the second equation, we have
  $\oxx[v]=\oxxc[v]\ge 1/K-\epsilon$,
contradicting the assumption that
  $\oxx[v]<\min\hspace{0.03cm}(\hspace{0.03cm}\oxx[v_1]
  -\oxx[v_2],1/K\hspace{0.03cm})-\epsilon\le 1/K-\epsilon$.
\end{proof}

\begin{proof}[Proof for \mbox{${G_<}$} Gates]
From (\ref{eqn:set1}), (\ref{eqn:set2}) and Figure~\ref{PART2}, we have
\begin{eqnarray*}
&\form{\xx}{\bb^\calS_{2k-1}}-\form{\xx}{\bb^\calS_{2k}}= \oxx[v_1]-\oxx[v_2],\ \ \ \text{and}&\\ [0.5ex]
&\form{\aa^\calS_{2k-1}}{\yy}-\form{\aa^\calS_{2k}}{\yy}= \big(\hspace{0.04cm}\oyyc[v]-\oyy[v]
\hspace{0.04cm}\big)-\oyy[v].&
\end{eqnarray*}
If $\oxx[v_1]<\oxx[v_2]-\epsilon$, then $\oyy[v]=0$ according to the
  first equation.
By the second equation, we have $\oxx[v]=\oxxc[v]=1/K\pm \epsilon$ and thus,
  $\oxx[v]=^{\hspace{0.06cm}\epsilon}_B 1$.

If $\oxx[v_1]>\oxx[v_2]+\epsilon$, then $\oyy[v]=\oyyc[v]$ according
  to the first equation.
By the second one, we
  have $\oxx[v]=0$ and thus, $\oxx[v]=^{\hspace{0.06cm}\epsilon}_B 0$.
\end{proof}

\begin{proof}[Proof for \mbox{${G_\lor}$} Gates]
From (\ref{eqn:set1}), (\ref{eqn:set2}) and Figure~\ref{PART2}, we have
\begin{eqnarray*}
&\form{\xx}{\bb^\calS_{2k-1}}-\form{\xx}{\bb^\calS_{2k}}= \oxx[v_1]+\oxx[v_2]-1/(2K),\ \ \ \text{and}&\\ [0.5ex]
&\form{\aa^\calS_{2k-1}}{\yy}-\form{\aa^\calS_{2k}}{\yy}= \oyy[v] - \big(\hspace{0.04cm}\oyyc[v]-\oyy[v]
\hspace{0.04cm}\big).&
\end{eqnarray*}
If $\oxx[v_1] \BOO= 1$ or $\oxx[v_2] \BOO= 1$, then
  $\oxx[v_1]+\oxx[v_2]\ge 1/K-\epsilon$.
By the first equation $\oyy[v]=\oyyc[v]$. By the
  second equation, we have
  $\oxx[v]=\oxxc[v]=1/K\pm \epsilon$ and thus,
  $\oxx[v]\BOO= 1$.

If $\oxx[v_1]\BOO=0$ and $\oxx[v_2]\BOO= 0$, then $\oxx[v_1]
  +\oxx[v_2]\le 2\epsilon$.
From the first equation, $\oyy[v]=0$. Then, the second equation implies
  $\oxx[v]\BOO= 0$.
\end{proof}

\begin{proof}[Proof for \mbox{${G_\land}$} Gates]
From (\ref{eqn:set1}), (\ref{eqn:set2}) and Figure~\ref{PART2}, we have
\begin{eqnarray*}
&\form{\xx}{\bb^\calS_{2k-1}}-\form{\xx}{\bb^\calS_{2k}}= \oxx[v_1]+\oxx[v_2]-3/(2K),\ \ \ \text{and}&\\ [0.5ex]
&\form{\aa^\calS_{2k-1}}{\yy}-\form{\aa^\calS_{2k}}{\yy}= \oyy[v] - \big(\hspace{0.04cm}\oyyc[v]-\oyy[v]
\hspace{0.04cm}\big).&
\end{eqnarray*}
If $\oxx[v_1]\BOO=0$ or $\oxx[v_2]\BOO=0$, then
  $\oxx[v_1]+\oxx[v_2]\le 1/K+2\epsilon$.
From the first equation, we have $\oyy[v]=0$. By the second equation, we have $\oxx[v]=0$ and thus,
  $\oxx[v]\BOO= 0$.

If $\oxx[v_1]\BOO=1$ and $\oxx[v_2]\BOO=1$, then
  $\oxx[v_1]+\oxx[v_2]\ge 2/K-2\epsilon$.
The first equation shows $\oyy[v]=\oyyc[v]$. By the second equation,
  $\oxx[v]=\oxxc[v]=1/K\pm \epsilon$ and thus, $\oxx[v]\BOO=1$.
\end{proof}

\begin{proof}[Proof for \mbox{${G_\lnot}$} Gates]
From (\ref{eqn:set1}), (\ref{eqn:set2}) and Figure~\ref{PART2}, we have
\begin{eqnarray*}
&\form{\xx}{\bb^\calS_{2k-1}}-\form{\xx}{\bb^\calS_{2k}}= \oxx[v_1]-\big(\hspace{0.04cm}\oxxc[v_1]-\oxx[v_1]
\hspace{0.04cm}\big),\ \ \ \text{and}&\\ [0.5ex] &\form{\aa^\calS_{2k-1}}{\yy}-\form{\aa^\calS_{2k}}{\yy}=
\big(\hspace{0.04cm}\oyyc[v]-\oyy[v] \hspace{0.04cm}\big)-\oyy[v].&
\end{eqnarray*}
If $\oxx[v_1]\BOO=1$, then by the first equation,
  $\oyy[v]=\oyyc[v]$. Then, by the second equation,
  we have $\oxx[v]=0$.

If $\oxx[v_1]\BOO=0$, then the first equation shows
  that $\oyy[v]=0$.
By the second equation, we have
  $\oxx[v]=\oxxc[v]$ and thus, $\oxx[v]\BOO=1$.
\end{proof}

\end{document}